\def\fermionuphalf{\begin{picture}(1,15)(0,0)
                         \put(0,0){\vector(0,1){7.5}}
                         \put(0,7.5){\line(0,1){7.5}}
                   \end{picture}
                  }

\def\fermiondown{\begin{picture}(1,30)(0,-30)
                       \put(0,0){\vector(0,-1){15}}
                       \put(0,-15){\line(0,-1){15}}
                 \end{picture}
                }
\def\fermiondownhalf{\begin{picture}(1,15)(0,-15)
                           \put(0,0){\vector(0,-1){7.5}}
                           \put(0,-7.5){\line(0,-1){7.5}}
                     \end{picture}
                    }

\def\fermionlefthalf{\begin{picture}(15,1)(0,0)
                           \put(15,0){\vector(-1,0){7.5}}
                           \put(7.5,0){\line(-1,0){7.5}}
                     \end{picture}
                    }

\def\fermionrighthalf{\begin{picture}(15,1)(0,0)
                            \put(0,0){\vector(1,0){7.5}}
                            \put(7.5,0){\line(1,0){7.5}}
                      \end{picture}
                     }
\def\fermionul{\begin{picture}(15,15)(0,0)
                        \put(0,0){\vector(-1,1){7.5}}
                        \put(-7.5,7.5){\line(-1,1){7.5}}
                  \end{picture}
                 }

\def\fermionur{\begin{picture}(15,15)(0,0)
                        \put(-15,-15){\vector(1,1){7.5}}
                        \put(-7.5,-7.5){\line(1,1){7.5}}
                  \end{picture}
                 }

\def\fermiondl{\begin{picture}(15,15)(0,0)
                        \put(15,15){\vector(-1,-1){7.5}}
                        \put(7.5,7.5){\line(-1,-1){7.5}}
                  \end{picture}
                 }

\def\fermiondr{\begin{picture}(15,15)(0,0)
                        \put(0,0){\vector(1,-1){7.5}}
                        \put(7.5,-7.5){\line(1,-1){7.5}}
                  \end{picture}
                 }

\def\gaugebosonright{\begin{picture}(30,1)(0,0)
                            \put(0,0){\line(1,0){0.75}}
                            \multiput(2.25,0)(3,0){9}{\line(1,0){1.5}}
                            \put(29.25,0){\line(1,0){0.75}}
                     \end{picture}
                    }
\def\gaugebosonrighthalf{\begin{picture}(15,1)(0,0)
                            \put(0,0){\line(1,0){0.75}}
                            \multiput(2.25,0)(3,0){4}{\line(1,0){1.5}}
                            \put(14.25,0){\line(1,0){0.75}}
                         \end{picture}
                        }
\def\gaugebosonup{\begin{picture}(1,30)(0,0)
                    \put(0,0){\line(0,1){0.75}}
                    \multiput(0,2.25)(0,3){9}{\line(0,1){1.5}}
                    \put(0,29.25){\line(0,1){0.75}}
                  \end{picture}
                 }
\def\gaugebosonuphalf{\begin{picture}(1,15)(0,0)
                            \put(0,0){\line(0,1){0.75}}
                            \multiput(0,2.25)(0,3){4}{\line(0,1){1.5}}
                            \put(0,14.25){\line(0,1){0.75}}
                      \end{picture}
                     }

\def\gaugebosondrhalf{\begin{picture}(15,15)(0,0)
                            \put(0,0){\line(1,-1){15}}
                  \end{picture}
                 }

\def\gaugebosondlhalf{\begin{picture}(15,15)(0,0)
                            \put(0,0){\line(-1,-1){15}}
                  \end{picture}
                 }

\newenvironment{Feynman}[3]{\begin{center}
                            \setlength{\unitlength}{#3 mm}
                            \begin{picture}(#1)(#2)
                            \thicklines
                           }{\end{picture} \end{center}}
\documentstyle[12pt,epsf]{article}
\newcommand{\SUBR}[1]{\bigskip\fbox{\tt #1}\bigskip}
\newcommand{\ALIBABA}{{\tt ALIBABA}}
\newcommand{\ZSHAPE}{{\tt ZSHAPE}}
\newcommand{\KORALZ}{{\tt KORALZ}}
\newcommand{\ZBIZON}{{\tt ZBIZON}}
\newcommand{\BHANG}{{\tt BHANG}}
\newcommand{\DIZET}{{\tt DIZET}}

\newcommand{\nn}{\noindent}

\newcommand{\bq}{\begin{equation}}
\newcommand{\eq}{\end{equation}}
\newcommand{\ba}{\begin{eqnarray}}
\newcommand{\ea}{\end{eqnarray}}
\newcommand{\naive}{na\"\i{}ve}
\newcommand{\LEPI}{LEP~I}
\newcommand {\zf}{$_{Z\!
F}\!I\!^{\textstyle T}\!\!T\!\!_
{{\textstyle E}\!R}$}

\newcommand{\ee}{$e^+e^-$}
\newcommand{\afb}{$A_{FB}$}
\newcommand{\st}{$\sigma_{T}\:$}
\newcommand{\oalf}{\mbox{${\cal O}(\alpha) \:$}}
\newcommand{\oalff}{\mbox{${\cal O}(\alpha)$}}
\newcommand{\oalz}{\mbox{${\cal O}(\alpha^2) \:$}}
\newcommand{\os}{\mbox{${\cal O}(\alpha \alpha_s) \:$}}
\newcommand{\oass}{\mbox{${\cal O}(\alpha_s^2) \:$}}
\newcommand{\ost}{\mbox{${\cal O}(\alpha \alpha_s m_t^2) \:$}}

\newcommand{\oaa}{\mbox{${\cal O}(\alpha^2 m_t^4) \:$}}

\newcommand{\BS}{\bigskip}

\newcommand{\Z}{$Z$}
\newcommand{\z}{$Z$}
\newcommand{\w}{$W$}
\newcommand{\zp}{$Z'$}

\newcommand{\IE}{ i.e. }
\newcommand{\EG}{{ e.g. }}
\newcommand{\SWE}{$
s_W^{2,{\mathrm{eff}}}
$}
\newcommand{\SWMS}{$\sin^2\theta_W^{\mathrm {\overline{MS}}}$}
\newcommand{\Sw}{$\sin^2\theta_W$}
\newcommand{\MZ}{$M_Z$}
\newcommand{\MT}{$m_t$}
\newcommand{\MH}{$M_H$}
\newcommand{\GAMZ}{$\Gamma_Z$}

\newcommand{\RS}{$\sqrt{s}$}
\newcommand{\BB}{$b\bar{b}$}
\newcommand{\FF}{$f\bar{f}$}

\newcommand{\MM}{$\mu^+\mu^-$}

\begin{document}
\def\theequation{\arabic{section}.\arabic{equation}}
%

%
%
%
%
\begin{titlepage}
\begin{flushright}
CERN-TH. 6443/92  \\
\end{flushright}
\begin{center}
\bigskip
{\LARGE \zf \\ \vspace*{.5cm}}
{\Large \bf
An Analytical Program for Fermion Pair \\
Production  in  \ee\  Annihilation
}

\bigskip\bigskip

D.~Bardin$^{1}$, \
M.~Bilenky$^{1,2,\dag}$, \
A.~Chizhov$^{1}$, \
O.~Fedorenko$^{3}$, \     
S.~Ganguli$^{4}$, \        \\
A.~Gurtu$^{4}$, \
M.~Lokajicek$^{1}$, \
G.~Mitselmakher$^{1}$, \                     
A.~Olshevsky$^{1}$, \
J.~Ridky$^{1}$, \            \\
S.~Riemann$^{5,\ddagger}$, \
T.~Riemann$^{5,6}$, \
M.~Sachwitz$^{5}$, \         
A.~Sazonov$^{1}$, \
A.D.~Schaile$^{7}$, \        \\
Yu.~Sedykh$^{1}$, \
I.~Sheer$^{8}$, \
L.~Vertogradov$^{1}$  \\
\BS
{\small
$^{1}$
Joint Institute for Nuclear Research, Dubna, Russia  \\
\medskip
$^{2}$
Universit\"at Bielefeld, Germany \\
\medskip
$^{3}$
Petrosavodsk State University, Petrosavodsk, Russia  \\
\medskip
$^{4}$
Tata Institute of Fundamental Research, Bombay, India \\
\medskip
$^{5}$
DESY -- Institut f\"{u}r Hochenergiephysik, Zeuthen, Germany \\
\medskip
$^{6}$
Theory Division, CERN, Geneva, Switzerland  \\
\medskip
$^{7}$
Albert-Ludwigs-Universit\"{a}t, Freiburg, Germany \\
\medskip
$^{8}$
University of California, San Diego, USA \\
} 
\bigskip
\end{center}
\vfill
\begin{center}
{\bf\large Abstract}
\end{center}
{\small
We describe how to use \zf, a program based on a semi-analytical
approach to fermion pair production in \ee\ annihilation and
Bhabha scattering.
A flexible treatment of complete ${\cal O}(\alpha)$
QED corrections, also including higher orders,
allows for three calculational {\bf chains} with
different realistic sets of
restrictions in the photon phase space.
\zf \ consists of several {\bf branches} with varying assumptions
on the underlying hard scattering process.
One includes complete ${\cal O}(\alpha)$ weak loop corrections with
a resummation of leading higher-order terms.  
Alternatively, an ansatz inspired from
S-matrix theory, or several model-independent effective Born cross
sections may be convoluted.
The program calculates cross sections, forward-backward asymmetries, and
for $\tau$~pair production also the final-state polarization.
Various {\bf interfaces} allow fits to be performed
with different sets of free parameters.
}  

\bigskip

\hrule width 6.cm
\vskip6pt
{\small
$^{\dag}$ Alexander-von-Humboldt Fellow

$^{\ddagger}$
Partly supported by the German Bundesministerium f\"ur Forschung
und Technologie} \hfill
\bigskip

\begin{flushleft}
  CERN-TH. 6443/92  \\
  May 1992      \\
\end{flushleft}
\end{titlepage}
%
%
\tableofcontents \newpage
\listoffigures  
\listoftables    \newpage
\newpage

\section
[Introduction]
{Introduction
\label{intro}}
\setcounter{equation}{0}
There is a growing demand for flexible programs to fit the very
precise data on fermion pair production
from experiments at the \ee  storage ring \LEPI:
\begin{equation}
e^+e^- \longrightarrow f {\bar f}({\mathrm{n}} \gamma),
\label{firsteq}
\end{equation}
including Bhabha scattering,
\begin{equation}
e^+e^- \longrightarrow e^+e^-(\mathrm{n}\gamma).
\label{seconeq}
\end{equation}
It is important that such programs allow for model-independent and
Standard Model~\cite{no:gws} fits to the data.
In addition, it is interesting to be able to fit the data to theories
that
go beyond the Standard Model.
Because experimental cuts tend to be more complicated than can be
realized with semi-analytic programs, typically Monte Carlo programs
are used to correct for such cuts and detector inefficiencies
before fitting.

In this article, we describe the subroutine package \zf.
This program
\cite{v4authors} is based on a semi-analytical approach to the radiative
corrections that
are needed for the analysis of reactions (\ref{firsteq}) and
(\ref{seconeq}).

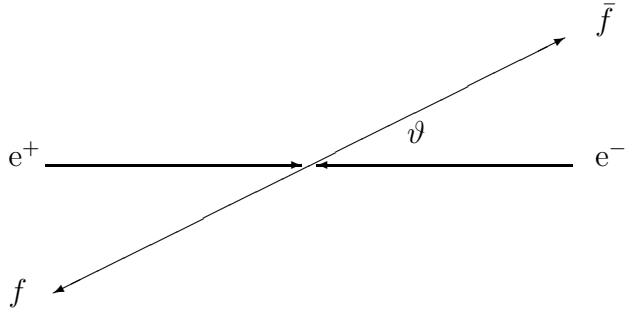
\begin{figure}[bhtp]
\setlength{\unitlength}{1mm}
\begin{picture}(140,45)(-20,0)
\put(30,20){\vector(1,0){34}}
\put(100,20){\vector(-1,0){34}}
\put(65,20.0){\vector(-2,-1){34}}
\put(65,20.0){\vector(2,1){34}}
\put(78,23){$\vartheta$}
\put(25,20){e$^+$}
\put(25,2){$f$}
\put(103,20){e$^-$}
\put(103,38){$\bar{f}$}
\end{picture}
\caption
[Scattering angle $\vartheta$ in \ee annihilation.]
{{\it Scattering angle $\vartheta$ in \ee annihilation.}}
\label{fig.theta}
\end{figure}

The \zf \ package employs an approach which relies on
formulae that are analytically integrated over a finite angular region
with respect to the scattering angle, as shown in fig.~\ref{fig.theta}.
The program directly calculates predictions for observable quantities
and {\em not} corrections to Born approximations.
The total cross section, $\sigma_T$,
and the forward-backward asymmetry, $A_{FB}$, may be calculated
in a non-symmetric angular interval,
$c_1 < \cos \vartheta <c_2$:
\bq
\sigma_T(c_1,c_2)
=  \int_{c_1}^{c_2} d\cos\vartheta
                     \frac{d\sigma}{d\cos\vartheta},
\label{sigmatn}
\eq
\bq
A_{FB}(c_1,c_2)
=  \frac{\sigma_{FB}(c_1,c_2)}{\sigma_T(c_1,c_2)},
\label{afbafb}
\eq
where
\bq
\sigma_{FB}(c_1,c_2)
=
\left[  \int_{0}^{c_2} d\cos\vartheta
      - \int_{c_1}^{0} d\cos\vartheta \right]
    \frac{d\sigma}{d\cos\vartheta}.
\label{sigafbn}
\eq
These expressions are constructed from the following integrals:
\bq
\sigma(0,c) \equiv
\int_{0}^{c} d\cos\vartheta     \frac{d\sigma}{d\cos\vartheta}
= \frac{1}{2} \left[ \sigma_T(c) + \sigma_{FB}(c) \right].
\label{binsec}
\eq
Here $\sigma_T(c)$ and $\sigma_{FB}(c)$ are two special cases
of $\sigma_A(c_1,c_2)$, which will be the basis of the discussion in the
following chapters:
\bq
\sigma_T(c)
=  \int_{-c}^{c} d\cos\vartheta
                     \frac{d\sigma}{d\cos\vartheta},
\label{sigmat}
\eq
\bq
\sigma_{FB}(c)
=
\left[  \int_{0}^{c} d\cos\vartheta
      - \int_{-c}^{0} d\cos\vartheta \right]
    \frac{d\sigma}{d\cos\vartheta}.
\label{sigafb}
\eq
By simple algebraic combinations of the above constructs,
one may derive various measurable cross sections and asymmetries.
One must, of course, take into account the
possible beam polarizations and final-state helicities
within the hard subprocess description ($\sigma^o_{T,FB}$)
as explained below.

For reaction (\ref{firsteq}), excluding  Bhabha scattering which will be
discussed in section \ref{subsec:bhabha},
both functions $\sigma_T$ and $\sigma_{FB}$ may be split into
different contributions
from initial-state radiation, $\sigma^{\mathrm{ini}}$,
final-state radiation, $\sigma^{\mathrm{fin}}$, and their
interference, $\sigma^{\mathrm{int}}$ ($A=T,FB$):
\bq
\sigma_A(c) =
\sigma_A^{\mathrm{ini}}(c)
+
\sigma_A^{\mathrm{ int}}(c)
            +
            \sigma_A^{\mathrm{ fin}}(c).
\label{eif}
\eq
{\em Common} soft photon exponentiation for initial- and
final-state radiation, which relies on
a more compact (but also more sophisticated)
formula, has been realized in \zf:
\bq
\sigma_A(c) =
\sigma_A^{\mathrm{ ini + fin}}(c)  + \sigma_A^{\mathrm{  int}}(c).
\label{efexp}
\eq
Alternatively,
the program allows the user to choose a simplified treatment of the
(small) contribution from final-state radiation:
\bq
\sigma_A(c) = \sigma_A^{\mathrm{ini}}(c)  \left( 1 + \frac{3}{4}
\frac{\alpha}{\pi} Q_f^2 \right)
  + \sigma_A^{\mathrm{ int}}(c).
\label{efast}
\eq
The expressions introduced in (\ref{efexp}) and in (\ref{efast})
are realized in \zf  \
as one-dimensional numeric integrations over a photon phase space
variable $s'$.

Photonic corrections to the cross sections and asymmetries are
implemented
by convoluting the Born cross sections $(\sigma^{a,o}_A)$ with radiator
functions $(R^a_A)$:
\bq
\sigma_A^{a}(c) = \frac{1}{d_A} \Re e  \int_0^{\Delta} dv
         \sigma_A^{a,o}(s,s') R_A^{a}(v,c),
\label{sigexa}
\eq
where $a$ = ini,ini+fin,int; $d_T=\frac{4}{3}, d_{FB}=1$;
$s'=(1-v)s$; and $v$ is the energy of the
radiated photon in units of the beam energy.
Further, $\sigma^o_A$ contains the dynamics of the basic
process to be studied, and the functions $R^a_A$ depend on the treatment
of the QED effects.
There are several ways to describe \zf.
It contains:
\begin{itemize}
\item three calculational {\em chains} with a different handling
      of QED corrections plus the Bhabha {\em chain},
\item four {\em branches} which differ by the
      theoretical description of the hard scattering process,
\item seven {\em interfaces} with different choices of
      input/output parameters.
\end{itemize}

\subsection
[\zf\ Chains]
{\zf\ Chains
\label{subsec:chains}  }
\bigskip\noindent{\bf
No cuts - a fast option }   \\
In this chain,
the cross sections are calculated with formulae that assume
that there are no cuts applied to the photon phase space.

\bigskip\noindent{\bf
Phase-space cut on the minimum invariant mass of the $f\bar{f}$ pair} \\
The underlying formulae may be found in \cite{ba:phe},\cite{ba:pl90}
and in references quoted therein. This chain allows for a cut on the
minimum invariant mass of the final-state $f\bar{f}$ pair, which can
be reinterpreted as a cut on the maximum of the allowed energy of the
bremsstrahlung photon.


\bigskip\noindent{\bf
Cuts on energies and acollinearity of final-state fermions} \\
This treatment of the photon phase space follows the basic lines of
that of the above chain.
The restriction on the maximal photon energy
is replaced by a simultaneous
cut on both the energies of the produced fermions
and on their acollinearity~\cite{bs:dp}.
This chain also allows the calculation of differential and
integrated cross sections for Bhabha scattering
using the {\tt BHANG} package~\cite{bhabha}, which
has been incorporated into \zf.

\bigskip

Furthermore, in both the latter chains
one can impose a restriction on the maximum production angle
of the outgoing antifermion\footnote{
As a matter of convention this cut is imposed on the antifermion only.
Because of CP invariance the cut could equally well be applied on the
fermion instead.}.

\subsection
[\zf\ Branches]
{\zf \, Branches
\label{subsec:branches}  }

\bigskip\noindent{\bf
Analytic Standard Model formulae with higher-order corrections} \\
This is the central branch of the program.
The calculations of the partial and total \Z \, and $W$ widths follow
\cite{zwidth} and \cite{wwidth} respectively.
The explicit formulae for the improved Born cross sections with
electroweak corrections are described in \cite{rokap}.
In addition, improvements have been realized in the program by
including various higher-order corrections, which will be described
in detail later.
The electroweak loop corrections are determined in \zf\ using
the {\tt DIZET} package~\cite{di:cpc} for all channels including
Bhabha scattering~\cite{mp:bhr}.

\bigskip\noindent{\bf
Model-independent ansatz using effective couplings} \\
This approach
assumes that the effective axial-vector and vector couplings
of fermions to the \Z\ are real, constant, process- and
energy-independent as in \cite{modind1,yrafb}.
It is known from comparisons with Standard Model
predictions that these assumptions allow for quite a good approximation.

\bigskip\noindent{\bf
Model-independent ansatz using partial decay widths} \\
Following general arguments of field theory, one can describe resonance
scattering with the help of the partial decay widths of the resonance.
This is particularly advantageous since measuring the \Z \, line shape
allows for a very precise determination of the partial decay widths.
A compact description of the underlying
formalism may be found in \cite{modind1}.

\bigskip\noindent{\bf
S-matrix ansatz} \\
The cross section ansatz due to general
S-matrix ideas as described in \cite{smatrix} has been implemented in the
program.
The main advantage of this branch is that it gives the mass and total
width of the \Z\ with minimal assumptions on the underlying dynamics.
As with the other branches
some additional degrees of freedom are available;
however, the physical
information, which can be extracted from them is limited.

\bigskip

The advantage of
the various model-independent branches is the simple picture
of the dynamics and the gain in flexibility compared with the Standard
Model.
Such model-independent approaches also allow for pragmatic checks of
Standard Model predictions and practical gains in computing time.
If different branches give statistically significant differences with
respect to the various parameters then perhaps a strong indication of
New Physics exists!

\bigskip

Thanks to the
flexibility of the convolution approach to QED corrections in \zf,
it is relatively easy
to make different assumptions on the hard scattering
process.
While the above branches
cover some of the most important theoretical tools for LEP I physics,
it is a straight-forward job to add new branches
so that predictions for New Physics can be made
within the \zf\ framework.
One such example is described in \cite{zefit}, where
the mixing of the \Z\ with an additional heavy $Z'$ is implemented.
In addition, other possibilities, which cover some New Physics
by extensions of the weak form factors will also be discussed.

\subsection
[\zf\ Interfaces]
{\zf \, Interfaces
\label{subsec:interfaces}  }

\BS\nn{\bf Subroutine ZUTHSM} \\
Calculation of Standard Model
{\em cross sections} and {\em forward--backward
asymmetries} as functions of $M_Z, m_t, M_H$, and $\alpha_s$.

\BS\nn{\bf Subroutine ZUTPSM} \\
Calculation of Standard Model
{\em $\tau$~polarization}, $A_{\mathrm{pol}}$, and
{\em $\tau$~polarization for\-ward-back\-ward
asymmetry, $A^{\mathrm{pol}}_{FB}$},
as functions of $M_Z, m_t, M_H$, and $\alpha_s$.

\BS\nn{\bf Subroutine ZUXSA} \\
Calculation of model-independent
{\em cross sections} and {\em asymmetries}
as functions of the normalization form factors ($\hat{\rho}$),
effective vector ($\hat{v}$) and axial-vector ($\hat{a}$) couplings,
respectively.

\BS\nn{\bf Subroutine ZUTAU} \\
Calculation of
model-independent {\em final-state polarization} in $\tau$~pair
production as functions of the normalization form factors, effective
vector and axial-vector couplings.

\BS\nn{\bf Subroutine ZUXSA2} \\
Calculation of
model-independent {\em cross sections} and {\em asymmetries}
as functions of the squares of the normalization form factors, effective
vector and axial-vector couplings.

\BS\nn{\bf Subroutine ZUXSEC} \\
Calculation of model-independent {\em cross sections} as functions of the
partial ($\Gamma_{f}$) and total \Z\ widths.

\BS\nn{\bf Subroutine ZUSMAT} \\
Calculation of model-independent {\em cross sections}, based on an
S-matrix inspired ansatz, as functions of $M_Z, \Gamma_Z$, etc.

\BS

The above interfaces have been designed with the analysis of LEP I data
in mind.
In fact, the accuracy of the Standard Model branch of \zf\ has been
optimized near the \Z\ pole.
Nevertheless, the Standard Model branch of the package can be used at
PETRA, TRISTAN, and linear collider energies without changes.
Many examples of the use of the \zf\ package exist in the
literature (see e.g.~\cite{pl:lep}.
We will thus make no attempt to describe
how to use \zf\ to fit data.

The organization of the article is as follows:  section~\ref{chains}
describes the treatment of photonic corrections;
the description
of the various theoretical treatments of the hard scattering
process is given in section~\ref{branches.1} for the Standard Model {\em
branch} and in section~\ref{branches.2} for the other {\em branches};
the search for effects of New Physics with \zf\ is commented in
section~\ref{subbeyo};
initialization is described in section~\ref{init};
section~\ref{interfaces} documents the interface structure;
finally section~\ref{compa0}
compares \zf\ results with those of other programs
for weak mixing angles
{}~\cite{statho1,dfs}
and  widths~\cite{statho2,ds},
and also cross sections in an energy range covering
both PETRA and LEP~I energies and beyond
({\tt ZSHAPE}~\cite{zshape,zshap2}, {\tt ALIBABA}~\cite{alibaba}).
The appendices contain a description of the contents of some of the
common blocks of \zf \ and an example of the use of the package.

\section
[Chains of \zf: Photonic Corrections with Different Cuts]
{Chains of \zf: \\
Photonic Corrections with Different Cuts
\label{chains}
}
\setcounter{equation}{0}

In this section, we will describe the functions $\sigma_T(c)$ and
$\sigma_{FB}(c)$, which were introduced in (\ref{sigmat}) and
(\ref{sigafb}).
A complete treatment of photonic corrections would also include
the running of the electromagnetic coupling
constant, which will be discussed in section \ref{branches.1}, on the
hard subprocess description.

\begin{figure}[bhtp]
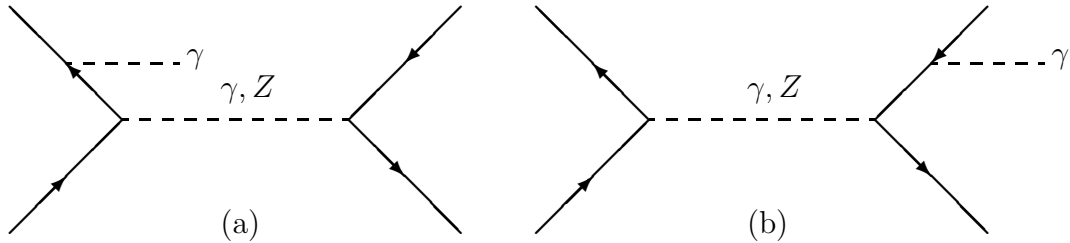

\begin{Feynman}{130,30}{0,0}{1.0}
\put(15,15){\fermionur}
\put(15,15){\fermionul}
\put(7.5,22.5){\gaugebosonrighthalf}
\put(23.5,22.5){$\gamma$}
\put(15,15){\gaugebosonright}
\put(28,18){$\gamma,Z$}
\put(45,15){\fermiondl}
\put(45,15){\fermiondr}
\put(28,0){(a)}
\put(85,15){\fermionur}
\put(85,15){\fermionul}
\put(85,15){\gaugebosonright}
\put(98,18){$\gamma,Z$}
\put(115,15){\fermiondl}
\put(115,15){\fermiondr}
\put(122.5,22.5){\gaugebosonrighthalf}
\put(138.5,22.5){$\gamma$}
\put(98,0){(b)}
\end{Feynman}
\caption
[
Real photon emission.]
{\label{fig1paw}
{\it
Real photon emission from initial (a) and final (b) states.}
}
\label{fig.realbr}
\end{figure}

\begin{figure}[thbp]
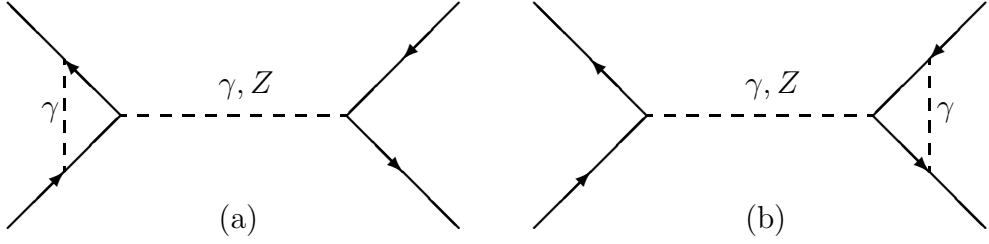

\begin{Feynman}{130,30}{0,0}{1.0}
\put(15,15){\fermionur}
\put(15,15){\fermionul}
\put(7.5,7.5){\gaugebosonuphalf}
\put(4.5,15){$\gamma$}
\put(15,15){\gaugebosonright}
\put(28,18){$\gamma,Z$}
\put(45,15){\fermiondl}
\put(45,15){\fermiondr}
\put(28,0){(a)}
\put(85,15){\fermionur}
\put(85,15){\fermionul}
\put(85,15){\gaugebosonright}
\put(98,18){$\gamma,Z$}
\put(115,15){\fermiondl}
\put(115,15){\fermiondr}
\put(122.5,7.5){\gaugebosonuphalf}
\put(123.5,15){$\gamma$}
\put(98,0){(b)}
\end{Feynman}
\caption[
Photonic vertex corrections.]
{\label{fig2paw}
{\it
The photonic vertex corrections for the initial (a) and final (b) states.
}}
\label{fig.virtph}
\end{figure}

\begin{figure}[thbp]
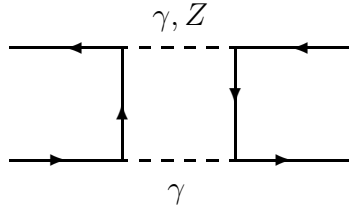

\begin{Feynman}{45,30}{0,0}{1.0}
\put(0,5){\fermionrighthalf}
\put(0,20){\fermionlefthalf}
\put(15,5){\fermionuphalf}
\put(15,5){\gaugebosonrighthalf}
\put(21,0){$\gamma$}
\put(15,20){\gaugebosonrighthalf}
\put(19,23){$\gamma,Z$}
\put(30,5){\fermiondownhalf}
\put(30,5){\fermionrighthalf}
\put(30,20){\fermionlefthalf}
\end{Feynman}
\caption[
Box diagrams with virtual photons.
]{\label{fig3paw}
{\it
Box diagrams with virtual photons, which combine with the
initial-final interference bremsstrahlung into finite and gauge-invariant
contributions.
}}
\label{fig.virtbx}
\end{figure}

In order to get a finite, gauge-invariant result,
real photon bremsstrahlung from the diagrams of fig.
\ref{fig.realbr} has to be combined with photonic vertex
corrections of fig.~\ref{fig.virtph}
for initial- or final-state radiation and
for their interference
with the box-diagram corrections of
fig.~\ref{fig.virtbx}.


\zf \ relies on the following numerical
integration for the contributions introduced in (\ref{efexp}).
The common soft photon exponentiation of initial- and
final-state radiation is taken into account with:
\bq
\sigma_A^{\mathrm{ini + fin}}(c) = \frac{1}{d_A} \Re e
\int_0^{\Delta} dv
         \sigma_A^o(s')
         R_A^{\mathrm{ ini}}(v,c)
        {\bar R}_A^{\mathrm{ fin}}(v),
\label{siginifin}
\eq
where
$s'=(1-v)s$.
Final-state radiation is described by
${\bar R}^{\mathrm{ fin}}_A$, which
is more complex than a simple angular integral
of $R_A^{\mathrm{fin}}(v,c)$; in~\cite{ba:phe}
it has been shown that ${\bar R}^{\mathrm{fin}}_A$
is almost completely angle-independent.
For each of the cross section parts,
the contributions from $\gamma$ and \Z \, exchange and
from their interference can be separated:
\bq
\sigma_A^o(s) = \sum_{m,n} \sigma_A^o(s;m,n) \equiv
\sigma_A^o(s;\gamma,\gamma) +
\sigma_A^o(s;\gamma,Z) +
\sigma_A^o(s;Z,Z).
\label{giz}
\eq
For the interference of initial and final states, this decomposition is
unavoidable.
This is due to the differences in the $\gamma\gamma$ and $\gamma Z$
boxes in fig.~\ref{fig.virtph}, which regularize the infared divergence:
\bq
\sigma_A^{\mathrm{ int}}(c) = \frac{1}{d_A} \Re e  \int_0^{\Delta} dv
         \sum_{m,n}  \sigma_A^o(s,s';m,n)
         R_A^{\mathrm{int}}(v,c;m,n),
\label{sigint}
\eq
where $m,n = \gamma,Z$.
The origin of the complex structure of the initial-final
interference bremsstrahlung contribution is two-fold.
First, the cross section
part originates from the interference of matrix elements with
emission of a photon {\em before} and one {\em after}
the hard-scattering
process.
This leads to the dependence of the hard-scattering cross
section, $\sigma_A^o$, on both $s$ and $s'$.
Secondly, the virtual corrections of initial- and final-state
radiation or of the interference have different structure
(fig.~\ref{fig.realbr}).
The simple vertex diagrams (fig.~\ref{fig.virtph}) of the former
factorize into the Born
cross section and a universal factor, while
the box diagrams (fig.~\ref{fig.virtbx})
with two-particle exchange do not.
This leads to a dependence of the interference radiator functions,
$R^{\mathrm{int}}_A$, on $m,n$.

The radiator functions $R_A^a(v,c)$ are the result of a three-fold
analytic integration of the corresponding photon phase space:
\bq
R_A^a (v, c, m, n) =
 \int dv_2  \, \int d\cos\vartheta \, \int d \phi_{\gamma} \,
\chi^a_A(s,v,v_2, \cos\vartheta, \phi_{\gamma}),
\label{symb}
\eq
where $\chi_A^a$ is the result of a Feynman diagram calculation.
Further, $s'=Rs =(1-v)s=M_{{\bar f} f}^2$ is the invariant
mass of the fermion
pair, $v_2=M_{{f}\gamma}^2/s$
and $\phi_{\gamma}$ is one of the photon angles in the
($\gamma,{f}$) rest system.
Two  treatments of the photon phase space are realized in \zf.
These are shown below in the Dalitz plots of figs.~\ref{dalitz1}
and~\ref{dalitz2}.
The variable $v_2$ has been integrated over analytically, while $R$
remains to be numerically integrated by \zf.
Note that the corner of the photon phase space, which corresponds to the
emission of a soft photon is located near $R=1$.

As may be seen from the definitions (\ref{sigmat}) and (\ref{sigafb}),
the angular acceptance cut, $c_1 \leq \cos \vartheta \leq c_2$,
limits the scattering angle $\vartheta$ of the final-state
antifermions (see fig.~\ref{fig.theta}).
In this case, the
scattering angle of the fermion  $f$  remains unrestricted if the other
cut(s) do not imply an implicit restriction (see section
  \ref{subsec:acol}).

In \zf, the QED contributions include the complete \oalf cor\-rect\-ions
and soft photon exponentiation.
It should be mentioned that the
radiator functions (flux factors), $R_A^a$,
{\em differ} for different
observables ($A=T, FB$) and for different bremsstrahlung origin
($a={\mathrm{ini,fin,int}}$).
In addition, the radiator functions for the integrated cross sections and
the differential cross sections are not the same~\cite{ba:phe}.
Only at LEP~I, around the \Z \,
resonance, do all the radiator functions agree approximately
{}~\cite{riwa,afb:pl}.
Some other semi-analytic programs use equal
radiator functions for the total cross section and for $A_{FB}$.
At LEP~I energies,
where hard photon emission is suppressed, and for loose cuts
(thus not enhancing the initial-final interference terms), this is
numerically acceptable.
\zf, however, uses
the {\em correct} radiator functions; the underlying
formalism thus allows an application of \zf \ at energies far away from
the \Z \, peak.
No part of our treatment of the bremsstrahlung
is specific to physics near the \Z\ resonance peak.

Higher order QED corrections have been implemented in \zf\ for
initial-state radiation contributions, besides the above-mentioned
soft photon exponentiation, to \st as in \cite{zshape} and to \afb\ as in
\cite{ringberg}.
For the two calculational chains, which involve an
angular acceptance cut, these higher-order corrections
are treated with an approximation that assumes a Born-like
angular behavior.

When no acceptance cut is applied, $c=1$,
the expressions for $\sigma_T(1)$ and  $A_{FB}(1)$ approach
well-known formulae for $\sigma_T$ \cite{boma,greco,zshape} and
\afb~\cite{afb:pl,ringberg}.

\subsection
[No Cuts]
{No Cuts
\label{subsec:nocut} }

Of the various calculational chains contained in the \zf\ package
the simplest to describe is the one where no cuts are allowed.
This chain has been realized with special formulae in order to make
it as computationally fast as possible.
Here the photon may have any energy, $\Delta$, up to the kinematic limit:
\bq
\Delta \equiv E_{\gamma}^{\max} / E_{\mathrm{beam}},
\hspace{1.cm} \Delta \leq  \Delta^{\max} =
                          1-4 m_f^2/s.
\label{cuts2}
\eq
Thus, the radiative corrections depend on the fermion masses even for the
light quarks and leptons.
This dependence can be important when total cross sections
are determined from experimental data, and {\it is} of special importance
when comparing results from other semi-analytic
programs.
The latter will be discussed in section \ref{compa2}.


\subsection
[Convolution Integral with Cut on the
Invariant Mass of the Outgoing Fermion
Pair ($s'_{\min}$)]
{Convolution Integral with Cut on the
Invariant Mass of the Outgoing Fermion
Pair ($s'_{\min})$
\label{subsec:delta}   }
\begin{figure}[htbp]
\begin{center} \mbox{
\epsfysize=5.0cm
\epsffile{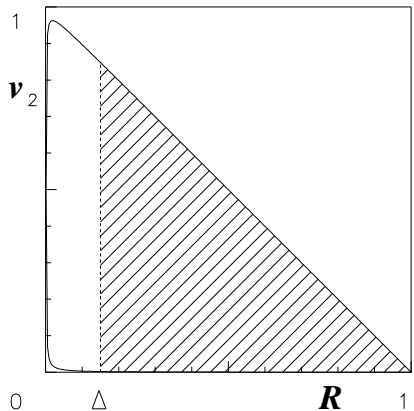}}
\end{center}
\caption[
Dalitz plot for the photon phase space with a cut on $s'$.
]{\label{fig4paw}
{\it
Dalitz plot for the photon phase space with a cut on $s'$.
}  }
\label{dalitz1}
\end{figure}

\zf\ allows for a constraint on the minimum allowed invariant mass of the
outgoing fermion pair, $s'_{\min}$:
\bq
s'_{\min} = (1 - \Delta) s.
\label{spmin}
\eq
This is easily re-interpreted as a cut on the maximum
allowed energy of the bremsstrahlung photon, $\Delta$.
In this calculational chain the $s'_{\min}$ cut may be combined with an
angular acceptance cut:
\bq
 c_1 \leq \cos \vartheta \leq c_2.
\label{cuts1}
\eq
The Dalitz plot
shown in fig.~\ref{dalitz1} corresponds to a $\Delta$ cut
of $\Delta=1-R^{\min}$.

\vspace{.5cm}

The allowed region is a triangle in the
ultra-relativistic limit.
Note that $v_2$ is not influenced by the cut.
This simplifies the analytical integration over $v_2$.

Explicit expressions for the radiator functions $R^a_A(v,c)$
discussed above may be found in the literature.
For initial-state radiation and initial-final interference,
they may be found in eqs.~(8), (18) in~\cite{ba:pl90}, respectively.
For final
state radiation, the angular dependence is relatively simple and
eqs.~(132-134) in~\cite{ba:phe} are valid.
The radiator functions for common exponentiation of initial-
 and
final-state
 soft-photon emission implemented in \zf\ are derived from
 eq.~(157) in~\cite{ba:phe}, as has been described in
 section~4 of~\cite{ba:pl90}.

\subsection
[Convolution Integral with Cuts on Fermion Energies
and  Acollinearity  ($E_f^{\min}$,$\xi^{\max}$)]
{Convolution Integral with Cuts on Fermion Energies
and \\
\mbox{Acollinearity} ($E_f^{\min}$,$\xi^{\max}$)
\label{subsec:acol}}

As an alternative to the $s'_{\min}$ cut, one can apply another
set of cuts on the outgoing $f\bar f$ pair~\cite{bs:dp}.
Cuts on the minimum energy, $E^{\min}_f$, and the
maximum acollinearity, $\xi^{\max}$, of the $f\bar f$ pair in addition to
angular acceptance cuts have been implemented in \zf.

\begin{figure}[thbp]
\begin{center} \mbox{
\epsfysize=5.0cm
\epsffile{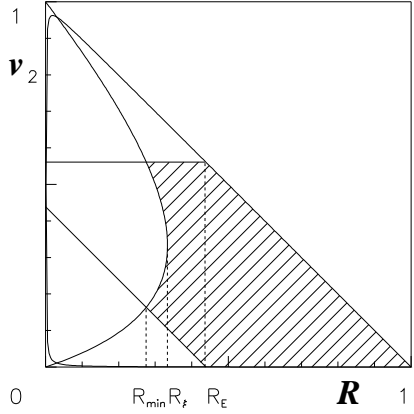}}
\end{center}
\caption[
Dalitz plot for the photon phase space with cuts on $E^f$ and $\xi$.
]{\label{fig5paw}
{\it
Dalitz plot for the photon phase space with cuts on $E^f$ and $\xi$
as explained in the text.
}   }
\label{dalitz2}
\end{figure}

Figure~\ref{dalitz2} shows a Dalitz plot of the allowed phase space
for the two energy variables $R$ and $v_2$, introduced above.

\vspace{.5cm}

The boundaries of the allowed phase-space region are
defined by the following conditions:
\bq
v_2^{\max} = 1-R_{\bar E},
\label{rmax}
\eq
\bq
v_2^{\min}(R) = R_E - R,
\label{rmin}
\eq
\bq
R^{\min}(v_2)
 = \frac  {4 R_{\xi} v_2 (1-v_2) }   { (1-R_{\xi})^2 + 4 R_{\xi} v_2},
\label{raco}
\eq
where
\bq
R_{\bar E} = \frac{2 E_{\bar f}^{\min}}{\sqrt{s}}, \hspace{1cm}
R_{E} = \frac{2 E_{f}^{\min}}{\sqrt{s}}, \hspace{1cm}
R_{\xi} = \frac{1-\sin(\xi^{\max}/2)} {1+\sin (\xi^{\max}/2)}.
\label{rvv}
\eq
The absolute minimum of $R$ is given
by\footnote{The current implementation of \zf\ assumes that
$E^{\min}_f = E^{\min}_{\bar f}$.}
\bq
R^{\min} =  \min \left( R_{E}, R_{\bar E} \right)
\left( 1 - \frac {\sin^2 (\xi^{\max}/2) }
                          { 1- R_E \cos^2 (\xi^{\max}/2)} \right).
\label{Rmin}
\eq
Further,
the upper integration limit in (\ref{sigexa}) becomes
\bq
\Delta = 1 - R^{\min}.
\label{deltaa}
\eq

The above relations are independent of the scattering angle and are,
thus,
compatible with an angular acceptance cut:
\bq
 c_1 \leq \cos \vartheta \leq c_2.
\label{cuts3}
\eq

The turning point, $P_t$ in fig.~\ref{dalitz2}, of the acollinearity
bound of the integration region is:
\bq
P_t \equiv
[R_t;v_{2,t}] =
\left[ R_\xi; \frac{1}{2}(1-R_\xi) \right].
\label{pt}
\eq
This is significant since it allows the user to apply
a reasonable approximation of the acollinearity
cut in terms of the  simpler $\Delta$ cut;
this can be achieved by using  $\Delta_\xi$ for the definition of the
integration limit (\ref{spmin}) in section \ref{subsec:delta}:
\bq
\Delta_{\xi}  \equiv 1 - R_\xi =
\frac {2 \sin(\xi^{\max}/2) }
                          { 1 + \sin (\xi^{\max}/2)}.
\label{delxi}
\eq
The quality of such an approximation depends critically on the values
of the
$E^f_{\min}$ cut and the $\xi^{\max}$ cut; for loose cuts it improves.

Because of the approximations that have been implemented in the QED
calculational chain, the user must be cautious in applying severe
cuts.
Since the approximation is ultra-relativistic one should restrict oneself
to the region of the phase space:
\bq
E^{\min} \gg m_f, \hspace{1cm} \xi^{\max}
 \ll \left( 1 - \frac{8m_f}{ \sqrt{s}} \right) \pi.
\label{limi}
\eq
Near the turning point $P_t$ introduced in (\ref{pt}) the validity of
the soft photon exponentiation approximation comes into
question\footnote{A more advanced exponentiation
procedure~\cite{jadachex} circumvents these limitations.}.
To avoid any such problems the following restrictions should be observed:
\bq
E^{\min} < 0.95 \, E_{\mathrm{beam}},
\hspace{1cm}
\xi^{\max}
 > 2^{\circ}.
\label{lime}
\eq
This last limitation guarantees that the second-order terms
$\left[ \beta \log(1-R_{\mathrm {cut}}) \right]^2$
 with $\beta = 2 (\alpha / \pi)
\linebreak[0]
 \left[ \log (s/m_e^2) - 1 \right] $ and
 $R_{\mathrm {cut}} = R_E, R_{\xi}$
do not become too large:
\bq
| \beta \log(1-R_{\mathrm {cut}}) | \ll 1.
\label{accon}
\eq
This corresponds to
\bq
E^{\min} \ll \frac {\exp(\beta^{-1}) - 1 } {\exp(\beta^{-1})}
E_{\mathrm{beam}},
\hspace{1.cm}
\xi^{\max}
 \ll \exp(-\beta^{-1}).
\label{acco1}
\eq


Finally, we would like to point out that the acollinearity cut
has an indirect influence on the acceptance cut.
It is easy to see that the maximal scattering
angle of the second fermion (which is unrestricted by the user's
acceptance cut) becomes limited by an acollinearity cut, i.e.
the scattering angle of the second fermion is limited to
$[-(\xi^{\max}+\vartheta^{\max}),(\xi^{\max}+\vartheta^{\max})]$.

\subsection
[Photonic Corrections for Bhabha Scattering]
{Photonic Corrections for Bhabha Scattering
\label{subsec:bhabha}}

The Bhabha scattering cross section (\ref{seconeq}) arises from
the sum of
s- and t-channel exchange cross sections and from their interference.
The s-channel part needs no further comment since it corresponds
completely to ordinary fermion pair production.
In the t channel the energy variables, which correspond to $(s,s')$ are
$(t,t')$, where:
\bq
t = -\frac{1}{2} s (1 - \cos \vartheta )
\label{t-prop}
\eq
and
\bq
t' = t \frac{s'}{s}.
\eq
The t-channel propagator for a massless photon is proportional to
$1/t$ or $1/t'$; it thus becomes
divergent in the forward direction, \IE\ as $\vartheta \rightarrow 0$:
\bq
\frac{d\sigma^{\mathrm{Bhabha}}}{d\cos\vartheta}
\sim  \frac{1}{\vartheta^4}.
\label{asympbha}
\eq
Such a divergence is common in calculations of the Bhabha scattering
cross section and it prevents
a reasonable definition of a total cross section without at least
an acceptance cut, even at the level of the Born approximation.

For {\em large-angle} Bhabha scattering~\cite{revbha},
$\vartheta \geq 10^{\circ}$,
this problem is absent.
Near the \Z\ peak, such a condition
guarantees that the photonic t-channel exchange contributions,
including the QED corrections (with an effective
$t' = s' (1 - \cos \vartheta) \leq t$ in the
hard-scattering process), are at most of the
same order as the non-resonating terms from the s channel.
Of course, an acceptance cut does not prevent $t'$ from becoming small
because of the emission of a hard initial-state photon, in which case the
t channel dominates.
This divergence can, however, be circumvented
by excluding very hard photons from the observed cross
section with an $E_f^{\min}$ cut.
In any case, the hard photon corrections to these contributions
must be carefully taken into account.

At LEP~I energies, terms with \Z~exchange in the
t channel are strongly suppressed owing to the form of the \Z propagator
$(\sim 1/(t' + M_Z^2))$ and
contribute less than 1\% to the cross section.
In summary, the contributions, which arise from photon exchange
in the t channel, compete with those of the s channel;
however, near the \Z\ resonance it is clear that the s channel must
dominate.

An explicit description of the QED corrections to Bhabha scattering
which have been implemented
in \zf\ will be presented elsewhere \cite{bhabha}.
In order to discuss some features of the present implementation,
we give an explicit example for the general structure of the
Bhabha cross section:
\bq
\frac{d\sigma^{\mathrm{Bhabha}}}{d\cos\vartheta}
= \frac{d\sigma^{(s)}(s,\cos \vartheta)}{d\cos\vartheta}
+ \int_0^{1-R^{\min}} dv  \sum_{a}
\sum_{V_1,V_2} \sigma^{a,o}(s,s';V_1,V_2)
R^a(v,\cos\vartheta;V_1,V_2).
\label{sigbhaqed}
\eq
Here the first term corresponds to the s-channel part.
The sum under the integral extends over $a$, denoting in the s-channel
diagrams initial- and final-state radiation, in the t-channel up  and
down radiation, and in the interferences the
corresponding combinations.
Further, a sum extends over ($V_1,V_2$),
the possible combinations of propagators $\gamma_s,
\gamma_t, Z_s, Z_t$ from the t channel and the interference.
In (\ref{sigbhaqed}),  all functions $R^a$ have the form
\bq
R^a(v,\cos\vartheta;V_1,V_2)
=
\delta(1-v)
\left[
1+\frac{\alpha}{\pi}S^a(s,\cos\vartheta ;V_1,V_2) \right]
\Delta^\beta
+\frac{\alpha}{\pi} H^a(v,\cos\vartheta ; V_1, V_2) ,
\label{radbha}
\eq
where
\bq
\Delta = 1 - R^{\min}, \hspace{1cm}
\beta=4\frac{\alpha}{\pi} \left( \log \frac{s}{m_e^2}
+ \log\frac{1-\cos\vartheta}{1+\cos\vartheta} \right) ,
\label{betbha}
\eq
and $R^{\min}$ was introduced in (\ref{deltaa}).

The functions $S^a$ in (\ref{radbha}) contain  the
soft photon (plus corresponding virtual) corrections, and $H^a$ the
complete \oalf hard photonic corrections.
In the s channel, the hard photon part depended  on $s'/s$ only,
while here, due to the t-channel propagators, it is also dependent
on $t'/s$ or on $t/s$.
As a consequence, it looses its universality and
depends also on the kind of bosons which are exchanged
(\IE\ $\gamma$ or \Z).
Further, the running QED coupling (if not assigned formally to the
radiator functions, it is contained in the hard-scattering
cross section $\sigma^{a,o}$) depends, in the t channel, on the
scattering angle as well.

These $(t,t')$ dependences have the far-reaching consequence that the
integrand in (\ref{sigbhaqed}) depends in a more non-trivial
way on the scattering angle compared with the s-channel case -- thus
preventing an analytic integration over $\cos \vartheta$, which
is the basis of the fast computing of \zf.

In the current implementation of {\tt BHANG}, the cut conditions
of section \ref{subsec:acol} are taken into account
in the functions (\ref{radbha})\footnote{
At present there is a limitation on the allowed value of
the scattering angle $\vartheta$; it must be larger than the
acollinearity $\xi^{\max}$.
This is due to purely technical reasons and this restriction will be
removed in successive versions of the code.}.

Further,
in (\ref{radbha}) the leading higher-order corrections due to soft and
hard collinear photon radiation with t-channel participation are
taken into account in an approximate way.
The cross section of the hard
process is considered to be independent of the
actual energy scale, \IE\ assuming $s'=s$ and $t'=t$.
At LEP~I, the error induced by this is definitely less than 1\%.
A simple improvement could be the choice of some better
scales for the effective s- and t-invariants in the
hard cross section, which effectively
take into account the change of kinematics due to radiation.

Hard photon radiation is considered
in the collinear approximation
for the cross section parts which
correspond to \Z~exchange in the t channel,
\IE\ the appropriate functions $H_A^a(v,\cos\vartheta,Z_t,Z_t)$ are set
to zero.

The user of \zf\ should be aware that the Bhabha cross section returned
with the aid of {\tt BHANG} is to a much larger extent adapted for
LEP~I physics than for the other fermion channels and contains
more approximations in the treatment of the QED corrections.

part.

\section
[The Hard Scattering Process:  (I) The Stan\-dard Mo\-del Branch]
{
The Hard Scattering Process: \\
 (I) The Stan\-dard Mo\-del Branch
\label{branches.1}
}
\setcounter{equation}{0}
We now describe some general features of the cross section formulae
for the hard-scattering subprocesses.
In all branches of \zf, we can denote:
\bq
\sigma_A^o(s,s';m,n)
= {\cal I}_A(m,n;s,s') \frac{1}{2}
\left[ {\cal K}_m(s') {\cal K}_n^{\ast}(s) + {\cal K}_m(s)
{\cal K}_n^{\ast}(s') \right].
\label{sigmao}
\eq
For initial-state radiation this simplifies to:
\bq
\sigma_A^o(s',s';m,n) \rightarrow
\sigma_A^o(s';m,n)
= {\cal I}_A(m,n;s')
            {\cal K}_m(s') {\cal K}_n^{\ast}(s').
\label{sigifo}
\eq
For final-state radiation $s'$ has to be replaced by $s$.
The propagator functions ${\cal K}_n(s)$ are:
\bq
{\cal K}_n(s) = \frac{s}{s-{\cal M}_n^2
+ i {\cal M}_n {\cal G}_n}.
\label{propagn}
\eq
Here, ${\cal M}_n$ are the masses and
${\cal G}_n$ are the widths of the intermediate gauge bosons.

In addition to the QED-corrected cross sections (\ref{sigexa}),
\zf\ can also return (improved or effective)
Born cross sections, $\sigma_A^{\mathrm{Born}}$.
These are constructed out of the expressions introduced above:
\bq
\sigma_A^{\mathrm{ Born}}(s,c) = D_A(c) \left\{
{\cal I}_A(\gamma,\gamma;s)
+ \Re e \left[  {\cal I}_A(\gamma,Z;s) {\cal K}_Z^{\ast}(s) \right]
+ {\cal I}_A(Z,Z;s) |{\cal K}_Z(s)|^2
\right \},
\label{bornik}
\eq
\bq
D_A(c)
= \left \{
\begin{array}{ll}
2 ( c + \frac{1}{3} c^3) & \mbox{\rm{for}} a=T   \\
2 c^2                      & \mbox{\rm{for}} A=FB. \\
\end{array}
\right.
\label{dcbo}
\eq
The functions ${\cal I}_A$ contain the underlying dynamics of the
hard-scattering process.
Often, but not necessarily, they are assumed to be inversely proportional
to $s,s'$.
The different branches
of \zf \ rely on various assumptions with respect to
${\cal I}_A$, as will be discussed later.

For the photon ($n=0$) the propagator becomes
${\cal K}_{\gamma} = 1$,
while for the \Z, various possibilities exist in \zf.
In recent years, much influenced by the discussions of
the 1989 workshop  on physics at LEP~I \cite{yr:alt},
it became common to use the following definitions:
\bq
{\cal K}_Z(s) = \frac{s}{s-{\cal M}_Z^2
+ i {\cal M}_Z {\cal G}_Z},
\label{propagz}
\eq
\bq
{\cal M}_Z = M_Z,
\label{mren}
\eq
\bq
{\cal G}_Z = \Gamma(s) \approx \frac {s} {M_Z^2} \Gamma_Z,
\label{mg}
\eq
where $M_Z$ and $\Gamma_Z$ are considered to be the mass and total
width of
the \Z.
This point of view
reflects the fact that in a quantum field theory such as
the Standard Model the \Z \, width is predicted as a result of
quantum corrections (self-energy insertions) and is, thus,
naturally s-dependent.
This s-dependence of ${\cal G}$ becomes important only
because the very narrow \Z \,
peak may be scanned with extreme precision, leading to errors
of a few MeV for mass and width of the \Z.

The definitions (\ref{propagz})-(\ref{mg}) may be related to an
alternate resonance description, which assumes a constant width:
\bq
{\bar {\cal K}}_Z(s) =
      \frac {s} {s- {\bar M}_Z^2 + i {\bar M}_Z {\bar \Gamma}_Z}.
\label{mg2}
\eq

The following
equality holds as long as the approximate relation in
(\ref{mg}) may be considered to be exact\footnote{
In the Standard Model,
this is the case if two conditions are fulfilled: (i)
there are no opening new \Z \,
decay channels (production thresholds) near $s=M_Z^2$; (ii)
radiative corrections to ${\cal G}$
are practically independent  of $s$ in a
region where ${\cal G}$ essentially influences the cross sections.}
 \cite{pl:blrs}:
\bq
G_{\mu} {\cal K}_Z(s) \equiv  {\bar G}_{\mu}
{\bar {\cal K}}_Z(s),
\label{equivmzmz}
\eq
Compared to (\ref{propagz})-(\ref{mg}),
(\ref{mg2})
 corresponds to another ansatz for mass, width, and coupling
constant:
\bq
{\bar{\cal M}} = {\bar M}_Z =
\left[ 1 + (\Gamma_Z / M_Z)^2 \right]^{-\frac{1}{2}}
M_Z \approx M_Z - \frac{1}{2} \frac{\Gamma_Z^2}{M_Z} \approx M_Z - 34\
{\mathrm{MeV}},
\label{mbar}
\eq
\bq
{\bar{\cal G}}
= {\bar \Gamma}_Z =
\left[ 1 + (\Gamma_Z / M_Z)^2 \right]^{-\frac{1}{2}}
\Gamma_Z \approx \Gamma_Z - \frac{1}{2} \frac{\Gamma_Z^3}{M_Z^2}
\approx \Gamma_Z - 1\ {\mathrm{MeV}},
\label{gbar}
\eq
\bq
{\bar G}_{\mu} =
\frac {G_{\mu}} {1+i \Gamma_Z / M_Z}.
\label{gmub}
\eq

A na\"\i{}ve
use of a constant width in (\ref{propagz})-(\ref{mg})
would lead to a wrong
determination of what has been introduced there to be the \Z\ mass.
In fact, one can put forward a completely different point of view
\cite{burgers}-\cite{smatrix} (see also references cited therein
and \cite{gounaris}-\cite{mward}).
There is no physical reason to consider (\ref{propagz})-(\ref{mg}) as the
final result of a perturbative calculation.
After so many formal
manipulations, including renormalization, one could consider
the transformations (\ref{mg2})-(\ref{gbar}) as part of the
renormalization procedure.
In doing so, one is in complete
agreement with the ideas of S-matrix theory: Unstable particles are
described by simple poles of the S-matrix in the complex energy plane
whose location is defined by the particle's mass (real part)
and its width (imaginary part).
Such an approach automatically anticipates the propagator
${\bar{\cal K}}_Z(s)$ with mass (\ref{mbar}) and width (\ref{gbar}).

The default mass and width definition for the S-matrix branch of the
\zf\ package is (\ref{mg2})-(\ref{gbar}),
while
(\ref{propagz})-(\ref{mg}) should be used
for all other branches.
The final choice of the definition of the \Z\ mass and width is
left to the
user (see flag {\tt GAMS}).

For the t channel of Bhabha scattering, the above discussion regarding
the propagators is also of some relevance.
Of course, here the width of the resonance is absent.
Furthermore, one must replace $s$ by $t$ and $s'$ by $t'$ in the
propagator functions.

For the corresponding Standard Model calculations,
\zf\ relies on the {\tt DIZET}~\cite{di:cpc} package.
The following parameters are passed to \DIZET:
\bq
\alpha, \alpha_s, G_\mu, M_Z, M_H, m_f,
\label{sminput}
\eq
which returns $M_W$; the total and partial \Z\  widths;
the weak form factors, etc.
Thus, one arrives at improved Born cross sections,
which are used as building blocks of the QED formulae discussed in the
foregoing sections;
\zf\ {\em does not} calculate bare Born cross sections
since definitions of the latter tend to be ambiguous.
The weak mixing angle also will not be considered as a quantity
of physical relevance (although one could do so), but will only be used
for book-keeping of intermediate results.

In the remainder of this section and in the next, we discuss the various
assumptions regarding the functions ${\cal I}_A(m,n;s)$ that have been
implemented in \zf.

\subsection
[\oalf Corrections to $\Delta r$]
{\oalf Corrections to $\Delta r$
\label{subdr}}

In the on-mass-shell renormalization scheme~\cite{bible2} that is used
in \zf, the weak mixing angle is defined uniquely through the gauge-boson
masses:
\bq
\sin^2\theta_W \equiv 1 - \frac{M_W^2}{M_Z^2},
\label{defsw2}
\eq
\bq
\sin^2\theta_W \, M_W^2 = \frac{\pi \alpha / (\sqrt{2} G_{\mu}) }
                            {1 - \Delta r}.
\label{sw2delr}
\eq
In subroutine {\tt SEARCH} of \DIZET, $\Delta r$ is calculated
to order \oalf as defined in~\cite{bible2}, where the heavy top
contribution is calculated as in \cite{zwidth}.
Recently, a careful comparison~\cite{carecomp}
of two independent \oalf calculations of (\ref{sw2delr}) showed
agreement in 12 digits.

\subsection
[\oalf Corrections to $\Gamma_Z$]
{\oalf Corrections to $\Gamma_Z$
\label{subgamma}}

Electroweak corrections to the \Z \,
width have been calculated to order
\oalf in \cite{zwidth}.
The partial decay
width of the \Z \, into fermions of
type $f$ is given by:
\begin{eqnarray}
{\Gamma}_f &=& \frac{G_{\mu}}{\sqrt{2}} \frac{M_Z^3}{12 \pi}
\, \mu \, R_{\mathrm {QED}} \, c_f \, R_{\mathrm {QCD}}(M_Z^2)  \,
\rho_f^Z \times \hfill \nonumber \\ & &
\left\{
           \left[ 1 - 4 |Q_f| \sin^2\theta_W \kappa_f^Z
                    + 8 (|Q_f| \sin^2\theta_W \kappa_f^Z)^2 \right]
\left(1+2\frac{m_f^2}{M_Z^2}\right) - 3 \frac{m_f^2}{M_Z^2}
\right\},
\nonumber \\
         &=& \frac{G_{\mu}}{\sqrt{2}} \frac{M_Z^3}{24 \pi}
\, \mu \, R_{\mathrm {QED}} \, c_f \, R_{\mathrm {QCD}}(M_Z^2)
\left\{
\left[ ({\bar v}_f^Z)^2 + ({\bar a}_f^Z)^2  \right]
\left(1 + 2 \frac{m_f^2}{M_Z^2}\right)
- 6 ({\bar a}_f^Z)^2 \frac{m_f^2}{M_Z^2}
\right\} \! \! .
\label{defzwidth}
\end{eqnarray}
The renormalized
vector and axial-vector couplings\footnote{The Born axial-vector
coupling is often defined to be equal to the weak isospin,
$a_f = I_3^L(f)$,
{    e.g.}\/ $a_e=-\frac{1}{2}$.}
are defined as follows:
\begin{eqnarray}
\bar a_f^Z = \sqrt{\rho_f^Z},
\label{axialz}
\end{eqnarray}
\begin{eqnarray}
\bar v_f^Z =\bar a_f^Z \left[ 1 - 4 |Q_f |\sin^2 \theta_W
                                        \kappa_f^Z
                    \right].
\label{vectcoup}
\end{eqnarray}
The bare Born vector and axial-vector couplings, $v$ and $a$, correspond
to
$\rho=\kappa=1$.
The weak form factors, $\rho$ and $\kappa$, are real, constant, and
depend
slightly on the decay channnel.
They contain the electroweak corrections to the process,
including the dependence on $m_t$ and $M_H$.
Since \zf \,
exactly follows~\cite{zwidth}, we need not go into details here.
However, it must be mentioned that the \Z \, width depends
on the choice of the definition of the \Z \, mass
(see the discussion presented at the beginning of this section).

Sometimes, the combination
\bq
s_W^{2,{\mathrm{eff}}} = \kappa_e^Z \,\sin^2 \theta_W
\label{sw2eff}
\eq
is used as a definition of the `effective' weak mixing angle;
see {    e.g.}~\cite{hofo,heraho} and the recent discussion on possible
alternatives in~\cite{l3:som}
and references quoted therein.
Such a definition can also rely on the other decay channels:
\bq
s_W^{2,{\mathrm{f}}} = \kappa_f^Z \,\sin^2 \theta_W.
\label{sw2f}
\eq

Several factors contain additional corrections:
\begin{eqnarray}
 \mu   \equiv \mu (M_Z^2),
\nonumber
\ea
\ba
\mu (s) =
 \sqrt{1-4m_{f}^{2}/s},
\label{defmu}
\end{eqnarray}
\bq
R_{\mathrm {QED}} =
1 + \frac{3}{4} \frac{\alpha}{\pi} Q_f^2 = 1 + 0.0017 Q_f^2,
\label{defrqed}
\eq
\bq
c_f = \left \{
\begin{array}{ll}
3 & \mbox{\rm{for quarks}}   \\
1 & \mbox{\rm{for leptons,}} \\
\end{array}
\right.
\label{rqed}
\eq
\bq
R_{\mathrm {QCD}}   = \left\{
\begin{array}{ll}
1+c_1(m_f^2) \displaystyle
{\frac{\alpha_{s}(M_{Z}^{2},\Lambda_{\overline{\mathrm{MS}}})}
                                      { \pi } } + \cdots
                            &   \mbox{{\rm{ for quarks}}}
\vspace{.1cm} \nonumber \\
1     & \mbox{{\rm{for leptons.}}} \\
\end{array}
\right.
\label{rqcd}
\eq
The corrections $R_{\mathrm {QED}}$ and $R_{\mathrm {QCD}}$
contain the photonic and gluonic corrections, respectively.
The factor $c_1$ is of relevance only for the b-quark channel.
The exact definition of $R_{\mathrm {QCD}}$ will be given in
section~\ref{subhigh}.

\begin{figure}[htbp]
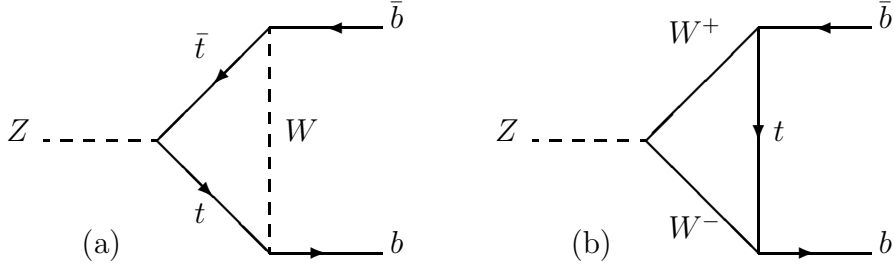

\begin{Feynman}{120,30}{0,0}{1.0}
\put(5,15){\gaugebosonrighthalf}
\put(0,15){$Z$}
\put(20,15){\fermiondr}
\put(25,4){$t$}
\put(35,0){\fermionrighthalf}
\put(51,0){$b$}
\put(35,0){\gaugebosonup}
\put(37,15){$W$}
\put(35,30){\fermionlefthalf}
\put(51,30){$\bar{b}$}
\put(20,15){\fermiondl}
\put(25,26){$\bar{t}$}
\put(10,0){(a)}
\put(70,15){\gaugebosonrighthalf}
\put(65,15){$Z$}
\put(85,15){\gaugebosondrhalf}
\put(88,2){$W^-$}
\put(100,0){\fermionrighthalf}
\put(116,0){$b$}
\put(100,0){\fermiondown}
\put(102,15){$t$}
\put(100,30){\fermionlefthalf}
\put(116,30){$\bar{b}$}
\put(100,30){\gaugebosondlhalf}
\put(88,28){$W^+$}
\put(75,0){(b)}
\end{Feynman}
\caption[
Top-quark exchange diagrams which contribute to $\Gamma_b$.
]{\label{fig6paw}  \it
Top quark exchange diagrams which contribute to $\Gamma_b$.
}
\label{figbdec}
\end{figure}

Owing to the large mass splitting between the t~and b~quark, there
are two vertex diagrams for the \Z\  decay into b~quarks
(fig.~\ref{figbdec}), which contribute
additional $m_t$-dependent corrections that are absent in the
cases of light quarks~\cite{zwidth,mannrie} (see also \cite{beho,pich}).
The matrix element for the decay of the \Z\ boson into d~and s~quarks
(and similarly, for u~and c~quarks) may be written as follows:
\bq
{\cal M}_d
  \sim
\sqrt{ \frac{G_{\mu}}{\sqrt{2}} M_Z^2 }
\epsilon^{\beta}
 {\bar u} \left[
 \gamma_{\beta}  \bar{v}_d^Z
+
 \gamma_{\beta} \gamma_5 {\bar a}_d^Z
\right] u.
\label{zdecds}
\eq
\clearpage
\begin{table}[htbp]
\begin{center}
\begin{tabular}{|c|c|c|c|c|c|c|c|c|c|c|c|}   \hline
& & & & & & & & & & & \\
\Sw           &     $\Gamma_{\nu}$ &
$\Gamma_{e}$ &        $\Gamma_{\mu}$ &
$\Gamma_{\tau}$ & $\Gamma_{u}$ &
$\Gamma_{d}$ &        $\Gamma_{c}$ &
$\Gamma_{s}$ &        $\Gamma_{t}$ &
$\Gamma_{b}$ &        $\Gamma_{\mathrm{tot}}$ \\
& & & & & & & & & & & \\  \hline
& & & & & & & & & & & \\
0.2282   & 166.6 &  83.6 &  83.6 &  83.4
& 296.6 & 382.9 & 296.2 & 382.9
& 0.    & 375.7 & 2484.7  \\
& & & & & & & & & & & \\ \hline
\end{tabular}
\end{center}
\caption[
Weak mixing angle and partial and total \Z\ widths.
]{\it
Weak mixing angle and partial and total \Z\ widths;
widths are given in MeV.}
\label{t1}
\end{table}
The corresponding matrix element for b~quarks has an additional
left-handed contribution:
\begin{eqnarray}
{\cal M}_b
 &\sim&
\sqrt{ \frac{G_{\mu}}{\sqrt{2}} M_Z^2 }
\epsilon^{\beta}
 {\bar u} \left[
 \gamma_{\beta}  \bar{v}_d^Z
+
 \gamma_{\beta} \gamma_5 {\bar a}_d^Z
+
 \Delta_b(m_t^2) \gamma_\beta (1+\gamma_5)
  \right] u
\nonumber \\
&\sim&
\sqrt{ \frac{G_{\mu}}{\sqrt{2}} M_Z^2 }
\epsilon^{\beta} \sqrt{\rho_b^Z}
  a_b
 {\bar u} \left[ \gamma_{\beta} (1+\gamma_5) - 4 \sin^2\theta_W
 \kappa_b^Z
\gamma_{\beta}
 \right] u.
\label{zmate}
\end{eqnarray}
Here, $\Delta_b$
vanishes
for $m_t \rightarrow 0$.
By simple algebra, one can show that
\bq
\rho_b^Z
 = \rho_d^Z
  - 2 \frac{\Delta_b(m_t^2)}{a_b},
\label{delrhob}
\eq
\bq
\kappa_b^Z
 = \kappa_d^Z
  +
  \frac{\Delta_b(m_t^2)}{a_b}.
\label{delkab}
\eq
These exact form factors have been implemented in \zf.
In the limit of large t-quark masses, the leading terms
are given by~\cite{zwidth}:
\bq
  \frac{\Delta_b(m_t^2)}{a_b}
 =
\frac{\alpha}{4 \pi \sin^2\theta_W}  |V_{tb}|^2 \frac{1}{2}
\left[ \frac{m_t^2}{M_W^2} + \left(\frac{8}{3}
+\frac{1}{6 \cos^2 \theta_W} \right) \log \frac{m_t^2}{M_W^2} \right],
\label{zbb}
\eq
where $V_{tb}$ is the $(t,b)$ Kobayashi-Maskawa mixing matrix element.
\zf\ calculations use the normalization $a_b=1$ in (\ref{zmate}).

The calculation of the $W$ width~\cite{wwidth}
follows the same principles as that of the
\Z\ width and
is realized in subroutine {\tt ZWRATE} of {\tt DIZET}.
Since the $W$ width is not that important for the description of
fermion pair production, we do not go into details.

The partial and total widths, returned by
\zf\footnote{In the examples we have taken
\MZ\ = 91.175~GeV, \MH\ = 300~GeV, \MT\ = 140~GeV, $\alpha_s$ = 0.12
and default flag values (see table~\ref{ta7}
in section~\ref{zuflag}) unless explicitly stated otherwise.},
are summarized in table~\ref{t1}.
In addition the weak mixing angle is given in the table.
\subsection
[\oalf Corrections to Fermion Pair Production]
{\oalf Corrections to Fermion Pair Production
\label{subfpair}}
Fermion pair production in the Standard Model gets
contributions from self-energy insertions, vertex corrections
and box diagrams.
We divide these into two gauge-invariant subsets.

\begin{figure}[hbtp]
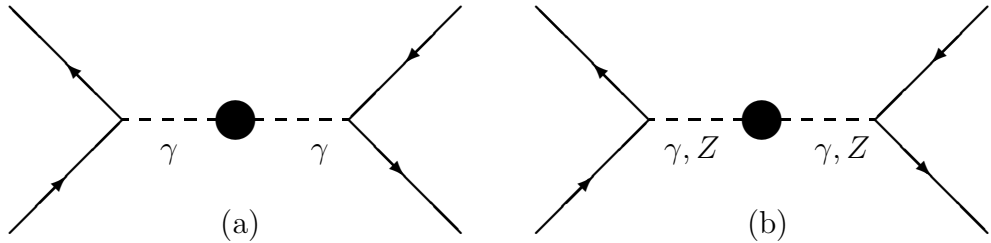

\begin{Feynman}{130,30}{0,0}{1.0}
\put(15,15){\fermionur}
\put(15,15){\fermionul}
\put(15,15){\gaugebosonright}
\put(30,15){\circle*{10}}
\put(20,10){$\gamma$}
\put(40,10){$\gamma$}
\put(45,15){\fermiondl}
\put(45,15){\fermiondr}
\put(28,0){(a)}
\put(85,15){\fermionur}
\put(85,15){\fermionul}
\put(85,15){\gaugebosonright}
\put(100,15){\circle*{10}}
\put(87,10){$\gamma,Z$}
\put(107,10){$\gamma,Z$}
\put(115,15){\fermiondl}
\put(115,15){\fermiondr}
\put(98,0){(b)}
\end{Feynman}
\caption[
Photon and \Z\ self energies.]
{\it
Photon (a) and \Z\ (b) self energies. In (b), the case
$\gamma,\gamma$ is not included.
\label{figsmgam}}
\end{figure}

Fermion loop insertions to the photon propagator
are summed together with the photonic Born diagram
(see fig.~\ref{figsmgam}a) to form
the matrix element ${\bf\cal M}_{\gamma}$.
In effect these corrections change $\alpha$ into $\alpha(s)$:
\begin{equation}
{\bf\cal M}_{\gamma}(s)  \sim \frac{1}{s} \alpha(s)  Q_e Q_f
         \gamma_{\beta} \otimes \gamma^{\beta},
\end{equation}
where the following short notation for bilinear combinations of spinors
$u_f$ is used:
\begin{equation}
A_{\beta} \otimes B^{\beta} = \left[ \bar u_e A_{\beta} u_e \right]
                          \cdot \left[ \bar u_f B^{\beta} u_f \right].
\end{equation}
After a Dyson summation of the fermion loop
insertions $\Delta\alpha(s)$ to the photon self energy,
the running electromagnetic coupling constant contains higher-order
corrections:
\bq
\alpha(s) = F_A(s) \alpha \equiv
            \frac{\alpha}{1-\Delta\alpha(s)}.
\label{dysfa}
\eq
The function {\tt XFOTF1} in
{\tt DIZET} is used to calculate $\Delta \alpha$.

Some numerical examples are given in
table~\ref{tab:fa} as a function of \RS.

\begin{figure}[htbp]
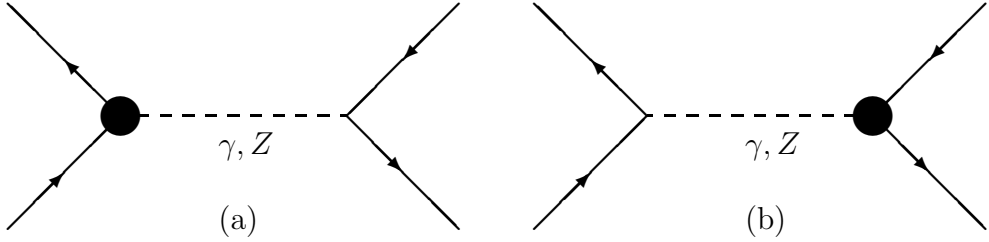

\begin{Feynman}{130,30}{0,0}{1.0}
\put(15,15){\fermionur}
\put(15,15){\fermionul}
\put(15,15){\gaugebosonright}
\put(15,15){\circle*{10}}
\put(28,10){$\gamma,Z$}
\put(45,15){\fermiondl}
\put(45,15){\fermiondr}
\put(28,0){(a)}
\put(85,15){\fermionur}
\put(85,15){\fermionul}
\put(85,15){\gaugebosonright}
\put(115,15){\circle*{10}}
\put(98,10){$\gamma,Z$}
\put(115,15){\fermiondl}
\put(115,15){\fermiondr}
\put(98,0){(b)}
\end{Feynman}
\caption
[Vertex corrections to the \Z\ matrix element.]
{\it
Vertex corrections to the \Z\ matrix element.
\label{figsmz}}
\end{figure}

\begin{figure}[hbtp]
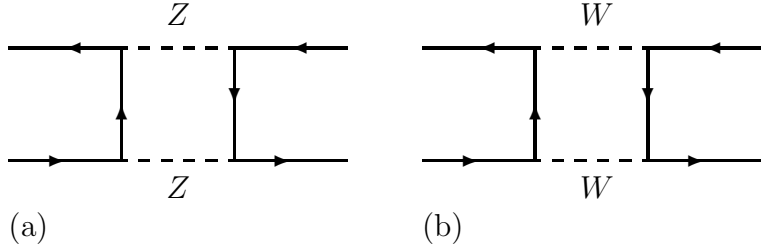

\begin{Feynman}{100,35}{0,0}{1.0}
\put(0,10){\fermionrighthalf}
\put(0,25){\fermionlefthalf}
\put(15,10){\fermionuphalf}
\put(15,10){\gaugebosonrighthalf}
\put(21,5){$Z$}
\put(15,25){\gaugebosonrighthalf}
\put(21,28){$Z$}
\put(30,10){\fermiondownhalf}
\put(30,10){\fermionrighthalf}
\put(30,25){\fermionlefthalf}
\put(0,0){(a)}
\put(55,10){\fermionrighthalf}
\put(55,25){\fermionlefthalf}
\put(70,10){\fermionuphalf}
\put(70,10){\gaugebosonrighthalf}
\put(76,5){$W$}
\put(70,25){\gaugebosonrighthalf}
\put(76,28){$W$}
\put(85,10){\fermiondownhalf}
\put(85,10){\fermionrighthalf}
\put(85,25){\fermionlefthalf}
\put(55,0){(b)}
\end{Feynman}
\caption[
Box diagrams contributing to the \Z\ matrix element.
]{ \it
Box diagrams contributing to the \Z\ matrix element.
\label{figsmbx}}
\end{figure}

\begin{table}[hbtp]\centering
\begin{tabular}{|c|c|c|c|c|c|c|c|}  \hline
$\sqrt{s}$ & 30 & 87 & 89 & 91 & 93 & 95 & 200 \\  \hline
 $F_A$             & 1.0504    & 1.0630   & 1.0633
                   & 1.0635    & 1.0638   & 1.0640   & 1.0723   \\
  & $-$ i.0186  & $-$ i.0188 & $-$ i.0188 & $-$ i.0188
  & $-$ i.0188  & $-$ i.0189 & $-$ i.0191  \\   \hline
\end{tabular}
\caption
[
The running QED coupling
as a function of the centre-of-mass energy.
]
{{\it
The running QED coupling, $F_A(s)=\alpha(s)/\alpha$,
as a function of the centre-of-mass energy ($\sqrt{s}$ in GeV).}
\label{tab:fa}
}
\end{table}
\clearpage
At \RS\ = \MZ\ the running coupling constant has the value:
\begin{equation}
\alpha(M_Z^2) = F_A(M_Z^2)\alpha \simeq \frac{1}{137} F_A(M_Z^2) \simeq
\frac{1}{128.8}.
\end{equation}
In addition to the running of $\alpha(s)$,
there are diagrams with additional internal \w\ and
\Z\ boson propagators to the photonic Born amplitude,
\EG\ a $W^{\pm}$-pair insertion or a vertex correction with a
\Z\ propagator.
These diagrams could be treated as corrections to the photon
amplitude as well. However, this would make ${\bf\cal M}_{\gamma}$
dependent on the gauge.
Diagrams of this type
form a gauge-invariant subset together with all the insertions
to the \z\  Born diagram as well as with $ZZ$ and $WW$ boxes.
So, any diagram with at least one additional massive gauge boson
will be combined with the \z~exchange Born diagram to form the
matrix element
${\bf\cal M}_{Z}$ (see figs.~\ref{figsmgam}b, \ref{figsmz}
and~\ref{figsmbx}).

The contributions to the corresponding matrix element  ${\cal M}_Z$
can be expressed in terms of four weak form factors $(\rho_{ef},
\kappa_{e}, \kappa_{f}, \kappa_{ef})$ as introduced to order \oalf
in~\cite{rokap}:
\begin{eqnarray}
{\bf\cal M}_{Z}(s,\cos \vartheta)  \sim
\frac
{G_{\mu} a_e a_f \rho_{ef}(s,\cos \vartheta) }
{s - M_Z^2 + i M_Z \Gamma_Z}
\left[ L_{\beta} \otimes L^{\beta}
- 4|Q_e| \sin^2 \theta_W
\kappa_e (s,\cos \vartheta ) \gamma_{\beta} \otimes L^{\beta}
\right. \nonumber  \\
\left.
- \,  4|Q_f| \sin^2 \theta_W
\kappa_f (s,\cos \vartheta ) L_{\beta} \otimes \gamma^{\beta}
    + 16 |Q_e Q_f| \sin^4 \theta_W
    \kappa_{ef}(s,\cos \vartheta) \gamma_{\beta} \otimes
                                             \gamma^{\beta}
\right],
\label{mzff}
\end{eqnarray}
\begin{eqnarray}
L_{\beta} = \gamma_{\beta}(1+\gamma_5 ),
\end{eqnarray}
where $L_{\beta}$ is the left-handed projector.
The $(\rho - 1)$ and $(\kappa - 1)$ are normalized with
the factor
$\alpha / (4 \pi \sin^2 \vartheta_W)$.

The matrix element may be rewritten in terms of renormalized
vector $(\bar{v})$ and axial-vector $(\bar{a})$ couplings:
\begin{eqnarray}
{\bf\cal M}_{Z}(s,\cos \vartheta)  \sim \frac{G_{\mu}}
{s - M_Z^2 + i M_Z \Gamma_Z}
      [ \bar a_e \bar a_f \gamma_{\beta} \gamma_5 \otimes
                                \gamma^{\beta} \gamma_5
      + \bar v_e \bar a_f  \gamma_{\beta} \otimes \gamma^{\beta} \gamma_5
              \nonumber \nopagebreak[4]
              \\        \nopagebreak[4]
      + \bar a_e \bar v_f  \gamma_{\beta} \gamma_5 \otimes \gamma^{\beta}
      +        \bar v_{ef} \gamma_{\beta} \otimes \gamma^{\beta} ],
\label{Mva}
\end{eqnarray}
\begin{eqnarray}
\bar a_f = \sqrt{\rho_{ef}(s,\cos \vartheta)} \: I_3^L (f),
\label{axial}
\end{eqnarray}
\begin{eqnarray}
\bar v_f = \bar a_f
\left[ 1 - 4 |Q_f | \sin^2 \theta_W  \kappa_f (s,\cos \vartheta)
                    \right],
\end{eqnarray}
\begin{eqnarray}
\bar v_{ef} = \bar a_e \bar v_f + \bar v_e \bar a_f - \bar a_e \bar a_f
 \left[ 1 - 16 |Q_e Q_f |\sin^4 \theta_W \kappa_{ef} (s,\cos \vartheta)
              \right].
\end{eqnarray}
The four form factors are the most general ansatz for the weak radiative
corrections.
In the Born approximation, $\rho=\kappa=1$, and $v_{ef}=
v_e v_f$.
The coupling $v_{ef}$ has no parallel in the Born approximation
and is, in principle, completely independent of $v_e$ and $v_f$.
Form factors are calculated in subroutine {\tt ROKANC} of
the {\tt DIZET} package.

In table~\ref{t2} we show the $s$ dependence of the
weak form factors for lepton production.

\begin{table}[thbp]\centering
\begin{tabular}{|c|c|c|c|c|c|c|c|}  \hline
$\sqrt{s}$ & 30  & 87  & 89  & 91  & 93  & 95  & 200  \\  \hline
 $\rho_{ef}      $ & 0.9992  & 1.0020  & 1.0021
                   & 1.0022  & 1.0022  & 1.0023  & 1.0102   \\
                   & $-$ i.0006  & $-$ i.0043 & $-$ i.0045
       & $-$ i.0047  & $-$ i.0048  & $-$ i.0050 & $-$ i.0283  \\  \hline
 $\kappa_e,\kappa_f$    & 1.0283  & 1.0227  & 1.0226
                   & 1.0226  & 1.0225  & 1.0225  & 1.0117   \\
                   & + i.0115  & + i.0134 & + i.0135
       & + i.0135  & + i.0136  & + i.0137 & + i.0329  \\  \hline
 $\kappa_{ef}$    & 1.0552   & 1.0459  & 1.0458
                   & 1.0456  & 1.0455  & 1.0454  & 1.0355   \\
                   & + i.0204  & + i.0265 & + i.0268
       & + i.0271  & + i.0273  & + i.0276 & + i.0560  \\  \hline
 $\kappa_e \kappa_f$    & 0.0020   & $-$0.0002  & $-$0.0002
                   & $-$0.0002   & $-$0.0002  & $-$0.0001  & $-$0.0010
                   \\
$- \kappa_{ef}$    & + i.0032  & + i.0008 & + i.0007
       & + i.0006  & + i.0005  & + i.0004 & + i.0110  \\  \hline
\end{tabular}
\caption
[
Leptonic form factors $\rho_{ef},
\kappa_e,\kappa_f,\kappa_{ef}$
as functions of the centre-of-mass energy.
 ]
{\it
Leptonic form factors $\rho_{ef},
\kappa_e,\kappa_f,\kappa_{ef}$
as functions of the centre-of-mass energy ($\sqrt{s}$
in GeV).
}
\label{t2}
\end{table}

The last row in the table shows how well $\kappa_{ef}$ can be
factorized in terms of $\kappa_e$ and $\kappa_f$.
The form factors shown in the table have been calculated without the
box diagrams of fig.~\ref{figsmbx}.
The program \DIZET\ allows for three options: inclusion of these
box diagrams into the weak form factors; calculation of them as an extra
cross-section piece,
$\sigma_{\mathrm{box}}$, to be added incoherently (see
\cite{rokap}); or neglecting them completely.
\zf\ allows for the last two options only.
This has been arranged in order to make the
weak form factors independent of the scattering angle.
Thus, the angular integration could be performed analytically.
Table~\ref{ta1} shows the influence of the box-diagram corrections on the
differential cross section for different angular bins at LEP~I
energies\footnote{
The contributions from box diagrams are non-resonant at
LEP~I energies.
\zf\ users should be aware that, off the \Z\ resonance peak, box
diagrams may not be neglected with respect to
              other radiative corrections.}.

\begin{table}[bhtp]\centering
\begin{tabular}{|c|c|c|c|c|c|c|}    \hline
$\vartheta$ region&\multicolumn{3}{c|}{$\sqrt{s}=M_{Z}$}
&\multicolumn{3}{c|}{$\sqrt{s}=94$ GeV}       \\
\cline{2-7}
          & $\sigma_{\mu}-\sigma_{\mathrm{box}}$ & $ \sigma_{\mu}$
          & $\sigma_{box} \times 10^7$
          & $\sigma_{\mu}-\sigma_{\mathrm{box}}$ & $ \sigma_{\mu}$
          & $\sigma_{\mathrm{box}} \times 10^7$ \\
\hline
    0$^{\circ}$ $-$
    30$^{\circ}$          & 0.1410419     & 0.1410410  &
                9           & 0.0512613     & 0.0512605  &   8  \\
   30$^{\circ}$ $-$
   60$^{\circ}$          & 0.2993644     & 0.2993631  &
               13           & 0.1042011     & 0.1041998  &  13  \\
      60$^{\circ}$ $-$
      90$^{\circ}$          & 0.3005794     & 0.3005789  &
                5           & 0.0961092     & 0.0961088  &   4  \\
     90$^{\circ}$ $-$
     120$^{\circ}$          & 0.3016901     & 0.3016900  &
                1           & 0.0800074     & 0.0800073  &    1  \\
  120$^{\circ}$ $-$
  150$^{\circ}$          & 0.3015250     & 0.3015249  &
                1           & 0.0721641     & 0.0721640  &    1  \\
150$^{\circ}$ $-$
180$^{\circ}$          & 0.1420456     & 0.1420456  &
                0           & 0.0353764     & 0.0353763  &    1  \\
\hline
\end{tabular}
\caption[
Differential cross sections in nb for \MM\ production both
with and without weak box contributions for selected LEP~I energies.
]{\it
Differential cross section in nb for \MM\ production both
with and without weak box contributions for selected LEP~I energies.
QED corrections are calculated following section~\ref{subsec:delta}
with $s'_{\min} = 4 m_f^2/s$.
}
\label{ta1}
\end{table}

Henceforth, we omit the possibility of an angular dependence of the form
factors.
Using an alternative parametrization, the axial-vector couplings
may be chosen such that they are unchanged by radiative corrections.
In this case, the Fermi constant absorbs the weak form
factor $\rho(s)$ and becomes dependent on
the process and its kinematics:
\begin{equation}
G_{\mu}  \rightarrow \bar G_{\mu} = \rho(s) G_{\mu} .
\label{gmu}
\end{equation}
The other form factors may be absorbed into various weak mixing angles:
\begin{equation}
\sin^2\theta_W  = 1-M_W^2/M_Z^2
                \rightarrow
                \left\{
                 \begin{array}{l}
                 \kappa_e(s) \sin^2 \theta_W \\
                 \kappa_f(s) \sin^2 \theta_W \\
                 \sqrt{\kappa_{ef}(s)} \sin^2 \theta_W.
                \end{array}
                \right.
\label{swkap}
\end{equation}
This is similar to~(\ref{sw2eff}), even though more involved because of
the
additional complications presented by the kinematics.
The above parametrization of ${\cal M}_Z$ allows for a Born-like
interpretation of all weak corrections.
In this respect, we differ in our intentions from many other definitions
of weak form factors and couplings, which try to perform
dedicated approximations.
Of course, such approximations may be
applied to our weak form factors or to quantities derived from them;
in sections~\ref{branches.2} and~\ref{compa1} such approximations
will be discussed; see also section~\ref{subgamma}.
At \LEPI\ energies, the approximate relations hold:
\bq
\left| \rho_{ef}(M_Z^2) \right|^2  \sim \rho_e^Z \rho_f^Z,
\label{rhefapp}
\eq
\bq
{\bar v}_f (M_Z^2) \sim {\bar v}_f^Z,
\label{vfapp}
\eq
where the second relation may be replaced by:
\bq
\kappa_f(M_Z^2) \sim \kappa_f^Z.
\label{kappapp}
\eq

So far, we have concentrated on s-channel kinematics, which depend on
$s$ and $\cos\vartheta$.
It should be noted that for the
t channel in Bhabha scattering the energy variable becomes
$t=-\frac{1}{2}s(1-\cos\vartheta)$
instead of $s$.

For b-quark production, unlike d- and s-quark production,
a special contribution to the weak form factors arises from diagrams
in fig.~\ref{figsmz}b, which contain as building blocks
the Feynman graphs of fig.~\ref{figbdec}.
This contribution may be
of special interest at a high luminosity version of LEP~I as is discussed
in~\cite{hilumi}.
In general, the correction is $s$-dependent. It
can be approximated near the \Z\ resonance by the corresponding
correction
$ \Delta_b(m_t^2) $
to the \Z\ width as introduced in (\ref{zmate}):
\bq
\rho_{eb} =
\rho_{ed} -
  \frac{
  \Delta_b(m_t^2)
  }{a_b},
\label{rsbb}
\eq
\bq
\kappa_b =
\kappa_d +
  \frac{\Delta_b(m_t^2)}{a_b},
\label{ksbb}
\eq
\bq
\kappa_{eb} =
\kappa_{ed} +
  \frac{\Delta_b(m_t^2)}{a_b},
\label{kseb}
\eq
with $\kappa_e$ unchanged.
This approximation has been implemented in \zf; it is
valid only near the \Z\ resonance and for $m_t > \sqrt{s} / 2$.
At other energies, since it would be difficult or impossible to measure
the effects of this tiny correction due to small cross sections, we
assume that the approximation holds there as well.

Higher order corrections in $R_{\mathrm {QCD}}$ have been
implemented in \zf.
These and the effects of the
higher-order corrections to $\rho$ and $\kappa$
associated with a potentially large t-quark mass
will be discussed in the next section.

We now come to the cross-section
formulae, which are calculated in subroutine {\tt BORN}.
Both $\sigma_T$ and $\sigma_{FB}$ are sums of three terms:
\begin{eqnarray}
\sigma_A^{o,\mathrm{SM}}(s) &=&
\sigma_A^{o,\mathrm{SM}}(s;\gamma,\gamma) +
\sigma_A^{o,\mathrm{SM}}(s;\gamma,Z)      +
\sigma_A^{o,\mathrm{SM}}(s;Z,Z)
           \nonumber \\
  &=&
{\cal I}_A^{\mathrm{SM}} (\gamma,\gamma;s)
+
 \Re e \left[ {\cal I}_A^{\mathrm{SM}} (\gamma,Z;s)
                      {\cal K}_Z^\ast(s) \right]
+
{\cal I}_A^{\mathrm{SM}} (Z,Z;s)
\left| {\cal K}_Z(s) \right|^2.
\label{ssmg}
\end{eqnarray}

Here, we have written the cross section in a form that is suitable for
initial- and final-state radiation.
However, the initial-final state interference cross section depends on
two different energy scales $(s,s')$.
The correct handling of the
propagators can be inferred from (\ref{sigmao}).
The generalized couplings ${\cal I}_A$ are
assumed to be dependent on $s$ in \zf\  with the exclusion of the
running QED coupling where the scale ($s$ or $s'$) can be chosen by
a flag\footnote{
In principle, with initial-state radiation, the
form factors depend on $s'$, with final-state radiation on $s$ and
with the small initial-final interference on
both  $s$ and $s'$.}.
This assumption speeds up the calculations with negligible loss
of accuracy.
In principle, one can take into account the $s$ and $s'$ dependence,
in a trivial way for the factorizing parts of the form factors, and
the rest with a little effort.

For unpolarized scattering, $\sigma_T$ can be expressed by
\ba
{\cal I}_T^{\mathrm{SM}} (\gamma,\gamma;s) =
 c_m
N_\gamma(s)
 |Q_e|^2 |Q_f|^2 |F_A|^2,
\label{itsmg}
\ea
\ba
{\cal I}_T^{\mathrm{SM}} (\gamma,Z;s) =
  2 c_m
N_\gamma(s) N_Z
|Q_e \, Q_f|
\hspace{0mm}  \left[
F_A^\ast \, \rho_{ef} \, {\bar v}_{ef}
\right],
\label{itsmi}
\ea
\ba
{\cal I}_T^{\mathrm{SM}} (Z,Z;s) =
N_\gamma(s) N_Z^2
\left[ c_m \left( 1 + |{\bar v}_e|^2
   + |{\bar v}_f|^2  + |{\bar v}_{ef}|^2 \right)
 - \frac{6m_f^2}{s}(1 + |{\bar v}_e|^2 )   \right]
    \, |\rho_{ef}|^2  ,
\label{itsmz}
\ea
where
\bq
N_\gamma(s) = \frac{\pi \alpha^2}{2 s} \mu(s) \, c_f
R_{\mathrm {QCD}}(s),
\label{ngam}
\eq
\bq
N_Z = \frac{G_\mu }{\sqrt{2}} \frac{M_Z^2}{8 \pi \alpha},
\label{nz}
\eq
\bq
 c_m   =  (1+2m_{f}^{2}/s).
 \label{cm}
\eq
The variables $\mu(s)$, $c_f$, and $R_{\mathrm {QCD}}$ have already been
introduced in section~\ref{subgamma}, and $F_A$ in
(\ref{dysfa}).

The corresponding
generalized couplings for the anti-symmetric cross section
$\sigma_{FB}$  are:

\bq
{\cal I}_{FB}^{\mathrm{SM}} (\gamma,\gamma;s) = 0,
\label{ifbsmg}
\eq
\bq
{\cal I}_{FB}^{\mathrm{SM}} (\gamma,Z;s) =
 2 \mu(s)
N_\gamma(s) N_Z
 |Q_{e} \, Q_{f}|
\hspace{0mm}  \rho_{ef} \, F_A^*,
\label{ifbsmi}
\eq
\bq
{\cal I}_{FB}^{\mathrm{SM}} (Z,Z;s) =
 4 \mu(s)
N_\gamma(s) N_Z^2
   (\bar{v}_e \bar{v}_f^* + \bar{v}_{ef}) |\rho_{ef} |^2.
\label{ifbsmz}
\eq
In $\sigma_{FB}$, the QCD-factor is set zero,
$R_{\mathrm {QCD}}=0$.

Helicities and polarizations
may be included in the Standard Model cross sections
$\sigma_A$ in a compact way for massless fermion
production~\cite{ba:phe,rokap}.
To do this, one must replace the above couplings, ${\cal I}_A(m,n;s)$,
with\footnote{
This is not rigorous with respect to $\bar{v}_{ef}$,
which has been assumed to factorize in order to simplify the notation.
The correct expression, implemented in \zf, can be obtained
by performing the multiplications in~( \ref{eq:c14}) and replacing
the product $\bar{v}_e(Z) \bar{v}_f(Z)$ with $\bar{v}_{ef}(Z)$.
This may be verified by explicitly  squaring the
matrix element (\ref{mzff}).}
\begin{eqnarray}
   \lefteqn{
 C_{T}(m,n;\lambda_{1},\lambda_{2},h_{1},h_{2}) = } \nonumber \\
&  \{\lambda_{1}[{\bar v}_{e}(m)
{\bar v}_{e}^{*}(n)   + {\bar a}_{e}(m)
{\bar a}_{e}^{*}(n)] \hspace{3mm} +
&  \lambda_{2}[{\bar v}_{e}(m)
{\bar a}_{e}^{*}(n)   +
{\bar v}_{e}^{*}(n)
{\bar a}_{e}(m)]\}  \hspace{.2cm} \times
    \nonumber \\
&  \{h_{1}[{\bar v}_{f}(m)
{\bar v}_{f}^{*}(n)   +
{\bar a}_{f}(m)
{\bar a}_{f}^{*}(n)]  \hspace{3mm}  +
&  h_{2}[{\bar v}_{f}(m)
{\bar a}_{f}^{*}(n)   +
{\bar v}_{f}^{*}(n)
{\bar a}_{f}(m)]\},
\label{eq:c14}
\end{eqnarray}
\begin{equation}
  C_{FB}(m,n;\lambda_{1},\lambda_{2},h_{1},h_{2}) =
  C_{T}(m,n;\lambda_{2},\lambda_{1},h_{2},h_{1}).
\label{c15}
\end{equation}
The vector  and  axial-vector
couplings  $\bar{v}_{f}(0)$ and $\bar{a}_{f}(0)$ of the fermion to the
photon
are:
\bq
   {\bar v}_f(0) = Q_f \, F_A(s), \hspace{1cm} {\bar a}_f(0) = 0.
\label{vaph}
\eq

Here, we introduce
the longitudinal polarizations of the electron
$(\lambda_-)$ and positron $(\lambda_+)$ and the helicities of the final
state fermions $h_\pm$ in the following combinations:
\begin{equation}
  \lambda_{1} = 1 - \lambda_{+}\lambda_{-}, \hspace{1cm}
  \lambda_{2} =  \lambda_{+} - \lambda_{-},
\end{equation}
\begin{equation}
  h_{1} = \frac{1}{4}(1 - h_{+}h_{-}), \hspace{1cm}
  h_{2} = \frac{1}{4}(h_{+} - h_{-}).
\label{eq:h217}
\end{equation}

The various parts of the cross section in
(\ref{sigmao}, \ref{ssmg}) now become:
\begin{equation}
   \sigma_{A}^{o}(s,s';m,n) =
\Re e \left\{
   C_{A}
   (m,n;\lambda_{1},\lambda_{2},h_{1},h_{2})  \frac{1}{2}
   [\chi_{m}(s')\chi_{n}^{*}(s) + \chi_{m}(s)\chi_{n}^{*}(s')]
\right\},
\label{eq:s10}
\end{equation}
\begin{equation}
   \chi_{\gamma}(s) =
\sqrt{N_{\gamma}(s)} {\cal K}_\gamma(s),
\label{eq:c212}
\end{equation}
\begin{equation}
   \chi_{Z}(s) =
   \sqrt{N_{\gamma}(s)} N_Z {\cal K}_Z(s),
\label{eq:c211}
\end{equation}
where again the QCD-factor $R_{\mathrm {QCD}}$ in $N_{\gamma}$ is set
equal zero for $\sigma_{FB}$.

If at least one incoming and one outgoing fermion are polarized, then the
contribution to the forward--backward anti-symmetric Standard Model cross
section from
pure photon exchange does {\em not} vanish as in (\ref{ifbsmg}).
This can be seen from formulae (\ref{eq:c14})-(\ref{eq:s10}).

We have not gone into details that are specific to Bhabha
scattering: this is done in~\cite{mp:bhr} and it represents a
straight-forward extension of the above formulae.

Asymmetries represent a clean and near systematic free measurement
with which to test various models.
In addition to the forward--backward asymmetry, several other asymmetries
are interesting.
It is useful to define a generic `spin' asymmetry, $A(h)$:
\begin{equation}
A(h) =
\frac{\sigma(h) - \sigma(-h)}
{\sigma(h) + \sigma(-h)},
\label{asym}
\end{equation}
where $h$ can denote the polarization of an incoming fermion
or the helicity of an outgoing one.

Choosing $h$ to be $h_+=+1$ the helicity
of a final-state $\tau^+$ and $\sigma(h)$ to be $\sigma_T(h_+)$,
$A(h)$ becomes the $\tau$~polarization,
$\lambda_\tau \equiv A_{\mathrm{pol}}$.
Similarly, one can define:
$A_{F}^{\mathrm{pol}}$,  $A_{B}^{\mathrm{pol}}$
as in~\cite{rstau} from (\ref{asym}) with
$\sigma(h) = \sigma_A(h_+), A= F,B,FB$, respectively.
The subscript $F$ ($B$) is used to indicate that only data from
the forward (backward) hemisphere are in the measurement; the
corresponding theoretical relations are given by (\ref{eq:s10}) and
(\ref{binsec}).
The forward--backward polarization asymmetry
$A_{FB}^{\mathrm{pol}}$
may be defined as follows:
\bq
A_{FB}^{\mathrm{pol}} =
\frac{\sigma_{FB}(h) - \sigma_{FB}(-h)}
{\sigma_T}.
\label{lafb}
\end{equation}

\subsection
[Higher-Order Corrections]
{Higher-Order Corrections
\label{subhigh}}

Here, we give a summary of treatment and common resummation
of some higher-order weak and QCD corrections in \zf.

Some higher-order terms are used to correct
$\Delta r$, $\rho$ and $\kappa$.
These terms are exclusively due to t-quark mass corrections.
\zf\ takes into account the following $m_t$-dependent corrections:
\begin{itemize}
\item complete $m_t$-dependent \oalf terms~\cite{zwidth},
\item leading \oaa terms~\cite{mp:hol,mp:bhr},
\item complete (either approximated as a function of energy or exact)
      \os terms~\cite{p1:kni,du:bc} 
      with leading part \ost.
\end{itemize}

For $\Delta r$ as introduced in (\ref{sw2delr}),
a common resummation of these leading terms may be performed
as follows~\cite{fs,dfs,ds,heraho}\footnote{
A detailed discussion of the \zf\ flags which control the
implementation of these corrections will be presented in
section~\ref{zuflag}.}:
\begin{equation}
\frac{1}{1-\Delta r} = \frac{1} {
\left[1 - \Delta \alpha(M_Z^2)\right]
( 1 +
{\displaystyle \frac {\cos^2 \theta_W}{\sin^2 \theta_W} }
\delta {\bar{\rho}}
) -
\Delta r_{\mathrm{rem}} },
\label{delr12}
\end{equation}
\bq
\Delta r_{\mathrm{rem}} =
\Delta r^{\mathrm{1loop}} -
\frac{\cos^2\theta_W}{\sin^2\theta_W} \Delta \rho
                     - \Delta \alpha(M_Z^2)
+ \Delta r_{\mathrm{rem}}^{\alpha \alpha_s},
\label{rem0rem}
\eq
\ba
\Delta \rho
&=&
\Delta \rho^{\alpha} +
\Delta \rho^{\alpha \alpha_s}
+ X_0  \nonumber \\
 &=& \displaystyle
 \frac{3 \alpha}{16 \pi \sin^2\theta_W \cos^2 \theta_W}
\frac{m_t^2}{M_Z^2}
\left[ 1 - \frac{2}{3} (1 + \frac{\pi^2}{3})
\frac
{\alpha_s(q^2,\Lambda_{\overline{\mathrm{MS}}})}{\pi}
\right] + X_0,
\label{dro0}
\ea
where
the $\Delta \alpha(s)$ is introduced in (\ref{dysfa})
and
\bq
X_0 = \Re e
\left[ \frac{ \Pi_Z(M_Z^2)}{M_Z^2} - \frac{ \Pi_W(M_W^2)}{M_W^2}
      \right]^{\mathrm{1loop}}_{\overline {\mathrm{MS}}}
      - \Delta \rho^{\alpha}.
\label{x0}
\eq
In $X_0$ the UV divergencies are removed according to the
${\overline{\mathrm{MS}}}$ renormalization scheme with
$\mu = M_Z$ \footnote{
In \cite{bible2} this corresponds
to a replacement of $M_W$ by $M_Z$ in the UV divergence
${\cal P}_{\mathrm{UV}}$.}.
The separation of $X_0$ is not uniquely
defined; it introduces a dependence of the resummation on the
renormalization procedure.
Further,
\ba
\delta {\bar{\rho}}
&=&
\delta {\bar{\rho}}^{\alpha} +
\delta {\bar{\rho}}^{\alpha^2} + \delta {\bar{\rho}}^{\alpha \alpha_s}
+ X    \nonumber \\
 &=& \displaystyle
3 {\cal T} \left[ 1 - (2 \pi^2 - 19) {\cal T}
- \frac{2}{3} (1+ \frac{\pi^2}{3})
\frac
{\alpha_s(q^2,\Lambda_{\overline{\mathrm{MS}}})}{\pi}
\right] + X,
\label{high2}
\ea
\begin{equation}
{\cal T} = \frac{G_{\mu}}{\sqrt{2}} \frac{m_t^2}{8 \pi^2},
\end{equation}
\label{calt}
\bq
X = 2 \sin^2\theta_W \cos^2\theta_W \frac{G_{\mu} M_Z^2}
{\sqrt{2} \pi \alpha} \left[1 - \Delta \alpha(M_Z^2) \right]
X_0.
\label{defx}
\eq
For the cross-section form factors, $\rho(s,\cos\vartheta)$ and
$\kappa(s,\cos\vartheta)$, and partial
\Z\ width form factors, $\rho^Z$ and
$\kappa^Z$, similar formulae hold:
\ba
\rho
=  \frac{\rho^{\mathrm{1loop}}
+ \rho_{\mathrm{rem}}^{\alpha \alpha_s}
- \Delta \rho}
{1-\delta \bar\rho},
\label{rho12z}
\ea
\bq
\kappa  =
\left( \kappa^{\mathrm{1loop}}
+ \kappa_{\mathrm{rem}}^{\alpha \alpha_s}
-
\frac {\cos^{2}\theta_{W}}  {\sin^{2}\theta_{W}}
 \Delta \rho \right)
\left(
1 + \frac {\cos^{2}\theta_{W}}  {\sin^{2}\theta_{W}}
 \delta \bar\rho
\right).
\label{kappa3z}
\eq
For the cross section alone, we have additionally:
\bq
\kappa_{ef}  =
\left( \kappa_{ef}^{\mathrm{1loop}}
+ \kappa_{ef,\mathrm{rem}}^{\alpha \alpha_s}
-
2 \frac {\cos^{2}\theta_{W}}  {\sin^{2}\theta_{W}}
 \Delta \rho \right)
\left(1 + \frac {\cos^{2}\theta_{W}}  {\sin^{2}\theta_{W}}
 \delta \bar\rho\right)^2.
\label{kapefz}
\eq

Some numerical examples of the effect of leading \oaa corrections
on the weak
mixing angle, the \MM\ total cross section and forward--backward
asymmetry at $s=M_Z^2$ are shown in table~\ref{ta6}.
\begin{table}[htbp]\centering
\begin{tabular}{|c|c|c|c|c|c|c|}
\hline
Observable & {\tt AMT4} & \multicolumn{5}{c|}{$m_t$ (GeV)} \\
\cline{2-7}
           &            &    100 &    150 &    200 &    250 &    300 \\
\hline
           &     0      &  .2324 &  .2269 &  .2199 &  .2107 &  .1986 \\
\Sw        &     1      &  .2324 &  .2271 &  .2206 &  .2128 &  .2040 \\
           &     2      &  .2324 &  .2270 &  .2204 &  .2123 &  .2025 \\
           &     3      &  .2324 &  .2270 &  .2205 &  .2125 &  .2029 \\
\hline
           &     0      & 1.4845 & 1.4861 & 1.4885 & 1.4920 & 1.4971 \\
$\sigma^{\mu}_T$&1      & 1.4847 & 1.4867 & 1.4898 & 1.4942 & 1.5004 \\
(nb)       &     2      & 1.4847 & 1.4867 & 1.4895 & 1.4935 & 1.4990 \\
           &     3      & 1.4847 & 1.4867 & 1.4896 & 1.4935 & 1.4990 \\
\hline
           &     0      & $-$.0043 & $-$.0024 & $-$.0004 &  .0015
           &  .0023 \\
$A^{\mu}_{FB}$&  1      & $-$.0045 & $-$.0024 &  .0004 &  .0040 &  .0078
 \\
           &     2      & $-$.0045 & $-$.0023 &  .0008 &  .0052 &  .0115
            \\
           &     3      & $-$.0045 & $-$.0024 &  .0007 &  .0050 &  .0113
            \\
\hline
\end{tabular}
\caption[
The weak mixing angle, muon pair production cross section
and asymmetry both with and without
leading ${\cal O}(\alpha^2 m_t^4)$ terms.
]{\it
The weak mixing angle, muon pair production cross section
and asymmetry both with (\mbox{\tt AMT4}$\neq$0) and without
(\mbox{\tt AMT4}=0)
leading ${\cal O}(\alpha^2 m_t^4)$ terms.
}
\label{ta6}
\end{table}
\normalsize

In section~\ref{subgamma} we introduced the QCD
correction factor, $R_{\mathrm {QCD}}$, in (\ref{rqcd}).
Its exact definition as implemented in \zf\ is given by~\cite{kataev}:
\begin{equation}
R_{\mathrm {QCD}}
= 1+\frac{\alpha_s}{\pi} +
    1.409\left(\frac{\alpha_s}{\pi}\right)^2 -
    12.805\left(\frac{\alpha_s}{\pi}\right)^3  \mbox{\tt QCD3}.
\label{qcdcor}
\end{equation}
For b~quarks, the
top- and bottom-quark mass-dependent QCD corrections
($c_1, c_2$) up to \oass have been taken from~\cite{as:ku}:
\begin{equation}
R_{\mathrm {QCD}}
 = 1+c_1(m_b)\frac{\alpha_s}{\pi} +
    c_2(m_b,m_t)\left(\frac{\alpha_s}{\pi}\right)^2 -
    12.805\left(\frac{\alpha_s}{\pi}\right)^3  \mbox{\tt QCD3}.
\label{qcdcob}
\end{equation}
Where {\tt QCD3} has the value 0 or 1 as required by the user.
In subroutine {\tt ZUWEAK},
$R_{\mathrm {QCD}}$ is calculated and the result is stored
in the variables {\tt QCDCOR} and {\tt QCDCOB}.

Depending on a flag,
the running strong interaction coupling constant~\cite{mar:pr}
$\alpha(q^2,\Lambda_{\overline{\mathrm{MS}}})$
is calculated with functions {\tt ALPHA4} or {\tt ALPHA5}~\cite{alpha4}:
\begin{equation}
\alpha_s(q^2,\Lambda_{\overline{\mathrm{MS}}}) =
\frac{4 \pi}{b_0 A}
 \left[ 1 - \frac{b_1}{b_0^2 A}\ln A
+ \left(\frac{b_1}{b_0^2 A}\right)^2 \left\{
\left(\ln A - \frac{1}{2}\right)^2
+ b_2 \frac{b_0}{b_1^2}            - \frac{5}{4} \right\} \right],
\label{eq:alfas}
\end{equation}
with
\bq
b_0 =  11 - \frac{2}{3}n_f,  \hspace{1cm}
b_1 =  102 - \frac{38}{3}n_f,\hspace{1cm}
b_2 = \frac{1}{2} \left[2857 - \frac{5033}{9}n_f + \frac{325}{27}n_f^2
\right],
\label{qcd2}
\eq
\bq
A
  = \ln \frac{q^2}{\Lambda_{\overline{\mathrm{MS}}}^{(n_f)2}}.
\label{qcd3}
\eq

Here $q^2$ represents the energy scale and $n_f$ the number of quark
flavors.
The corresponding definition of
$\Lambda_{\overline{\mathrm{MS}}}$ may be found in
table~\ref{tablam}.  \\
\begin{table}[thbp]\centering
\begin{tabular}{|c|c|c|}
  \hline
                 &              &               \\
 $q^2$           & $n_f$         & $\Lambda_{\overline{\mathrm{MS}}}$  \\
                 &              &               \\ \hline
                &              &               \\
$\leq m_c^2$    & 3            &
$\Lambda^{(4)}_{\overline{\mathrm{MS}}}
\left(\frac{m_c}{\Lambda^{(4)}_
{\overline{\mathrm{MS}}}}\right)^{\frac{2}{27}}
\ln \left(\frac{m_c^2}{(\Lambda^{(4)}_{\overline{\mathrm{MS}}})^2}\right)
^
{\frac{107}{2025}}$ \\
                &              &               \\
$\leq m_b^2$    & 4            &
$\Lambda^{(4)}_{\overline{MS}}$ \\
                &              &               \\
$\leq m_t^2$     &  5            &
$\Lambda^{(4)}_{\overline{MS}}
\left(\frac{\Lambda^{(4)}_{\overline{\mathrm{MS}}}}{m_b}\right)
^{\frac{2}{23}}
\ln \left(\frac{m_b^2}{(\Lambda^{(4)}_{\overline{\mathrm{MS}}})^2}\right)
^{\frac{-963}{13225}}$ \\
                 &              &               \\
$ > m_t^2$       &  6            &
$\Lambda^{(5)}_{\overline{\mathrm{MS}}}
\left(\frac{\Lambda^{(5)}_
{\overline{\mathrm{MS}}}}{m_t}\right)^{\frac{2}{21}}
\ln \left(\frac{m_t^2}{(\Lambda^{(5)}_{\overline{\mathrm{MS}}})^2}\right)
                      ^{\frac{-107}{1127}}$ \\
                 &              &               \\
  \hline
\end{tabular}
\caption
[
$\Lambda_{{\mathrm{MS}}}$ for different energy regions.]
{\it
$\Lambda_{\overline{\mathrm{MS}}}$ for different energy regions.}
\label{tablam}
\end{table}

In $R_{\mathrm {QCD}}$, the $\alpha_s$ is calculated with $q^2=s$
for cross sections or $q^2=M_Z^2$ for the \Z\ width.
In the \os corrections the scale is chosen to be
$q^2 = M_Z^2$ for light quarks and
$q^2 = \max \{M_Z^2,m_t^2 \}$ for the tb doublet.
One can see that in the LEP energy region the difference between the
approximate\footnote{The approximation realized in \zf\ is a Taylor
series expansion in $s/m_t^2$ for $s$ smaller than $m_t^2$.
The leading term of this expansion is included in (\ref{high2}).}
and exact treatment of the \os corrections is            minor.

Table~\ref{truns} shows the
$\alpha_s(q^2,\Lambda_{\overline{\mathrm{MS}}})$
as a function of $q^2$.
\begin{table}[htbp]\centering
\begin{tabular}{|c|c|c|c|c|c|c|}  \hline
 $|q|$ (GeV)  & 50 & 100 & 150 & 200 & 250 & 300 \\  \hline
 $\alpha_s$ & .1193 & .1079 & .1022 & .0985 & .0958 & .0937  \\   \hline
\end{tabular}
\caption
[Running $\alpha_s(q^2)$ versus $|q|$.]
{\it
Running $\alpha_s(q^2)$ versus $|q|$.
\label{truns}}
\end{table}

The dependence of weak parameters on $\alpha_s$
is shown in table~\ref{trokap} as a function of how the \os corrections
are applied.  
\vfill
\footnotesize
\begin{table}[bhtp]\centering
\begin{tabular}{|c|c|c|c|c|}    \hline
 &                &              &               &
\\ 
{\tt QCDC} &  $\rho_{ef}$  &  $\kappa_e $   & $\kappa_f$ &
$\kappa_{ef}$
\\ 
 &            &                &            &
 \\   \hline
 \multicolumn{5}{|c|}{$\nu\bar{\nu}$ final state}
\\   \hline
  &                &                &                &                \\
0 & 1.004 $-$ i0.002 & 1.025 + i0.013 &                &
\\
1 & 1.004 $-$ i0.002 & 1.023 + i0.014 &                &
\\
2 & 1.004 $-$ i0.002 & 1.022 + i0.014 &                &
\\
  &                &                &                &
  \\   \hline
 \multicolumn{5}{|c|}{$l\bar{l}$ final state} \\
\hline
  &                &                &                &                \\
0 & 1.003 $-$ i0.005 & 1.025 + i0.013 & 1.025 + i0.013 & 1.050 + i0.026
\\
1 & 1.002 $-$ i0.005 & 1.023 + i0.014 & 1.023 + i0.014 & 1.046 + i0.027
\\
2 & 1.002 $-$ i0.005 & 1.022 + i0.014 & 1.022 + i0.014 & 1.045 + i0.027
\\
  &                &                &                &      \\   \hline
 \multicolumn{5}{|c|}{$u\bar{u}$ final state} \\
\hline
  &                &                &                &                \\
0 & 1.003 $-$ i0.004 & 1.025 + i0.013 & 1.024 + i0.012 & 1.050 + i0.026
\\
1 & 1.003 $-$ i0.004 & 1.023 + i0.014 & 1.022 + i0.013 & 1.045 + i0.026
\\
2 & 1.003 $-$ i0.004 & 1.022 + i0.014 & 1.022 + i0.013 & 1.045 + i0.026
\\
  &                &                &                &      \\   \hline
 \multicolumn{5}{|c|}{$d\bar{d}$ final state} \\
\hline
 &                &                &                &                \\
0 & 1.003 $-$ i0.003 & 1.025 + i0.013 & 1.024 + i0.012 & 1.049 + i0.025
\\
1 & 1.003 $-$ i0.003 & 1.023 + i0.014 & 1.022 + i0.012 & 1.045 + i0.026
\\
2 & 1.003 $-$ i0.003 & 1.022 + i0.014 & 1.022 + i0.012 & 1.044 + i0.026
\\
 &                &                &                &      \\
\hline
 \multicolumn{5}{|c|}{$b\bar{b}$ final state} \\
\hline
 &                &                &                &                \\
0 & 0.999 $-$ i0.003 & 1.025 + i0.013 & 1.028 + i0.012 & 1.054 + i0.025
\\
1 & 0.999 $-$ i0.003 & 1.023 + i0.014 & 1.026 + i0.012 & 1.049 + i0.026
\\
2 & 0.999 $-$ i0.003 & 1.022 + i0.014 & 1.026 + i0.012 & 1.049 + i0.026
\\
  &                &                &                & \\ \hline
\end{tabular}
\caption
[
The dependence of the form factors $\rho_{ef}$ and
$\kappa_e, \kappa_f, \kappa_{ef}$ at the \Z\ peak on different
treatments of \os corrections.]
{\it
The dependence of the form factors $\rho_{ef}$ and
$\kappa_e, \kappa_f, \kappa_{ef}$ at the \Z\ peak on {\mbox{different}}
treatments of \os corrections
(\mbox{\tt QCDC=0} - no,
\mbox{\tt 1} - approximate
and \mbox{\tt 2} - exact \os corrections are applied).
}
\label{trokap}
\end{table}
\normalsize

\clearpage
\section
[The Hard-Scattering Process: (II) Model-In\-de\-pen\-dent Bran\-ches]
{
The Hard-Scattering Process:  \\
  (II) Model-In\-de\-pen\-dent Bran\-ches
\label{branches.2}}
\setcounter{equation}{0}

\subsection
[Effective Couplings]
{Effective Couplings
\label{subeff}}

\begin{figure}[h]
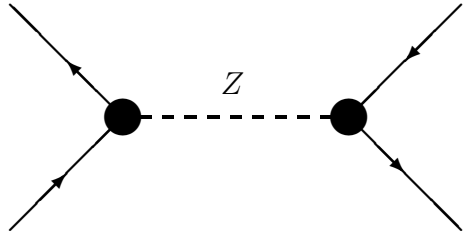

\begin{Feynman}{60,30}{0,0}{1.0}
\put(15,15){\fermionur}
\put(15,15){\fermionul}
\put(15,15){\circle*{5}}
\put(15,15){\gaugebosonright}
\put(28,18){$Z$}
\put(45,15){\circle*{5}}
\put(45,15){\fermiondl}
\put(45,15){\fermiondr}
\end{Feynman}
\caption[
Matrix element in the approach with effective \Z\ couplings.
]{\it
Matrix element in the approach with effective \Z\ couplings.
\label{figmeff}}
\end{figure}

In a simple quantum mechanical approach, the \Z\
boson may be assumed to
have real constant vector ($\hat{v}_f$) and axial-vector
($\hat{a}_f$) couplings to fermions ($f$).
This ansatz may be realized by a replacement
of the renormalized effective couplings as predicted from  the Standard
Model by \naive\ effective couplings in the cross section, see
fig.~\ref{figmeff}.    \\

The cross sections for the hard-scattering process are:
\begin{eqnarray}
\sigma_A^{o,\mathrm{eff}}(s) &=&
\sigma_A^{o,\mathrm{eff}}(s;\gamma,\gamma) +
\sigma_A^{o,\mathrm{eff}}(s;\gamma,Z)      +
\sigma_A^{o,\mathrm{eff}}(s;Z,Z)                 \nonumber \\
  &=&
{\cal I}_A^{\mathrm{eff}} (\gamma,\gamma;s)
+
 \Re e \left[ {\cal I}_A^{\mathrm{eff}} (\gamma,Z;s)
                      {\cal K}_Z^\ast(s) \right]
+
{\cal I}_A^{\mathrm{eff}} (Z,Z;s)
\left| {\cal K}_Z(s) \right|^2.
\label{sefg}
\end{eqnarray}
Here, the generalized couplings for the total cross section are:
\bq
{\cal I}_T^{\mathrm{eff}} (\gamma,\gamma;s) =
c_m  N_\gamma(s)  \,
|F_A(s)|^2 \, Q_e^2 Q_f^2,
\label{itggeff}
\eq
\bq
{\cal I}_T^{\mathrm{eff}} (\gamma,Z;s) =
2 c_m             N_Z                  N_\gamma(s)
F_A(s) \, |Q_e Q_f| \, {\hat v}_e {\hat v}_f,
\label{itgzeff}
\eq
\bq
{\cal I}_T^{\mathrm{eff}} (Z,Z;s) =
N_Z^2  N_\gamma(s)
\left[
c_m
( {\hat v}_e^2 + {\hat a}_e^2)
( {\hat v}_f^2 + {\hat a}_f^2)
-
\frac{6 m_f^2}{s}  ( {\hat v}_e^2 + {\hat a}_e^2) {\hat a}_f^2
\right],
\label{itzzeff}
\eq
where $N_\gamma(s)$ and $N_Z$ are defined in (\ref{ngam})-(\ref{nz}),
$c_m$
in (\ref{cm}), and $F_A$ in (\ref{dysfa}).

The asymmetric cross-section part is defined by:
\bq
{\cal I}_{FB}^{\mathrm{eff}} (\gamma,\gamma;s) =  0,
\label{iaggeff}
\eq
\bq
{\cal I}_{FB}^{\mathrm{eff}} (\gamma,Z;s) =
2 \mu(s) N_Z N_\gamma(s)
F_A(s) \, |Q_e Q_f| \, {\hat a}_e {\hat a}_f,
\label{iagzeff}
\eq
\bq
{\cal I}_{FB}^{\mathrm{eff}} (Z,Z;s) =
8 \mu(s) N_Z   N_\gamma(s) \,
{\hat v}_e {\hat a}_e {\hat v}_f {\hat a}_f,
\label{iazzeff}
\eq
where $\mu(s)$ is defined in (\ref{defmu}).

One may interpret effective couplings as approximations to the weak form
factors of the Standard Model e.g.:
\bq
{\hat a}_f \equiv 2 \hat{g}_a^f \sim \Re e \sqrt{\rho_{ef}(M_Z^2)} a_f,
\label{hatasm}
\eq
\bq
{\hat v}_f \equiv 2 \hat{g}_v^f \sim
\Re e \left[ \sqrt{\rho_{ef}(M_Z^2)}
                                     {\bar v}_f(M_Z^2) \right],
\label{hatvsm}
\eq
where the alternate notation ($\hat{g}_a^f,\hat{g}_v^f$) is favored by
the
LEP experiments.
We neglect here possible dependences on the scattering angle.
In \zf\ the normalization $a_f$=1 is used for all fermions.
In addition, one may choose an alternative parametrization in terms of
the effective
weak neutral current amplitude normalization, $\hat{\rho}_f$:
\bq
\hat{\rho}_f \equiv \frac{\hat{g}_a^f}{g_a^f},
\label{hatrhoa}
\eq
\bq
\hat{\rho}_e \hat{\rho}_f \sim \Re e \sqrt{\rho_{ef}(M_Z^2)},
\label{hatrho2}
\eq
\bq
{\hat v}_f \sim \Re e \left[ \sqrt{\rho_{ef}(M_Z^2)} {\bar v}_f(M_Z^2)
\right].
\label{hatvsma}
\eq

In the present approach, one
can leave $\Gamma_Z$ as a free fundamental parameter of the ansatz.
Alternatively, one may define it through the second line of
(\ref{defzwidth}), replacing there the renormalized ($\bar{v}^Z,
\bar{a}^Z$)
couplings by effective ($\hat{v}^Z, \hat{a}^Z$) ones.

In either case, one must realize that the normalization of the \Z\ width
may change depending on the definition of $M_Z$
(see earlier discussion in connection with
equations (\ref{mg}) and (\ref{gbar})).
For additional general comments on cross sections and asymmetries we
refer to section~\ref{branches.1}.

In principle, this branch is completely model independent.
However, \zf\ users should be aware that the current implementation
contains small Standard Model contributions in the form of imaginary
parts of weak form factors (see table~\ref{trokap}).

\subsection
[Partial \Z\ Widths]
{Partial \Z\ Widths
\label{subparti}}

\begin{figure}[hb]
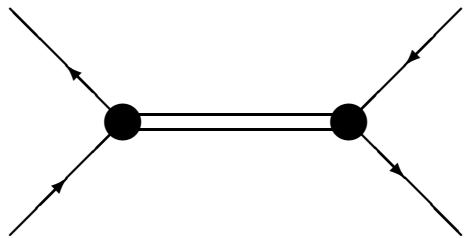

\begin{Feynman}{60,30}{0,0}{1.0}
\put(15,15){\fermionur}
\put(15,15){\fermionul}
\put(15,15){\circle*{5}}
\put(15,16){\line(1,0){30}}
\put(15,14){\line(1,0){30}}
\put(45,15){\circle*{5}}
\put(45,15){\fermiondl}
\put(45,15){\fermiondr}
\end{Feynman}
\caption[
Matrix element for the \Z\ resonance scattering.
]{\it
Matrix element for the \Z\ resonance scattering.
\label{figmgamz}}
\end{figure}

This approach to determining cross sections relies on the assumption
that scattering through the \Z\ boson may be considered as subsequent
formation and decay of a resonance, see fig.~\ref{figmgamz}.
The corresponding net cross section is (see {    e.g.}~\cite{modind1}
and references therein):
\bq
\sigma_T^{o,\Gamma}(s) =
\sigma_T^{o,\mathrm{SM}}(s;\gamma,\gamma)
+   I^{\Gamma}   +
\sigma_T^{o,\Gamma}(s;Z,Z),
\label{sgamt}
\eq
\bq
\sigma_T^{o,\Gamma}(s;Z,Z)
= {\cal I}_T^{\Gamma}(Z,Z;s) \, |{\cal K}_Z(s)|^2,
\label{sgamz}
\eq
\bq
{\cal I}_T^{\Gamma}(Z,Z;s) = \frac{3}{8 s}
\frac{12 \pi c_m \Gamma_e \Gamma_f}{\mu (M_Z R_{\mathrm{QED}})^2}.
\label{defitg}
\eq
The photonic
contribution to the cross section, $\sigma_T^{o,{\mathrm{SM}}}$,
is given in (\ref{itsmg})-(\ref{itsmi}).

One of the attractive features of the partial widths approach is the
treatment of hadron production.
Here one simply replaces
$\Gamma_f$ with $\Gamma_{\mathrm{had}}$ in (\ref{defitg}).
In case of hadron production, we have to use the quark language
for the photonic
cross-section contribution in (\ref{sgamt}), which is taken
over from (\ref{ssmg}) in section \ref{subfpair} without changes.

The interference term in (\ref{sgamt}) presents some complications with
this approach.
There are at least four different ways dealing with this term:
an exact calculation, a simple parametrization,
ignoring the term completely, or assuming the Standard Model prediction.

A correct handling would
rely on partial
decay widths into specific helicity states, $\Gamma_\pm(f)$,
as proposed in~\cite{modind1}:
\bq
 I^{\Gamma} \equiv \sigma_T^{o,\Gamma}(s;\gamma,Z)
= {\cal I}_T^{\Gamma}(\gamma,Z;s) \, \Re e {\cal K}_Z(s),
\label{sgami}
\eq
\bq
{\cal I}_T^{\Gamma}(\gamma,Z;s) = \pm \frac{3}{8 s}
\frac{
4 \pi Q_e Q_f \alpha(s)
c_m 
}
{M_Z R_{\mathrm{QED}} }
\left[ \Gamma_+^{\frac{1}{2}}(e)  - \Gamma_-^{\frac{1}{2}}(e) \right]
\, \left[\Gamma_+^{\frac{1}{2}}(f) - \Gamma_-^{\frac{1}{2}}(f)  \right].
\label{defiti}
\eq
In practice, however, experimental measurements
seem not to deliver a sufficiently high number of degrees of
freedom to use this formula.

A simple parametrization of the interference term, as has been
realized in the S-matrix approach, could be used.
This, however, leads to large uncertainties in the \Z\ mass
determination~\cite{smatrix}.

Ignoring the interference term completely (\IE\ assuming that it is
identically zero) is another possibility,
since this term is expected to be small.
In addition this removes one degree of freedom.

The last alternative is to assume the interference
term from the Standard Model, as defined in (\ref{ssmg}), (\ref{itsmi}):
\bq
I^{\Gamma}   =
\sigma_T^{o,\mathrm{SM}}(s;\gamma,Z).
\label{sgamint}
\eq
This is the approach that has been implemented in \zf.

In addition to the partial decay widths of the \Z, the total \Z\ width
and mass  are free parameters with this approach.
There
is an ambiguity in the normalization of the total and partial widths
which is due to
the different choices for the definition of the \Z \, propagator
(see discussion at the beginning of section~\ref{branches.1}).
The energy-dependent total \Z\ width, $\Gamma_Z(s)$, is related to the
constant total \Z\ width, ${\bar\Gamma}_Z$, by (\ref{mg}), (\ref{gbar}).
In relating the two approaches to the resonance definition,
there is no explicit constraint on the partial widths.

They may be related as follows.
In the Standard Model branch, the residua of the resonance functions
were normalized by the Fermi constant.
With this approach the actual residua are contained in the partial
widths.
Comparing the two,
one can derive the relation between the partial widths
for the two different definitions of the \Z\ propagator
from (\ref{gmub}):
\bq
{\bar \Gamma}_f = \frac{\Gamma_f}{\sqrt{1+(\Gamma_Z/M_Z)^2}}.
\label{relgamf}
\eq
This relation is in full accordance with the corresponding relation
for the total width~(\ref{gbar}).

\subsection
[S-Matrix]
{S-Matrix
\label{subsma}}

Besides the approach to the effective Born cross section based on the
Standard Model or on one of its extensions, there is only one accurate
model-independent approach to the $Z$ line shape. One can derive this
{\em rigorous} model-independent formula either starting from an
analysis of the Standard Model results~\cite{borrelli}
or from S-matrix theory.

\begin{figure}[h]
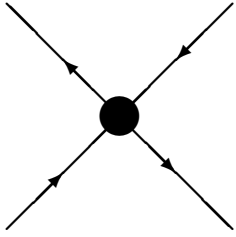

\begin{Feynman}{30,30}{0,0}{1.0}
\put(15,15){\fermionur}
\put(15,15){\fermionul}
\put(15,15){\circle*{15}}
\put(15,15){\fermiondl}
\put(15,15){\fermiondr}
\end{Feynman}
\caption[
Scattering in the S-matrix approach.
]
{\it
Scattering in the S-matrix approach.
\label{figsmat}}
\end{figure}

An early application of the S-matrix formalism to LEP~I physics
may be found in \cite{amartin}.
Recently, it has been proposed to use S-matrix theory
for a global description of the hard-scattering process
\cite{stuart1,smatrix}.
Such an ansatz
has the advantage that it contains no special assumptions on the dynamics
beyond general principles and
the existence of both photon and \Z \, boson.
In \cite{smatrix} the
necessary formalism has been described.
One starts from the incoherent sum of four
squared matrix elements for the scattering of helicity fermion
states ($e^-_Le^+_R \rightarrow f_L\bar{f}_R,\
e^-_Le^+_R \rightarrow f_R\bar{f}_L,\
e^-_Re^+_L \rightarrow f_R\bar{f}_L,\
e^-_Re^+_L \rightarrow f_L\bar{f}_R$) as seen in
fig.~\ref{figsmat}.

All have the following structure:
\begin{equation}
{\cal M}_i(s) = \frac {R_{\gamma}}{s} + \frac {R_{Z}^i}{s-s_{Z}}
+ F_i(s_Z)
+ (s-s_Z) F_i'(s_Z) + (s-s_Z)^2  F_i''(s_Z) + \ldots
\label{defsmatrix}
\end{equation}
The location of the \Z-boson pole is given by $s_Z$ in the complex
energy plane, and $R_{\gamma}$ and $R_Z^i$ represent constant complex
residuals.
The cross section is:
\begin{equation}
\sigma_T^{o,{\mathrm S}}(s) =  \frac{1}{4} \sum_{i=1}^4
            s |{\cal M}^i (s)|^2.
\label{sigma}
\end{equation}
In order to fit the cross section, it is perhaps useful to decompose
the above into a series of real-valued
terms with rising powers of $(s-M_Z^2)$
\footnote{
Each of the coefficients
itself is then a series in the variable $\Gamma_Z/M_Z$.
}:   
\begin{equation}
\displaystyle{\sigma_T^{o,{\mathrm S}} (s) =
\frac{r_{\gamma}}{s} +
\frac{ M_Z^2 R +  \left(s - M_Z^2 \right) I } {\left| s - s_Z \right|^2}
+ \frac{r_0}{M_Z^2}
+ (s-M_Z^2) \frac{r_1}{M_Z^4} + (s-M_Z^2)^2 \frac{r_2}{M_Z^6}  + \ldots}
\label{smasigfin}
\end{equation}
where we defined the constants to be dimensionless and
\bq
r_{\gamma} =  |R_{\gamma}|^2,
\label{rgamma}
\eq
\bq
R_{\gamma}   = \left\{
\begin{array}{ll}
Q_e \, \sqrt{3 Q_d^2 + 2 Q_u^2}
\sqrt{ \frac{4\pi}{3} c_f
R_{\mathrm{QCD}}
}
       \,            \alpha(M_Z^2)
                            &   \mbox{{\rm{for hadrons at LEP~I}}}
 \hfill
\nonumber \\
Q_e Q_f \, \sqrt{ \frac{4\pi}{3}} \, \alpha(M_Z^2)
                            & \mbox{{\rm{for leptons.}}} \hfill
                            \\
\end{array}
\right.
\label{rgammas}
\eq
Here $r_{\gamma}$
depends only on the dynamics of the photon, while the other parameters
($M_Z, \Gamma_Z, R, I, \linebreak[0] r_0, \linebreak[0]
\ldots$) depend also on the \Z.
The ansatz (\ref{smasigfin}) may be compared with the notation used
in the foregoing sections:
\begin{eqnarray}
\sigma_T^{o,\rm S}(s) &=&
\sigma_T^{o,\rm S}(s;\gamma,\gamma) +
\sigma_T^{o,\rm S}(s;\gamma,Z)      +
\sigma_T^{o,\rm S}(s;Z,Z)
           \nonumber \\
  &=&
{\cal I}_T^{\mathrm S} (\gamma,\gamma;s)
+
 \Re e \left[ {\cal I}_T^{\mathrm S} (\gamma,Z;s)
                      {\cal K}_Z(s) \right]
+
{\cal I}_T^{\mathrm S} (Z,Z;s)
\left| {\cal K}_Z(s) \right|^2.
\label{ssg}
\end{eqnarray}
Here,
\begin{eqnarray}
{\cal I}_T^{\mathrm S} (\gamma,\gamma;s) &=& \frac{r_{\gamma}}{s} +
\ldots,
\\
{\cal I}_T^{\mathrm S} (\gamma,Z;s) &\equiv& \frac{\cal J}{s} =
\frac{I - R}{s} + \ldots,
\\
{\cal I}_T^{\mathrm S} (Z,Z;s) &=& \frac{R}{s} + \ldots
\label{safcon}
\end{eqnarray}
The dots indicate contributions from the Taylor coefficients;
for instance,
  the photon-exchange contribution collects not only the $r_{\gamma}$
but also small additional terms due to the dependence of
the running QED coupling on $s$ or
$s'$
\footnote{
Strictly speaking,
the residuum of the photon pole (\ref{rgammas}) is
not $\alpha(M_Z^2)$ but
the QED coupling constant $\alpha$ at zero momentum;
the difference is related to non-leading terms
and may be absorbed by a redefinition of $r_i$.}.
The leading contribution to the $\gamma Z$ interference comes from
the combination ${\cal J}=I-R$;  in order to simplify the $s$ dependence
of the ansatz, $R$ has been introduced in (\ref{smasigfin}) with a
coefficient of $M_Z^2$ instead of $s$.

In order to get an intuitive feeling for the meaning of the S-matrix
parameters, it may be helpful to contrast this approach with that of the
effective couplings discussed in section~\ref{subeff}.
The cross section for muon production (\ref{smasigfin}) in this
{\em approximation} is given by fixing:
\bq
R = c^2 ({\hat v}_e^2+{\hat a}_e^2)
            ({\hat v}_\mu^2+{\hat a}_\mu^2),
\label{Rsma}
\eq
\bq
I = R + {\cal J},
\label{Isma}
\eq
\bq
{\cal J} = 2 c R_{\gamma}  {\hat v}_e {\hat v}_\mu,
\label{Jsma}
\eq
\bq
r_i = 0, \,\,\, i = 0,1,\ldots
\label{rasm}
\eq
with
\bq
c = \sqrt{\frac{4\pi}{3}}
    \frac{G_{\mu}}{\sqrt2} \frac{M_Z^2}{8\pi}.
\label{smapred}
\eq

Contrasting the S-matrix to the partial
width approach, one obtains instead:
\bq
R = 12 \pi \Gamma_e \Gamma_f + \ldots,
\label{rygam}
\eq
and the $\gamma Z$ interference part has to be fixed by a relation
analogous to (\ref{Isma})
(see the lengthy discussion in section~\ref{subparti}).
An exact treatment would enlarge the number of parameters to be fitted,
and weaken the numerical result. 

The general form of the above parameters $(R, I, r_0, \ldots)$
may be found in \cite{smatrix} (they were not made dimensionless
there as is done here).
In a quantum field theory, the constants $(r_0,r_1,\ldots)$
are non-vanishing, owing
 due to non-resonating quantum corrections.
A careful analysis of their calculation in accordance with
the S-matrix properties has been performed in~\cite{stuart1}.

An ansatz quite similar to~(\ref{smasigfin}) has been derived
in~\cite{borrelli}, starting from an on-mass-shell renormalization of the
Standard Model; for the production of flavor $f$:
\bq
\displaystyle{
\sigma_T^{o,{\mathrm S}} (s) =
\frac
{12 \pi \Gamma_e \Gamma_f} {\left| s - s_Z \right|^2}
\left\{
\frac{s}{M_Z^2}  +   {\cal R}_f \frac{s-M_Z^2}{M_Z^2}
+ {\cal I}_f \frac{\Gamma_Z}{M_Z} + \ldots
\right\}
+ \sigma_{\mathrm{QED}}^f,
}   
\label{boreli}
\eq
where terms of higher order in $(s-M_Z^2)/M_Z^2$ and in
$\Gamma_Z / M_Z$ are dropped.
There is a simple one--to--one correspondence to the terms
in~(\ref{smasigfin}), with exclusion of the ${\cal I}_f$;
the dominating part of this correction is due
to the imaginary part of the running QED coupling $\alpha(s)$.
The corresponding contribution in our notations may be found in
the exact definition of $R$:
\bq
R = \frac{1}{4} \sum_i |R_Z^i|^2 +
2 \frac{\Gamma_Z}{M_Z} \Im m R_{\gamma}^{*}
\left( \frac{1}{4} \sum_i R_Z^i \right) + \ldots
\label{Rexa}
\eq

As in the aforementioned branches,
the definitions of mass and width of the \Z\ boson
are correlated and deserve special attention.

 The possibility to describe asymmetries is mentioned in~\cite{smatrix}.

\section
[Beyond the Standard Model]
{Beyond the Standard Model
\label{subbeyo}}

In recent years, many searches for
possible effects from {\em New Physics} have been undertaken
in precision experiments at
\LEPI.
Rewievs of the present status and of the literature
may be found in \cite{je:ihvp1b}--\cite{altar}.
Here, we would like to restrict ourselves to some hints on the
possible use of \zf\ for corresponding searches.

To start with, let us assume that some more general theory leads to
predictions for the scattering of two fermions into two fermions.
This may be described by an additional
matrix element
${\bf\cal M}_{E}$, to be added to ${\bf\cal M}_{\gamma}$ and
${\bf\cal M}_{Z}$:
\begin{eqnarray}
{\bf\cal M}_{E}(s,\cos \vartheta)  \sim C_E
      [ u_{\rho} L_{\beta} \otimes L^{\beta}
      + u_e \gamma_{\beta} \otimes L^{\beta}
      + u_f L_{\beta} \otimes \gamma^{\beta}
      + u_{ef} \gamma_{\beta} \otimes \gamma^{\beta} ],
\label{ME}
\end{eqnarray}
and the $C_E, u_a$ can depend on $s$ and $\cos \vartheta$.
An instructive example for new physics at the Born level
is an additional heavy neutral gauge boson $Z'$ with
mass $M_{Z'}$, width $\Gamma_{Z'}$, vector and axial-vector couplings
$v'_f, a'_f$, and coupling constant $g_{Z'}$.
As long as this $Z'$ does
not mix with the ordinary $Z$, the influence on the scattering
process is due to ${\bf\cal M}_{E}$ and thus limited to higher
energy~\cite{zplinac}:
\begin{eqnarray}
C_E = C_{Z'} \equiv
 \frac{g_{Z'}^2}{s-M_{Z'}^2 + i M_{Z'} \Gamma_{Z'}},
\label{CE}
\end{eqnarray}
\begin{eqnarray}
u_{\rho} = a'_e a'_f \rho'_{ef},
\hspace{0.5cm}
u_f      = (v'_f-a'_f) \rho'_{ef} \kappa'_f,
\hspace{0.5cm}
u_{ef}   = (v'_e-a'_e)(v'_f-a'_f) \rho'_{ef} \kappa'_{ef},
\label{UR}
\end{eqnarray}
with $\rho' = \kappa' = 1$ for \zp\ Born physics.

In general, the corrections $u_a$ may also be due to some loop insertions
to the $Z$ matrix element
from a generalized renormalizable theory, or even simply due to
some Standard Model corrections
not yet included in the definitions of the weak form factors. An example
of the latter case had been given in the $Z$-vertex corrections from
t-quark exchange, see~(\ref{rsbb})-(\ref{kseb}).
Usually, the additional loop corrections are incorporated into the
$Z$ matrix element (\ref{Mva}):
\begin{equation}
{\cal M}(Z,E) = {\cal M}_{Z}  +  {\cal M}_{E},
\label{MZE}
\end{equation}
\begin{eqnarray}
{\bf\cal M}(Z,E)(s,\cos \vartheta)  \sim  C_Z
\biggl\{
      \left[  \bar a_e \bar a_f + \frac{C_E}{C_Z} u_{\rho} \right]
      \gamma_{\beta} \gamma_5 \otimes
                                \gamma^{\beta} \gamma_5
      + \left[
      \bar v_e \bar a_f + \frac{C_E}{C_Z} (u_{\rho} + u_e) \right]
      \gamma_{\beta} \otimes \gamma^{\beta} \gamma_5
\nonumber \\
      + \,
      \left[ \bar a_e \bar v_f + \frac{C_E}{C_Z}(u_{\rho}+u_f)\right]
      \gamma_{\beta} \gamma_5 \otimes \gamma^{\beta}
+ \left[ \bar v_{ef} + \frac{C_E}{C_Z} (u_{\rho} + u_e + u_f + u_{ef})
\right]
\gamma_{\beta} \otimes \gamma^{\beta}
\biggr\},
\label{MVAE}
\end{eqnarray}
\bq
C_Z = \frac{G_{\mu}}{s - M_Z^2 + i M_Z \Gamma_Z}.
\label{cz}
\eq
In case of loop corrections to the $Z$ propagator, the ratio
$C_E/C_Z$ is free of the resonating $s$ dependence around the $Z$ peak,
and in some scenarios the corrections are even constant.

In general, however, this is not the
case.
Coming back to the example of an additional $Z'$, there
is evidently a potentially remarkable $s$ dependence of the insertions,
being even resonating near the $Z'$ peak.

In any case, one can go a step further and include the non-standard
corrections into the form factors introduced in (\ref{mzff})
by the following replacements:
\begin{equation}
\rho_{ef} \rightarrow \rho_{ef}(Z,E)
= \rho_{ef} \left( 1 + \frac{C_E}{C_Z}
                  \frac {u_{\rho}} {\rho_{ef} a_e a_f} \right),
\label{rhoze}
\end{equation}
\begin{equation}
\kappa_f \rightarrow \kappa_f(Z,E) = \frac{\rho_{ef}} {\rho_{ef}(Z,E)}
\kappa_f
   \left[1 + \frac{C_E}{C_Z} \frac{u_f}
   {\rho_{ef}
    \kappa_f a_e(v_f-a_f)} \right],
\label{kafze}
\end{equation}
\begin{equation}
\kappa_{ef} \rightarrow
\kappa_{ef}(Z,E) =
\frac{\rho_{ef}}{\rho_{ef}(Z,E)}\kappa_{ef}
\left[1+ \frac{C_E}{C_Z} \frac{u_{ef}} {\rho_{ef}\kappa_{ef}
  (v_e-a_e)(v_f-a_f)} \right].
\label{kaefze}
\end{equation}
The above replacements ensure the interpretation of the weak
form factors as
finite renormalizations of Fermi constant and weak mixing angle; see
(\ref{gmu}), (\ref{swkap}).
They can, however,
drastically change the numerical behaviour of the form factors,
which now need no longer  be small.
The advantage of the above formulae is two-fold.
Besides the compact
notation and simple interpretation, they may be used not only for the
description of the fermion pair production process~(\ref{firsteq}).
Without changes, they describe
also the effects of new physics in Bhabha scattering or in the crossed
channel, i.e. $ep$ scattering.

Besides  the neutral current amplitude
${\bf\cal M}_{E}$, new physics may show up also in other phenomena,
thus influencing fermion pair production in an indirect way.
It is well-known that e.g. a $Z'$, which mixes with the ordinary $Z$
boson, may influence the $Z$ and $W$ mass ratio and the $Z$ vector
and axial-vector couplings -- it is these effects, which may be searched
for at \LEPI.
How they can be covered in the language of form factors
has been explained in the references quoted above.
In addition, a careful
derivation of the weak form factors following the notations used in
the present paper may be found in~\cite{zefit}, where the use of
\zf\ for a $Z'$ search is explained\footnote{
See also the package {\tt ZEFIT}~\cite{zefit} at ZFITTER@CERNVM.
}. The main consequences are
contained in the following replacements in the definitions of
weak form factors $\rho_f^Z, \kappa_f^Z$
for partial widths and $\rho_{ef}, \kappa_f, \kappa_{ef}$
for the cross sections:
\ba
\rho_{f}^Z&\rightarrow&\rho_{\mathrm{mix}} (1-y_f)^2 \rho_f^Z,
\nonumber \\
\kappa_f^Z&\rightarrow&(1-x_f) \kappa_f^Z,
\\
\rho_{ef}&\rightarrow&\rho_{\mathrm{mix}} (1-y_e)(1-y_f) \rho_{ef},
\nonumber \\
\kappa_f&\rightarrow&(1-x_f) \kappa_f,
\nonumber \\
\kappa_{ef}&\rightarrow&(1-x_e)(1-x_f) \kappa_{ef}.
\label{rkzp}
\ea
Here, $x_f, y_f$ are small corrections to the $Z$-boson couplings
due to the $Z,Z'$ mixing, and $\rho_{\mathrm{mix}}$ is due to
the related slight $Z$ mass shift:
\begin{eqnarray}
\rho_{\mathrm{mix}}=
\frac{M_W^2}{M_Z^2 \cos^2 \theta_W} =
\frac{M_0^2}{M_Z^2}
=
1 + \sin^2 \theta_{M} \left( \frac{M_{Z'}^2}{M_Z^2} - 1 \right).
\label{rhomix}
\end{eqnarray}
The parameter $M_0$ is the $Z$ mass of the standard model without
$Z,Z'$ mixing.
The $\rho_{\mathrm{mix}}$ influences the widths and
cross sections directly, since we have replaced in~(\ref{defzwidth})
and in~(\ref{mzff}),(\ref{Mva}) the coupling
constant $\alpha$ of the on-mass-shell scheme by the Fermi constant
$G_{\mu}$. These are related as follows (see~(\ref{sw2delr})):
\begin{equation}
\frac{\pi \alpha}{2 \sin^2 \theta_W \cos^2 \theta_W}
= \frac{G_\mu}{\sqrt{2}}
M_0^2 (1 - \Delta r)
= \frac{G_\mu}{\sqrt{2}}
M_Z^2 \rho_{\mathrm{mix}}(1 - \Delta r).
\label{gmualf}
\end{equation}
In the same way as $(1-\Delta r)$ becomes part of $\rho_f^Z$ and
$\rho_{ef}$ without mixing,
the factor $\rho_{\mathrm{mix}}(1-\Delta r)$
becomes part of the form factors when $Z$ and $Z'$ mix.

Another, completely different source of deviations from the standard
model are self-energy corrections
$\Pi$ due to new physics, which may lead to the following
changes of the weak form factors~\cite{heraho}:
\bq
\Delta \rho_{ef}(s) = \Delta \rho(0) +
\Pi_{ZZ}(M_Z^2) -
\frac{ s \Pi_{ZZ}(s) - M_Z^2 \Pi_{ZZ}(M_Z^2)} {s - M_Z^2},
\label{holdro}
\eq
\bq
\Delta \rho_W(s) =
\Pi_{WW}(M_W^2) -
\frac{ s \Pi_{WW}(s) - M_W^2 \Pi_{WW}(M_W^2)} {s - M_W^2},
\label{holdrw}
\eq
\bq
\Delta \kappa(s) = \Delta \kappa(M_Z^2)-
\frac{ \cos \theta_W}{\sin \theta_W}
\left[ \Pi_{\gamma Z}(s) - \Pi_{\gamma Z}(M_Z^2) \right].
\label{holdrk}
\eq
For completeness, the corrections for the charged-current form factor
have been added.
The corrections at \LEPI\ may be obtained from the above expressions
by setting $s=M_Z^2$.

A different starting point has been used e.g.
in~\cite{yr:bj,altar}.
There it is studied how one can disentangle new physics from the
possibly large, unknown t-quark corrections of leading order
$G_{\mu} m_t^2$; see section~\ref{subhigh}.
For this purpose one can introduce three new
parameters:
\bq
\epsilon_1 = \Delta \rho,
\label{aleps1}
\eq
\bq
\epsilon_2 = c_0^2 \Delta \rho + \frac{s_0^2 \Delta r_W}
{(c_0^2 - s_0^2)} - 2 s_0^2 \Delta k',
\label{aleps2}
\eq
\bq
\epsilon_3 = c_0^2 \Delta \rho + (c_0^2 - s_0^2) \Delta k'.
\label{aleps3}
\eq
The quantities $\Delta \rho, \Delta r_W, \Delta \kappa', s_0^2$
may be identified with quantities used in \zf\ and introduced
above\footnote{Here, we exactly follow the notations of~\cite{altar};
see also the package {\tt ZFEPSLON}~\cite{zfepslon} at ZFITTER@CERNVM.
}:
\bq
\Delta r_W = 1 - (1-\Delta r) \frac{\alpha(M_Z^2)}{\alpha},
\label{rw}
\eq
\bq
g_a = - \frac{\sqrt{\rho}}{2} = - \frac{1}{2}
(1 + \frac{1}{2} \Delta \rho ),
\label{Drho}
\eq
\bq
\frac{g_v}{g_a} = 1 - 4 s_w^{2,\mathrm{eff}}= 1 - 4 (1+\Delta \kappa')
s_0^2,
\label{Dka}
\eq
\bq
s_0^2 c_0^2 = \frac{ \pi \alpha(M_Z^2)} {\sqrt{2} G_{\mu} M_Z^2}.
\label{s00}
\eq
It is up to the user of \zf\ to decide
which of the various coupling definitions
available in the program and described in sections~\ref{subgamma},
\ref{subfpair}, \ref{branches.2}  are used as couplings
$g_v, g_a$ in the above definitions.
The running QED coupling is defined in~(\ref{dysfa}).
    Thus, while it may appear that the $\epsilon$ parameters
    are merely rearrangements of previously defined quantities,
    their merit lies in separating out the $m_t$-dependent
    effects in $\epsilon_1$ and other (Higgs) effects in
    $\epsilon_3$. Furthermore, for an analysis of LEP I data
    alone, the $\epsilon_2$ parameter may be ignored.

Besides the notations introduced so far, there are several similar
ones used in the literature, often in quite a different
context. As one important example, we quote the following notation
which introduces again some self-energy corrections, but now
calculated before $\gamma,Z$ mixing~\cite{gavela}:
\bq
\alpha S \approx - 4 e^2 \frac{d}{dq^2}
\left[ \Pi_{30}(q^2)
\right]
|_{q^2=0},
\label{S}
\eq
\bq
\alpha T \approx \frac{e^2}{\sin^2 \theta_W M_W^2}
\left[ \Pi_{11}(0) - \Pi_{33}(0) \right],
\label{T}
\eq
\bq
\alpha U \approx 4 e^2 \frac{d}{dq^2} 
\left[
 \Pi_{11}(q^2) - \Pi_{33}(q^2) \right] 
 |_{q^2=0}.
\label{U}
\eq
The relation to the above notations is~\cite{ellis}:
\bq
\epsilon_1 = \alpha T, \hspace{1cm}
\epsilon_2 = -\frac{\alpha}{4 \sin^2 \theta_W} U, \hspace{1cm}
\epsilon_3 =  \frac{\alpha}{4 \sin^2 \theta_W} S.
\label{epsil}
\eq
Further relations between different notations may be found
in~\cite{je:ihvp1b}.

\section
[Initialization of \zf]
{Initialization of \zf
\label{init}
}
\setcounter{equation}{0}

\begin{figure}
\setlength{\unitlength}{1mm}
\begin{picture}(140,210)(-7.,30.)
\put(65,234){\dashbox(20,8){ZFTEST}}
\put(75,234){\vector(0,-1){4}}
\put(65,222){\framebox(20,8){ZUINIT}}
\put(75,222){\vector(0,-1){4}}
\put(65,210){\framebox(20,8){ZUFLAG}}
\put(75,210){\vector(0,-1){4}}
\put(65,198){\framebox(20,8){ZUWEAK}}
\put(85,202){\vector(1,0){19}}
\put(104,198){\framebox(22,8){ZUINFO}}
\put(75,198){\vector(0,-1){4}}
\put(03,198){\framebox(20,8){{\tt DIZET}}}
\put(65,202){\vector(-1,0){42}}
\put(65,186){\framebox(20,8){ZUCUTS}}
\put(75,186){\vector(0,-1){7}}             
\put(23,179){\line(1,0){84.5}}
\put(13,170){\vector(0,1){28}}
\put(13,198){\vector(0,-1){28}}
\put(23,179){\vector(0,-1){09}}            
\put(107.5,179){\vector(0,-1){09}}         
\put(0,154){\framebox(46,16){ }}
\put(23,165){\makebox(0,0){\it Standard Model}}
\put(23,160){\makebox(0,0){\it interfaces}}
\put(0,134){\framebox(46,20){ }}
\put(23,150){\makebox(0,0){ZUTHSM}}
\put(23,144){\makebox(0,0){ZUTPSM}}
\put(23,134){\vector(0,-1){09}}      
\put(52,162){\framebox(100,8){ }}
\put(101.5,166){\makebox(0,0){\it Model-independent interfaces}}
\put(52,154){\framebox(100,8){ }}
\put(107.5,158){\makebox(0,0){\it Partial widths}}
\put(71.5,158){\makebox(0,0){\it Effective couplings}}
\put(137.5,158){\makebox(0,0){\it S-matrix}}
\put(91,134){\line(0,1){28}}          
\put(124,134){\line(0,1){28}}         
\put(52,134){\framebox(100,20){ }}
\put(107.5,150){\makebox(0,0){ZUXSEC}}
\put(107.5,134){\vector(0,-1){09}}       
\put(71.5,150){\makebox(0,0){ZUXSA}}
\put(71.5,144){\makebox(0,0){ZUXSA2}}
\put(71.5,138){\makebox(0,0){ZUTAU}}
\put(71.5,134){\vector(0,-1){09}}       
\put(137.5,150){\makebox(0,0){ZUSMAT}}
\put(137.5,134){\vector(0,-1){09}}       
\put(23,125){\line(1,0){114.5}}       
\put(50,125){\vector(0,-1){16}}
\put(100,125){\vector(0,-1){16}}     
\put(35.5,111){INDF=}
\put(51,111){11}
\put(40,101){\framebox(20,8){\tt BHANG}}
\put(50,101){\vector(0,-1){12}}         
\put(90,101){\framebox(20,8){ZCUT}}
\put(100,101){\vector(0,-1){12}}         
\put(90,81){\framebox(20,8){EWINIT}}
\put(100,81){\vector(0,-1){12}}
\put(90,61){\framebox(22,8){EWCOUP}}
\put(100,61){\vector(0,-1){12}}
\put(85.5,51){ICUT=}
\put(101,51){$-$1,0,1}
\put(90,41){\framebox(20,8){SCUT}}
\put(40,81){\framebox(20,8){BHAINI}}
\put(50,81){\vector(0,-1){12}}
\put(40,61){\framebox(22,8){BHACOU}}
\put(50,61){\vector(0,-1){12}}
\put(35.5,51){ICUT=}
\put(51,51){1}
\put(40,41){\framebox(20,8){BHACUT}}
\end{picture}
\caption[
The structure of \zf.
]{\it
The structure of \zf.
{\tt ICUT=-1} gives observables without any cuts,
{\tt ICUT=0,1} with cuts.
}
\label{fi:eins}
\end{figure}
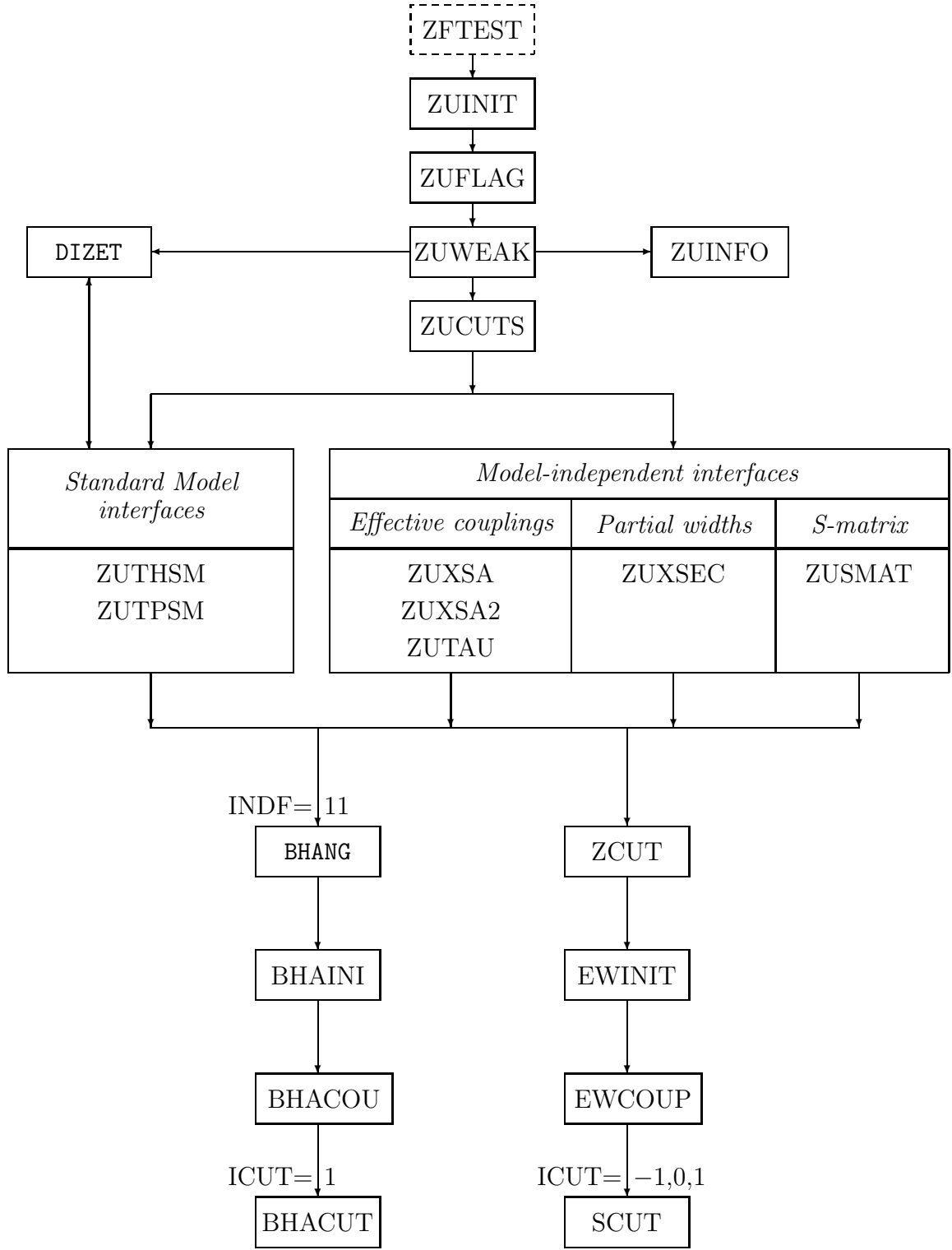

\zf\ is coded in {\tt FORTRAN 77} and it has been implemented
on IBM, IBM PC, VAX, and APOLLO.
It must be used with {\tt DIZET} and {\tt BHANG}.
Double-precision
variables have been used throughout the program in order to
obtain maximum
accuracy, which is especially important for resonance physics.
In all, the package (\zf, \DIZET\ and \BHANG) contains about
11500 lines of FORTRAN code.
A block diagram of \zf\ is shown in fig.~\ref{fi:eins}.

%

The following routines are normally called in the initialization phase
of programs using the \zf\ package.
Normally they are called in the order listed below.

\subsection
[Subroutine ZUINIT]
{Subroutine ZUINIT
\label{zuinit}}

Subroutine {\tt ZUINIT} is used to initialize variables to
their default values.
This routine {\em must} be called before any other \zf\ routine.

\SUBR{CALL ZUINIT}

\subsection
[Subroutine ZUFLAG]
{Subroutine ZUFLAG
\label{zuflag}}

Subroutine {\tt ZUFLAG} is used to modify the default
values of flags
which control various \zf\ options.

\SUBR{CALL ZUFLAG(CHFLAG,IVALUE)}

\newpage
\noindent \underline{Input Arguments:}
\smallskip
\begin{description}
  \item[\tt CHFLAG] is the character identifier of a \zf\ flag
   (see  table~\ref{ta7}).
  \item[\tt IVALUE] is the value of the flag.
\end{description}

\begin{table}[bthp]\centering
\begin{tabular}{|c|c|c|c|c|c|c|c|c|}  \hline
{    CHFLAG}&\multicolumn{2}{c|}{    IVALUE}&
{    CHFLAG}&\multicolumn{2}{c|}{    IVALUE}&
\mbox
{    CHFLAG}&\multicolumn{2}{c|}{    IVALUE}\\
\hline
{\tt AFBC} & \ 1 \  & 1 & {\tt ALPH} & \ 0 \ & 0 & {\tt ALST} & \ 1 \
& 0\\
{\tt AMT4} & 3 & 3 & {\tt BORN} & 0 & 0 & {\tt BOXD} & 0 & 1\\
{\tt CONV} & 0 & 1 & {\tt FINR} & 1 & 1 & {\tt FOT2} & 1 & 2\\
{\tt GAMS} & 1 & 1 & {\tt INCL} & 1 &0;1& {\tt INTF} & 1 & 1 \\
{\tt PART} & 0 & 0 & {\tt POWR} & 1 & 1 & {\tt PRNT} & 0 & - \\
{\tt QCDC} & 1 & 2 & {\tt QCD3} & 1 & 1 & {\tt VPOL} & 3 & 3 \\
{\tt WEAK} & 1 & 1 &            &   &   &            &   & \\
\hline
\end{tabular}
\caption[
Flag settings for \zf.
]
{\it
Flag settings for \zf; the values shown are:
in the first column  the default, recommended settings optimized for
\LEPI\ physics and in the second  the `best' settings, recommended
for use in a broader energy region.
}
\label{ta7}
\end{table}

\noindent Possible
combinations of {\tt CHFLAG} and {\tt IVALUE} are listed
below:
\begin{description}

  \item[\tt CHFLAG=`AFBC']
  Controls the calculation of the forward
  backward asymmetry for interface {\tt ZUTHSM}.
  \begin{description}
    \item[\tt IVALUE=0]
    Asymmetry calculation is inhibited (can speed up the program
    if asymmetries are not desired).
    \item[\tt IVALUE=1]
    (default) Asymmetry calculation is done.
  \end{description}

  \item[\tt CHFLAG=`ALPH']
  Controls the calculation of
  $\alpha_s(q^2,\Lambda_{\overline{\mathrm{MS}}})$ for
  $R_{\mathrm {QCD}}$~(\ref{rqcd}, \ref{qcdcor})-(\ref{qcd3})
  and for the \os corrections (section~\ref{subhigh}); see also flag
  {\tt ALST}.
  \begin{description}
    \item[\tt IVALUE=0]
    (default) {\tt ALPHA4} is used to calculate $\alpha_s$, where
    $\Lambda_{\overline{\mathrm{MS}}}$ is calculated according to
    table~\ref{tablam}; user input is defined by {\tt ALST}.
    \item[\tt IVALUE=1]
    {\tt ALPHA5} relies on 5 quark flavors and  is used to calculate
    $\alpha_s$.
  \end{description}

  \item[\tt CHFLAG=`ALST']
  Determines how the strong coupling constant
  $\alpha_s$  runs as a function of $s$
  in $R_{\mathrm{QCD}}$, (\ref{rqcd}).
  Form factor corrections   \os are calculated
  corresponding to {\tt ALST=0}.
  {\tt ALFAS} is input by the user in calls to {\tt ZUWEAK, ZUTHSM}
  and {\tt ZUTPSM}.
  \begin{description}
    \item[\tt IVALUE=0]
    Calculation of
    $\alpha_s=${\tt ALPHAn}$(q^2,\Lambda_{\overline{\mathrm{MS}}})$
    corresponding to~(\ref{eq:alfas})
    with $q^2$ = $M_Z^2$ for the \Z\ width and
    with $q^2$ = $s$ for the cross sections.
    Where {\tt n} is determined according to the flag {\tt ALPH}.
    For $n=4$,
    $\Lambda_{\overline{\mathrm{MS}}}$ is calculated from the input
    $\Lambda_{\overline{\mathrm{MS}}}^{(4)}$={\tt ALFAS},
    while for $n=5$, $\Lambda_{\overline{\mathrm{MS}}}$ $\equiv$
    $\Lambda_{\overline{\mathrm{MS}}}^{(5)}$ = {\tt ALFAS}.
    \item[\tt IVALUE=1]
    (default)
    In $R_{\mathrm{QCD}}$,
    the strong coupling constant runs as follows:
    $\alpha_s(q^2)= \alpha_s(M_Z^2)$
    $[$
    {\tt{ALPHAn}}($q^2$,
    $\Lambda_{\overline{\mathrm{MS}}}^{(n)}$)
    $/$
    {\tt{ALPHAn}}
    $(M_Z^2, \Lambda_{\overline{\mathrm{MS}}}^{(n)})$
    $]$, with
    $\alpha_s(M_Z^2)=${\tt{ALFAS}}.
    Here, $\Lambda_{\overline{\mathrm{MS}}}^{(n)}$ is fixed
    at 185 MeV for $n=4$ and at 122 MeV for $n=5$.

  \end{description}
  \item[\tt CHFLAG=`AMT4']
  Controls calculation of leading \oaa terms
  for $\Delta r$ and the weak form factors as discussed in
  section~\ref{subhigh} \footnote{
  To reproduce results presented in~\cite{yr:alt}, one has to choose
  {\tt AMT4}=-1 together with {\tt VPOL}=2. This combination of flags
  gives a reasonable result only at LEP~I energies.}.
  \begin{description}
    \item[\tt IVALUE=0]
    No resummation. $\Delta r, \rho, \kappa$ as introduced
    in sections \ref{subdr}, \ref{subgamma}, \ref{subfpair}
    are calculated to order \oalf, with possible inclusion
    of the \os corrections depending on another flag;
    \item[\tt IVALUE=1]
    Leading \oaa corrections are included
    in $\Delta \rho$ and $\bar \delta \rho$,
    (\ref{dro0}), (\ref{high2}), while the \os \- terms
    and $X_0,X$ are neglected there.
    The latter remain in the
    remainder part of the form factors;
    \item[\tt IVALUE=2]
    Common resummation of leading \oaa and \os terms;
    only the $X_{0},X$  are neglected;
    \item[\tt IVALUE=3]
    (default) Resummation as described in~\cite{fs,dfs,ds};
    $\Delta \rho$ and $\bar \delta \rho$ as defined in
    (\ref{dro0}), (\ref{high2}).
  \end{description}

  \item[\tt CHFLAG=`BORN']
  Controls calculation of QED and Born observables.
  \begin{description}
    \item[\tt IVALUE=0]
    (default) QED convoluted observables.
    \item[\tt IVALUE=1]
    Non-convoluted `effective' Born observables.
  \end{description}

  \item[\tt CHFLAG=`BOXD']
  Determines if the $ZZ$ and $WW$ box
  contributions (see fig.~\ref{figsmbx}) are calculated.
  \begin{description}
    \item[\tt IVALUE=0]
    (default) No box contributions are calculated.
    \item[\tt IVALUE=1]
    $ZZ$ and $WW$ box diagrams are calculated.
  \end{description}

  \item[\tt CHFLAG=`CONV']
  Controls the energy scale of running $\alpha$.
  \begin{description}
    \item[\tt IVALUE=0]
    (default) For {\tt WEAK}=1, $\alpha$ is calculated at the
    energy scale
    $s$ and for {\tt WEAK}=0 it is not running at all.
    \item[\tt IVALUE=1]
    $\alpha$ is calculated at the
    energy scale $s'$ and convoluted.
  \end{description}

  \item[\tt CHFLAG=`FINR']
  Controls the calculation of final-state radiation.
  \begin{description}
    \item[\tt IVALUE=0]
    Final-state radiation is included as in (\ref{efast}).
    \item[\tt IVALUE=1]
    (default) Include complete treatment of final-state radiation
    with common soft-photon
    exponentiation as in (\ref{efexp}), (\ref{siginifin}).
  \end{description}

  \item[\tt CHFLAG=`FOT2']
  Controls second-order leading
  log and next-to-leading log QED corrections.
  \begin{description}
    \item[\tt IVALUE=0]
    Second-order QED corrections are not included.
    \item[\tt IVALUE=1]
    (default) Second-order QED corrections are included as described
    in~\cite{zshape,ringberg}; constant terms   \oalz omitted.
    \item[\tt IVALUE=2]
    Second-order QED corrections are included as described
    in~\cite{zshape,ringberg}.
  \end{description}

  \item[\tt CHFLAG=`GAMS']
  Controls the $s$ dependence of ${\cal G}_Z$,
  the \Z-width function, introduced in (\ref{propagn}).
  \begin{description}
    \item[\tt IVALUE=0]
    Forces ${\cal G}_Z$ to be constant.  Propagator definition
    (\ref{mg2}) is used.
    \item[\tt IVALUE=1]
    (default) Allows ${\cal G}_Z$
    to vary as a function of $s$ as in (\ref{propagz})-(\ref{mg}).
  \end{description}

  \item[\tt CHFLAG=`INCL']
  Influences the treatment of final-state bremsstrahlung
  {\em exclusively} for quarks and hadrons.
  \begin{description}
    \item[\tt IVALUE=0] Same as {\tt FINR}=1.
    \item[\tt IVALUE=1] (default) For quarks and hadrons,
    final-state bremsstrahlung is treated as with
    {\tt FINR}={\tt INTF}=0.
  \end{description}

  \item[\tt CHFLAG=`INTF']
  Determines if the ${\cal O} (\alpha)$ initial-final state
  QED interference terms are calculated.
  These terms are very small near the \Z\ peak; however, they
  can become significant if severe kinematic cuts are applied.
  \begin{description}
    \item[\tt IVALUE=0]
    The interference term is ignored.
    \item[\tt IVALUE=1]
    (default) The interference term is included.
  \end{description}

  \item[\tt CHFLAG=`PART']
  Controls the calculation of various parts of Bhabha scattering.
  \begin{description}
    \item[\tt IVALUE=0]
    (default) Calculation of full Bhabha cross section and asymmetry.
    \item[\tt IVALUE=1]
    Only s channel.
    \item[\tt IVALUE=2]
    Only t channel.
    \item[\tt IVALUE=3]
    Only s-t~interference.
  \end{description}

  \item[\tt CHFLAG=`PRNT']
  Controls {\tt ZUWEAK} printing.
  \begin{description}
    \item[\tt IVALUE=0]
    (default) Printing by subroutine {\tt ZUWEAK} is suppressed.
    \item[\tt IVALUE=1]
    Each call to {\tt ZUWEAK} produces some output.
  \end{description}

  \item[\tt CHFLAG=`POWR']
  Controls inclusion of final-state masses 
  in kinematical factors.
  \begin{description}
    \item[\tt IVALUE=0]
    Lepton and light-quark masses are set to zero in
    (\ref{defzwidth}),
    (\ref{defmu}), (\ref{cm}), (\ref{itsmz}), (\ref{itzzeff});
    $m_c = 1.5, m_b = 4.5$ GeV.
    \item[\tt IVALUE=1]
    (default) Lepton and
    light-quark masses as taken in the calculation of
    vacuum polarization. In combination with {\tt VPOL}=1:
    $m_u=.062, m_d=.083, m_s=.215$ GeV, and with {\tt VPOL}=2,3:
    $m_u=.04145, m_d=.04146, m_s=.15$ GeV.

  \end{description}

  \item[\tt CHFLAG=`QCDC']
  Controls how
  ${\cal O}(\alpha \alpha_s)$\  corrections related to the t-quark mass
  are treated within weak form factors.
  The leading ${\cal O}(\alpha \alpha_s m_t^2)$ term of this QCD
  correction is expli\-citly given in (\ref{high2}).
  \begin{description}
    \item[\tt IVALUE=0]
    ${\cal O}(\alpha \alpha_s)$\ corrections to weak form factors
    are not calculated.
    This setting must be used for numerical
    comparisons with the tables shown
    in~\cite{yr:alt,mp:bhr,ds}.
    \item[\tt IVALUE=1]
    (default) They are included and
    determined with a fast approximate
    calculation as described in~\cite{du:bc}.
    It should be noted that this approximation is only valid
    for $s < m_t^2$.
    \item[\tt IVALUE=2]
    Same as 1 except that exact calculations of the Feynman
    diagrams are performed as in~\cite{du:bc}\footnote{
An alternative 
calculation~\cite{p2:kni} 
agrees numerically to 12 digits.}.
    No restriction on $s$.
  \end{description}

  \item[\tt CHFLAG=`QCD3']
  Controls the inclusion of the
  ${\cal O}(\alpha_s^3)$ in (\ref{qcdcor}), (\ref{qcdcob}).
  \begin{description}
    \item[\tt IVALUE=0]
    (default) This term is not included.
    \item[\tt IVALUE=1]
    The calculation is made to ${\cal O}(\alpha_s^3)$.
  \end{description}

  \item[\tt CHFLAG=`WEAK']
  Determines if weak loop calculations are to be performed.
  \begin{description}
    \item[\tt IVALUE=0]
    No weak loop corrections to the cross sections are calculated
    and weak parameters are forced to their Born values,  i.e.
    $\rho_{ef} = \kappa_{e,f,ef} = 1$.
    \item[\tt IVALUE=1]
    (default) Weak loop corrections to the cross sections are
    calculated.
  \end{description}

  \item[\tt CHFLAG=`VPOL']
  Controls, which parametrization of
  the hadronic vacuum polarization contribution
  $\alpha_{\mathrm{had}}$ to the photon propagator
  (\ref{dysfa}) is used.
  Three different parametrizations are available.
  \begin{description}
    \item[\tt IVALUE=1]
    Selects a parametrization taken from~\cite{je:ihvp1,je:ihvp1b}.
    \item[\tt IVALUE=2]
    Quarks are treated like leptons and their effective masses
    are as in the second set quoted in the description of the
    flag {\tt POWR}. This choice was used to obtain the results
    presented in \cite{yr:alt}\footnote{In addition, the
    value of the {\tt  AMT4} flag was set to $-$1.}.
    \item[\tt IVALUE=3]
    (default) Selects a parametrization that
    uses the hadronic vacuum polarization calculations described
    in~\cite{buje}.
  \end{description}

\end{description}

\subsection
[Subroutine ZUWEAK]
{Subroutine ZUWEAK
\label{zuweak}}

Subroutine {\tt ZUWEAK} is used to perform the weak sector calculations.
These are done internally with {\tt DIZET}~\cite{di:cpc}.
The routine calculates a number of important electroweak parameters
(\IE\ \Sw, the partial \Z\ widths, fermionic vacuum polarization, $F_A$,
and weak form factors for the cross section), which
are stored in common blocks for later use (see appendix A).
If any \zf\ flags are to be modified this must be done before
calling {\tt ZUWEAK}.

\SUBR{CALL ZUWEAK(ZMASS,TMASS,HMASS,ALFAS)}

\BS
\noindent \underline{Input Arguments:}
\smallskip
\begin{description}
  \item[\tt ZMASS] is the \Z\  mass (\MZ) in GeV.
  \item[\tt TMASS] is the top quark mass (\MT) in GeV, [10-400].
  \item[\tt HMASS] is the Higgs mass (\MH) in GeV, [10-1000].
  \item[\tt ALFAS] is the value of the strong coupling constant
    ($\alpha_s$) at $q^2 = M_Z^2$ (see also flag {\tt ALST}).
\end{description}

A tremendous saving in computing time can be realized by performing
weak sector calculations only once during initialization of the \zf\
package.
This is possible because weak parameters are nearly
independent of $s$ near the \Z\ peak, \EG\ $\sim \ln s/M_Z^2$.

\subsection
[Subroutine ZUCUTS]
{Subroutine ZUCUTS
\label{zucuts}}

Subroutine {\tt ZUCUTS} is used to define kinematic and geometric cuts
for each fermion channel.
In terms of the internal structure of \zf, this routine is used to select
the appropriate QED calculational {\em chain}.

\SUBR{CALL ZUCUTS(INDF,ICUT,ACOL,EMIN,S\_PR,ANG0,ANG1)}

\BS
\noindent \underline{Input Arguments:}
\smallskip
\begin{description}
  \item[\tt INDF] is the fermion index (see table~\ref{indf}).
\begin{table}[bthp]
\begin{center}
\begin{tabular}{|c|l|}
\hline
             & Final-    \\
{\tt INDF}   & state     \\
             & fermions  \\ \hline
 0 \hfill    & $\nu{\bar{\nu}}$   \\
 1 \hfill    & $e^+e^-$   \\
 2 \hfill    & $\mu^+\mu^-$       \\
 3 \hfill    & $\tau^+\tau^-$        \\
 4 \hfill    & $u\bar{u}$   \\
 5 \hfill    & $d\bar{d}$        \\
 6 \hfill    & $c\bar{c}$      \\
 7 \hfill    & $s\bar{s}$     \\
 8 \hfill    & $t\bar{t}$  \\
 9 \hfill    & $b\bar{b}$      \\
10 \hfill    & hadrons     \\
11 \hfill    & Bhabha      \\
\hline
\end{tabular}
\end{center}
\caption[
Indices used by \zf\ interface routines to select the
final-state fermion pair.
]{\it
Indices used by \zf\ interface routines to select the
final-state fermion pair.
Note that \mbox{\tt INDF=1} returns only s-channel observables,
\mbox{\tt INDF=8} always returns zero,  and
\mbox{\tt INDF=10} indicates a sum over all open quark channels.
}
\label{indf}
\end{table}
  \item[\tt ICUT] controls the kinds of cuts ({\em chain}) to be used.
    \begin{itemize}
      \item[\tt =-1:] (default) no cuts at all are to be used (fastest).
      \item[\tt = 0:] allows for a cut on the acollinearity of the
      \FF\ pair
       and the minimum energy of both fermion and antifermion.
      \item[\tt = 1:] allows for a cut on the minimum invariant mass of
      the
      \FF\ pair.
    \end{itemize}
  \item[\tt ACOL] is the maximum acollinearity angle $(\xi^{\max})$ of
  the
    \FF\ pair in degrees ({\tt ICUT} = 0).
  \item[\tt EMIN] is the minimum energy $(E^{\min}_f)$ of the fermion and
    antifermion in GeV ({\tt ICUT} = 0).
  \item[\tt S\_PR] is minimum allowed invariant \FF\ mass $(s')$ in GeV
    ({\tt ICUT} = 1).
    This is related to the maximum photon energy by
    (\ref{cuts2}), (\ref{spmin}).
  \item[\tt ANG0] (default = $0^{\circ}$) is the minimum polar angle
  ($\vartheta$) in degrees of the final-state antifermion.
  \item[\tt ANG1] (default = $180^{\circ}$) is the maximum polar angle
    ($\vartheta$) in degrees of the final-state antifermion.
\end{description}

\subsection
[Subroutine ZUINFO]
{Subroutine ZUINFO
\label{zuinfo}}

Subroutine {\tt ZUINFO} prints the values of  \zf\ flags and cuts.

\SUBR{CALL ZUINFO(MODE)}

\BS
\noindent \underline{Input Argument:}
\smallskip

\begin{description}
  \item[\tt MODE] controls the printing of \zf\ flag and cut values.
\begin{itemize}
  \item[\tt =0:] Prints all flag values.
  \item[\tt =1:] Prints all cut values.
\end{itemize}
\end{description}


\section
[Interface Routines of \zf]
{Interface Routines of \zf\
\label{interfaces} }
\setcounter{equation}{0}

Each calculational branch of \zf\ has corresponding interfaces.
These interfaces will be described below.
For the Standard Model branch the cross section and asymmetry interface
is subroutine {\tt ZUTHSM}, while for the tau polarization it is
subroutine {\tt ZUTPSM}.
Subroutines {\tt ZUXSA},
{\tt ZUXSA2} and {\tt ZUTAU} are interfaces for the
effective coupling's branch.
The interfaces for the partial widths  and S-matrix branches are
{\tt ZUXSEC} and {\tt ZUSMAT}, respectively.

Note that subroutine {\tt ZUWEAK} must be called prior to any of the
interfaces to be described below. As a consequence, flags used in this
subroutine can influence the calculation of cross sections and
asymmetries in the interfaces described now.

\subsection
[Subroutine ZUTHSM]
{Subroutine ZUTHSM
\label{zuthsm}}

Subroutine ZUTHSM is used to calculate Standard Model cross sections and
forward--back\-ward asymmetries as described in section~\ref{branches.1}.

\SUBR{CALL ZUTHSM(INDF,SQRS,ZMASS,TMASS,HMASS,ALFAS,XS*,AFB*)}

\BS
\noindent \underline{Input Arguments:}
\smallskip
\begin{description}
  \item[\tt INDF] is the fermion index (see table~\ref{indf}).
  \item[\tt SQRS] is the centre-of-mass energy (\RS) in GeV.
  \item[\tt ZMASS] is the \Z\ mass (\MZ) in GeV.
  \item[\tt TMASS] is the top quark mass (\MT) in GeV, [10-400].
  \item[\tt HMASS] is the Higgs mass (\MH) in GeV, [10-1000].
  \item[\tt ALFAS] is the value of the strong coupling constant
    ($\alpha_s$) at $q^2 = M_Z^2$ (see also flag {\tt ALST}).
\end{description}

\nn \underline{Output Arguments}:\footnote{An asterisk (*) following an
argument in a calling sequence is used to denote an output argument.}
\smallskip
\begin{description}
  \item[\tt XS] is the total cross section ($\sigma_T$) in nb.
  \item[\tt AFB] is the forward--backward asymmetry (\afb).
\end{description}

\subsection
[Subroutine ZUTPSM]
{Subroutine ZUTPSM
\label{zutpsm}}

Subroutine {\tt ZUTHSM} is used to calculate the Standard Model tau
polarization and tau polarization asymmetry as described in
section~\ref{branches.1}.

\SUBR{CALL ZUTPSM(SQRS,ZMASS,TMASS,HMASS,ALFAS,TAUPOL*,TAUAFB*)}

\BS
\noindent \underline{Input Arguments:}
\smallskip
\begin{description}
  \item[\tt SQRS] is the centre-of-mass energy (\RS) in GeV.
  \item[\tt ZMASS] is the \Z\ mass (\MZ) in GeV.
  \item[\tt TMASS] is the top quark mass (\MT) in GeV, [40-300].
  \item[\tt HMASS] is the Higgs mass (\MH) in GeV, [10-1000].
  \item[\tt ALFAS] is the value of the strong coupling constant
    ($\alpha_s$) at $q^2 = M_Z^2$ (see also flag {\tt ALST}).
\end{description}

\nn \underline{Output Arguments}:
\smallskip
\begin{description}
  \item[\tt TAUPOL] is the tau polarization ($A_{\mathrm{pol}}$)
  of (\ref{asym}).
  \item[\tt TAUAFB] is the tau polarization forward--backward
  asymmetry
  ($A_{FB}^{\mathrm{pol}}$) as defined in (\ref{lafb}).
\end{description}

\subsection
[Subroutine ZUXSA]
{Subroutine ZUXSA
\label{zuxsa}}

Subroutine {\tt ZUXSA} is used to calculate the cross
section and asymmetry described in section~\ref{subeff} as a function
of \RS, \MZ, \GAMZ, and the weak couplings
(\ref{hatasm}), (\ref{hatvsm}), (\ref{hatrhoa}).

\SUBR{CALL ZUXSA(INDF,SQRS,ZMASS,GAMZ,MODE,GVE,XE,GVF,XF,XS*,AFB*)}

\BS
\noindent \underline{Input Arguments:}
\smallskip
\begin{description}
  \item[\tt INDF] is the fermion index (see table~\ref{indf}), (1:9,11).
  \item[\tt SQRS] is the centre-of-mass energy (\RS) in GeV.
  \item[\tt ZMASS] is the \Z\ mass (\MZ) in GeV.
  \item[\tt GAMZ] is the total \Z\ width (\GAMZ) in GeV.
  \item[\tt MODE] determines which weak couplings are used:
  \begin{itemize}
    \item[\tt =0:] {\tt XE} ({\tt XF}) is the effective
    axial-vector coupling
    ($\hat{g}_a$) for electrons (final-state fermions).
    \item[\tt =1:] {\tt XE} ({\tt XF}) is the effective weak
    neutral-current
    amplitude normalization ($\hat{\rho}$) for electrons (final-state
    fermions).
  \end{itemize}
  \item[\tt GVE] is the effective vector coupling for electrons
  ($\hat{g}_v^e$).
  \item[\tt XE] is  the effective axial-vector coupling ($\hat{g}_a^e$)
  or
    weak neutral-current amplitude normalization ($\hat{\rho}_e$)
    for electrons (see {\tt MODE}).
  \item[\tt GVF] is the effective vector coupling for the final-state
    fermions ($\hat{g}_v^f$).
  \item[\tt XF] is  the effective axial-vector coupling ($\hat{g}_a^f$)
  or
    weak neutral-current amplitude normalization ($\hat{\rho}_f$)
    for the final-state fermions (see {\tt MODE}).
\end{description}

\noindent \underline{Output Arguments:}
\smallskip
\begin{description}
  \item[\tt XS] is the cross section $(\sigma_T)$ in nb.
  \item[\tt AFB] is the forward--backward asymmetry (\afb).
\end{description}

\subsection
[Subroutine ZUXSA2]
{Subroutine ZUXSA2
\label{zuxsa2}}

Subroutine {\tt ZUXSA2} is used to calculate the lepton cross
section and asymmetry as a function of \RS, \MZ, \GAMZ, and the weak
couplings {\em assuming lepton universality}.
This routine is similar to {\tt ZUXSA} except that the couplings
are squared.

\SUBR{CALL ZUXSA2(INDF,SQRS,ZMASS,GAMZ,MODE,GV2,X2,XS*,AFB*)}

\BS
\noindent \underline{Input Arguments:}
\smallskip
\begin{description}
  \item[\tt INDF] is the fermion index (see table~\ref{indf}) (1-3,11).
  \item[\tt SQRS] is the centre-of-mass energy (\RS) in GeV.
  \item[\tt ZMASS] is the \Z\ mass (\MZ) in GeV.
  \item[\tt GAMZ] is the total \Z\ width (\GAMZ) in GeV.
  \item[\tt MODE] determines which weak couplings are used:
  \begin{itemize}
    \item[\tt =0:] {\tt X2} is the square of the effective
    axial-vector coupling ($\hat{g}_a^l$) for leptons.
    \item[\tt =1:] {\tt X2} is the square of the effective
    neutral-current
      amplitude normalization ($\hat{\rho}_l$) for leptons.
  \end{itemize}
  \item[\tt GV2] is the square of the effective vector coupling
    ($\hat{g}_v^l$) for leptons.
  \item[\tt X2] is the square of the effective axial-vector coupling
    ($\hat{g}_a^l$) or neutral-current amplitude normalization
    ($\hat{\rho}_l$) for leptons (see {\tt MODE}).
\end{description}

\noindent \underline{Output Arguments:}
\smallskip
\begin{description}
  \item[\tt XS] is the cross section ($\sigma_T$) in nb.
  \item[\tt AFB] is the forward--backward asymmetry (\afb).
\end{description}

\subsection
[Subroutine ZUTAU]
{Subroutine ZUTAU
\label{zutau}}

Subroutine {\tt ZUTAU}  is used to calculate the $\tau^+$ polarization
as a
function of \RS, \MZ, \GAMZ, and the weak couplings (see discussion in
section~\ref{subeff}).

\SUBR{CALL ZUTAU(SQRS,ZMASS,GAMZ,MODE,GVE,XE,GVF,XF,TAUPOL*,TAUAFB*)}

\BS
\noindent \underline{Input Arguments:}
\smallskip
\begin{description}
  \item[\tt SQRS] is the centre-of-mass energy (\RS) in GeV.
  \item[\tt ZMASS] is the \Z\ mass (\MZ) in GeV.
  \item[\tt GAMZ] is the total \Z\ width (\GAMZ) in GeV.
  \item[\tt MODE] determines which weak couplings are used:
  \begin{itemize}
    \item[\tt =0:] {\tt XE} ({\tt XF}) is the effective
    axial-vector coupling ($\hat{g}_a$) for electrons (final-state
    fermions).
    \item[\tt =1:] {\tt XE} ({\tt XF}) is the effective weak
    neutral-current
    amplitude normalization ($\hat{\rho}$) for electrons (final-state
    fermions).
  \end{itemize}
  \item[\tt GVE] is the effective vector coupling for electrons
    ($\hat{g}_v^e$).
  \item[\tt XE] is  the effective axial-vector coupling ($\hat{g}_a^e$)
  or
    weak neutral-current amplitude normalization ($\hat{\rho}_e$)
    for electrons (see {\tt MODE}).
  \item[\tt GVF] is the effective vector coupling for the final-state
   fermions ($\hat{g}_v^f$).
  \item[\tt XF] is  the effective axial-vector coupling ($\hat{g}_a^f$)
  or
    weak neutral-current amplitude normalization ($\hat{\rho}_f$)
    for the final-state fermions (see {\tt MODE}).
\end{description}

\noindent \underline{Output Arguments:}
\smallskip
\begin{description}
  \item[\tt TAUPOL] is the tau polarization ($\lambda_{\tau}$) defined in
    (\ref{asym}).
  \item[\tt TAUAFB] is the forward--backward asymmetry for polarized
  tau's
    ($A_{FB}^{\mathrm{pol}}$) as defined in (\ref{lafb}).
\end{description}

\subsection
[Subroutine ZUXSEC]
{Subroutine ZUXSEC
\label{zuxsec}}

Subroutine {\tt ZUXSEC} is  used to calculate the cross
section as a function of \RS, \MZ, \GAMZ, $\Gamma_e$ and $\Gamma_f$
as was described in section~\ref{subparti}.

\SUBR{CALL ZUXSEC(INDF,SQRS,ZMASS,GAMZ0,GAMEE,GAMFF,XS*)}

\BS
\noindent \underline{Input Arguments:}
\smallskip
\begin{description}
  \item[\tt INDF] is the fermion index (see table~\ref{indf}).
  \item[\tt SQRS] is the centre-of-mass energy (\RS) in GeV.
  \item[\tt ZMASS] is the \Z\ mass (\MZ) in GeV.
  \item[\tt GAMZ0] is the total \Z\ width (\GAMZ) in GeV.
  \item[\tt GAMEE] is the partial \Z\ decay width ($\Gamma_e$) in GeV.
  \item[\tt GAMFF] is the partial \Z\ decay width ($\Gamma_f$) in GeV;
  if {\tt INDF}=10, {\tt GAMFF}=$\Gamma_{h}$.
\end{description}

\noindent \underline{Output Argument:}
\smallskip
\begin{description}
  \item[\tt XS] is the cross section ($\sigma_T$)in nb.
\end{description}

\subsection
[Subroutine ZUSMAT]
{Subroutine ZUSMAT
\label{zusmat}}

Subroutine ZUSMAT is used to calculate the cross section
from the S-matrix approach (see section~\ref{subsma}).

\SUBR{CALL ZUSMAT(INDF,SQRS,ZMASS,GAMZ,RR,RI,R0,R1,R2,RG,XS*)}

\BS
\noindent \underline{Input Arguments:}
\smallskip
\begin{description}
  \item[\tt INDF] is the fermion index (see table~\ref{indf}), [2,10].
  \item[\tt SQRS] is the centre-of-mass energy (\RS) in GeV.
  \item[\tt ZMASS] is the \Z\ mass (\MZ) in GeV.
  \item[\tt GAMZ] is the total \Z\ width (\GAMZ) in GeV.
  \item[\tt RR-R2] six parameters in S-matrix approach,
  (RR, RI, RG, R0, $\ldots$) = $(R,I,r_{\gamma},r_0,\ldots)$,
    introduced in (\ref{smasigfin}).
\end{description}

\noindent \underline{Output Argument:}
\smallskip
\begin{description}
  \item[\tt XS] is the cross section ($\sigma_T$) in nb.
\end{description}

Note that the default \Z\ mass and width definitions correspond to
(\ref{mg2}) and thus differ from those of the other \zf\ interfaces.

%
%
\section{
Comparisons
\label{compa0}
}
\setcounter{equation}{0}

In this section, we compare the predictions of the Standard Model branch
of \zf\ with other programs.
For this comparison we use the following parameter values, unless
explicitly stated otherwise: \MZ\ = 91.18, \MT\ = 150, \MH\ = 100 GeV,
and
$\alpha_s = 0.12$.
The section is broken up into two parts:
\begin{itemize}
  \item[1.] A comparison of the weak mixing angle (with its various
    definitions) as well as the partial and total \Z\ widths.
  \item[2.] A comparison of total cross sections and forward--backward
    asymmetries.
\end{itemize}

In the past many comparisons of this sort have been made.
In particular an earlier version of \zf, the
{\tt ZBIZON}~\cite{zb:bar}
package, was used in the 1989 Workshop on Z Physics at
LEP~1~\cite{yr:alt} for numerous comparisons~\cite{modind1,yrafb,zb:bar}.
In addition to these comparisons, others did not
explicitly include the {\tt ZBIZON} code~\cite{compyb}.
At that time, the predictions from all of these
programs agreed to within 0.5\%.
Since then, new codes have been developed and
the quality of several of the existing programs has been improved;
among these is the \zf\ package.

\subsection
[Weak Mixing Angles and Partial \Z\ Widths]
{Weak Mixing Angles and Partial \Z\ Widths
\label{compa1}
}

Throughout the Standard Model branch of \zf, we use the on-shell
definition (\ref{defsw2}) of the weak mixing angle \Sw.
For the sake of this comparison, we take into
account two additional definitions of the weak mixing angle:
the `effective' weak mixing angle, \SWE, introduced in (\ref{sw2eff}),
and the weak mixing angle of the ${\overline {\mathrm{MS}}}$
renormalization
scheme given below:
\bq
\sin^2\theta_W = 1 - \frac{M_W^2}{M_Z^2},
\label{cmpsw2}
\eq
\bq
s_W^{2,{\mathrm{eff}}} =
\kappa_e^Z \,\sin^2 \theta_W,
\label{cmpeff}
\eq
\ba
\sin^2 \theta_W^{\overline{\mathrm{MS}}} =   \left[
1 + \frac{\cos^2\theta_W}{\sin^2\theta_W}
 \delta {\bar \rho}  \right]
\,\sin^2 \theta_W,
\label{sw2ms}
\ea
where $\delta {\bar \rho}$ has been introduced in~(\ref{high2}).
In table~\ref{holbar:sin}, we compare predictions from \zf\ with those
obtained by W. Hollik~\cite{statho1} and
G. Degrassi, S. Fanchiotti, A. Sirlin~\cite{dfs}.

The agreement of the different calculations
for these three cases is impressive.
{}From the table it is apparent that the mixing angles \SWE\ and \SWMS\
depend to a lesser
extent on the unknown top and Higgs masses than does \Sw.
For a detailed discussion of the different approaches see for instance
{}~\cite{pl:as,yr:bj,fs,ds,pich,heraho}.

\footnotesize
\begin{table}[htbp]
\begin{center}
\begin{tabular}{|c|c|crc|crc|crc|}   \hline
     &       & \hspace{.5cm} &
     & \hspace{.5cm} &
               \hspace{.5cm} &
               & \hspace{.5cm} &
               \hspace{.5cm} &
               & \hspace{.5cm} \\
\MT  & \MH   &               &            \Sw
&               &
                             &            \SWE
                             &               &
               \multicolumn{3}{c|}
               {$\sin^2\theta_W^{\overline{\mathrm{MS}}}(M_Z^2)$}
                 \\
 & & & & & & & & & & \\   \hline
100  &  100 &  &    0.23056  &  & &    0.23362 & & &    0.23351 & \\
     &      &  &         44  &  & &         52 & & &          7 & \\
     &      &  &         62  &  & &         66 & & &         42 & \\
\cline{2-11}
     &  500 &  &    0.23266  &  & &    0.23447 & & &    0.23438 & \\
     &      &  &         53  &  & &         37 & & &         43 & \\
     &      &  &         73  &  & &         51 & & &         29 & \\
\cline{2-11}
     & 1000 &  &    0.23371  &  & &    0.23485 & & &    0.23477 & \\
     &      &  &         60  &  & &         77 & & &         83 & \\
     &      &  &         81  &  & &         91 & & &          0 & \\
\hline
150  &  100 &  &    0.22483  &  & &    0.23217 & & &    0.23213 & \\
     &      &  &         74  &  & &         07 & & &         25 & \\
     &      &  &         87  &  & &         23 & & &          0 & \\
\cline{2-11}
      & 500  &  &    0.22690  &  & &    0.23300 & & &   0.23299 & \\
      &      &  &         81  &  & &        291 & & &       309 & \\
      &      &  &          5  &  & &          6 & & &         5 & \\
\cline{2-11}
      & 1000 &  &    0.22794  &  & &    0.23337 & & &   0.23337 & \\
      &      &  &         88  &  & &          0 & & &        49 & \\
      &      &  &        802  &  & &         45 & & &         5 & \\
\hline
200   &  100 &  &    0.21782  &  & &    0.23025 & & &   0.23024 & \\
      &      &  &         78  &  & &         17 & & &        41 & \\
      &      &  &          7  &  & &         35 & & &         8 & \\
\cline{2-11}
      &  500 &  &    0.21985  &  & &    0.23106 & & &   0.23108 & \\
      &      &  &          3  &  & &        099 & & &        24 & \\
      &      &  &         91  &  & &         16 & & &        10 & \\
\cline{2-11}
      & 1000 &  &    0.22088  &  & &    0.23142 & & &   0.23144 & \\
      &      &  &          7  &  & &         37 & & &        63 & \\
      &      &  &         96  &  & &         54 & & &         9 & \\
\hline
250   &  100 &  &    0.20919  &  & &    0.22786 & & &   0.22785 & \\
      &      &  &         21  &  & &          1 & & &       808 & \\
      &      &  &         25  &  & &        800 & & &        96 & \\
\cline{2-11}
      &  500 &  &    0.21118  &  & &    0.22865 & & &   0.22866 & \\
      &      &  &         22  &  & &          0 & & &        89 & \\
      &      &  &         26  &  & &         79 & & &        76 & \\
\cline{2-11}
      & 1000 &  &    0.21217  &  & &    0.22899 & & &   0.22901 & \\
      &      &  &         25  &  & &          7 & & &        26 & \\
      &      &  &         28  &  & &        916 & & &        13 & \\
\hline
\end{tabular}
\end{center}
\caption
[
Comparison of \Sw, \SWE\, and $\sin^2\theta_W^{\mathrm{MS}}$
as calculated by \zf\ and by programs from Hollik
and Degrassi, Fanchiotti, Sirlin.]
{\it
Comparison of \Sw, \SWE\ and $\sin^2\theta_W^{\overline{\mathrm{MS}}}
(M_Z^2)$ from \zf\ (first line), Hollik~\cite{statho1} (second line),
and Degrassi, Fanchiotti, Sirlin~\cite{dfs} (third line);
with flags
\mbox{\tt AMT4=3},
\mbox{\tt QCDC=0},
\mbox{\tt QCD3=0},
\MZ=91.170 {\rm{GeV}}, \MT\ and \MH\ in {\rm{GeV}}.
}

\label{holbar:sin}
\end{table}
\normalsize

When using data to determine an effective weak mixing angle, one
must be careful, since measurements of mixing angles from different
observables may yield results that cannot be directly compared.
This delicate point was addressed in sections~\ref{subgamma} and
\ref{subeff} [see also (\ref{rhefapp})-(\ref{kappapp})]
and has been discussed in detail in~\cite{l3:som}.
It has been demonstrated that a proper
formulation of the hard-scattering subprocess and a correct unfolding
of the leptonic forward--backward, b-quark forward--backward,
 and tau
polarization asymmetries lead to results which are very close to
each other and to the value of \SWE\ expected from $\Gamma_e$, as
defined in
(\ref{sw2eff}).


In table~\ref{holbar:wid}, we compare partial, hadronic and total \Z\
widths with numbers of other authors: as W. Hollik~\cite{statho2} and
G. Degrassi, A. Sirlin~\cite{ds}.
Shown in the second and third lines are the digits which
differ from \zf\ -- as one can see the deviations are very small.

\footnotesize
\begin{table}[htbp]
\begin{center}
\begin{tabular}{|c|r|r|r|r|r|r|r|r|r|r|r|r|}   \hline
& & & & & & & & & & & & \\
\MT                & \MH                  &
$\Gamma_{\nu}$ & $\Gamma_{e}$        &
$\Gamma_{\mu}$ & $\Gamma_{\tau}$ &
$\Gamma_{u}$      & $\Gamma_{d}$        &
$\Gamma_{c}$      & $\Gamma_{s}$        &
$\Gamma_{b}$      & $\Gamma_{\mathrm{had}}$       &
$\Gamma_{\mathrm{tot}}$    \\
& & & & & & & & & & & & \\  \hline
 100    & 100   & 166.3 & 83.42 & 83.42  & 83.23  & 296.0 &
 382.2  & 295.6 & 382.2 & 377.5 & 1733.5 & 2482.4 \\
        &       &       &     4 &     4  &    -   &       &
     1  &   -   &   -   &     6 &    -   &    -   \\
        &       &       &    4* &        &    -   &       &
        &   -   &   -   &   -   &    -   &    -   \\ \cline{2-13}

        & 500   & 166.1 & 83.29 & 83.29  & 83.10  & 295.3 &
 381.4  & 294.8 & 381.4 & 376.7 & 1729.5 & 2477.5 \\
        &       &       &    30 &    30  &    -   &     2 &
     2  &   -   &   -   &       &    -   &    -   \\
        &       &       &    3* &    -   &    -   &     2 &
     3  &   -   &   -   &   -   &    -   &    -   \\ \cline{2-13}

        & 1000  & 166.0 & 83.22 & 83.22  & 83.03  & 294.8 &
 380.9  & 294.4 & 380.9 & 376.2 & 1727.3 & 2474.8 \\
        &       &       &     3 &     3  &    -   &     7 &
     7  &   -   &   -   &     3 &    -   &    -   \\
        &       &       &    2* &    -   &    -   &       &
    8   &   -   &   -   &   -   &    -   &    -   \\ \hline

 150    & 100   & 166.9 & 83.81 & 83.81  & 83.62  & 298.0 &
 384.4  & 297.5 & 384.4 & 376.5 & 1740.8 & 2492.9 \\
        &       &       &     3 &     3  &    -   &   7.9 &
        &   -   &   -   &     6 &    -   &    -   \\
        &       &   7.0 &    8* &    -   &    -   &   7.9 &
        &   -   &   -   &   -   &    -   &    -   \\ \cline{2-13}

        & 500   & 166.8 & 83.68 & 83.68  & 83.49  & 297.2 &
 383.6  & 296.8 & 383.6 & 375.7 & 1736.9 & 2488.0 \\
        &       &       &     9 &     9  &    -   &       &
      5 &   -   &   -   &     8 &    -   &    -   \\
        &       &       &    7* &    -   &    -   &       &
        &   -   &   -   &   -   &    -   &    -   \\ \cline{2-13}

        & 1000  & 166.6 & 83.61 & 83.61  & 83.42  & 296.8 &
 383.1  & 296.4 & 383.1 & 375.3 & 1734.7 & 2485.3 \\
        &       &       &       &        &    -   &     7 &
     0  &   -   &   -   &       &    -   &    -   \\
        &       &     7 &    6* &    -   &    -   &     7 &
        &   -   &   -   &   -   &    -   &    -   \\ \hline

 200    & 100   & 167.9 & 84.37 & 84.37  & 84.18  & 300.6 &
 387.4  & 300.2 & 387.4 & 375.3 & 1750.9 & 2507.4 \\
        &       &     8 &     8 &      8 &    -   &     5 &
        &   -   &   -   &       &    -   &    -   \\
        &       &       &    4* &    -   &    -   &       &
        &   -   &   -   &   -   &    -   &    -   \\ \cline{2-13}

        & 500   & 167.7 & 84.24 & 84.24  & 84.04  & 299.9 &
 386.6  & 299.4 & 386.6 & 374.5 & 1747.1 & 2502.6 \\
        &       &       &       &        &    -   &     8 &
        &   -   &   -   &       &    -   &    -   \\
        &       &       &    3* &    -   &    -   &     8 &
        &   -   &   -   &   -   &    -   &    -   \\ \cline{2-13}

        & 1000  & 167.6 & 84.16 & 84.16  & 83.97  & 299.5 &
  386.2 & 299.0 & 386.2 & 374.1 & 1745.0 & 2499.9 \\
        &       &     5 &       &        &    -   &     4 &
      1 &   -   &   -   &       &    -   &    -   \\
        &       &       &    2* &    -   &    -   &     4 &
      1 &   -   &   -   &   -   &    -   &    -   \\ \hline

 250    & 100   & 169.0 & 85.10 & 85.10  & 84.91  & 304.0 &
 391.4  & 303.6 & 391.4 & 373.7 & 1764.0 & 2526.3 \\
        &       &       &     1 &     1  &    -   &   3.8 &
        &   -   &   -   &     5 &    -   &    -   \\
        &       &     1 &    1* &    -   &    -   &       &
     3  &   -   &   -   &   -   &    -   &    -   \\ \cline{2-13}

        & 500   & 168.9 & 84.96 & 84.96  & 84.77  & 303.3 &
  390.6 & 302.9 & 390.6 & 373.0 & 1760.2 & 2521.5 \\
        &       &     8 &     7 &     7  &    -   &     1 &
        &   -   &   -   &   2.8 &    -   &    -   \\
        &       &       &  5.0* &    -   &    -   &       &
      5 &   -   &   -   &   -   &    -   &    -   \\ \cline{2-13}

        & 1000  & 168.8 & 84.89 & 84.89  & 84.69  & 302.9 &
 390.1  & 302.5 & 390.1 & 372.6 & 1758.2 & 2518.9 \\
        &       &     7 &       &        &    -   &     7 &
        &   -   &   -   &     4 &    -   &    -   \\
        &       &       &    9* &    -   &    -   &     8 &
        &   -   &   -   &   -   &    -   &    -   \\  \hline
\end{tabular}

\end{center}

\caption
[
Partial and total widths of the \Z\ boson from \zf,
Hollik, and Degrassi, Sirlin.]
{\it
Partial and total widths of the \Z\ boson in MeV from \zf \,
(first line),
Hollik~\cite{statho2} (second line) and Degrassi, Sirlin~\cite{ds}
(third line).
Shown are only the digits which differ, a dash means no entry, an
asterisk no digit available.
Flags as in  table~\ref{holbar:sin},
\MZ=91.170 {\rm{GeV}}, \MT\ and \MH\ in {\rm{GeV}}.
}
\label{holbar:wid}
\end{table}
\normalsize
\clearpage

\subsection
[Cross Sections and Asymmetries]
{Cross Sections and Asymmetries
\label{compa2}
}

In this section the cross sections and asymmetries for processes
(\ref{firsteq}) and (\ref{seconeq}) predicted by \zf\ are compared
with those of \ZSHAPE\ 2.0~\cite{zshape,zshapeyb} and
of \ALIBABA\ 2.0~\cite{alibaba}.
Earlier comparisons of \zf with the Cahn package \cite{cahn} can be found
in \cite{ganguli}; with \ALIBABA\ in \cite{alibabaafb}; and with
\ZSHAPE, \ALIBABA, and \KORALZ~3.8~\cite{koralz} in \cite{notes}.

For (\ref{firsteq}),
all packages include weak corrections of at least \oalff.
In \ZSHAPE, QED contributions of ${\cal O}(\alpha^2)$ to the
initial-state are calculated exactly, while in the other programs
a leading-log approximation is used.
In addition,
all programs include final-state radiation corrections to
${\cal O}(\alpha)$
and common exponentiation of initial- and final-state soft-photon
emission.
Initial-final interference is contained only in \ALIBABA\ and \zf.
The additional t-channel terms (including s-t interference),
which are necessary in order to calculate Bhabha scattering
(\ref{seconeq}), are available in both \ALIBABA\ and \zf.
In the latter, this is done via the \BHANG\ package, which
only contains some of the higher-order t-channel QED corrections
that have been implemented in \ALIBABA.
On the other hand, \zf\ contains higher-order weak and QCD
corrections as explained in section \ref{subhigh}, which are not
available in the other two packages.
As can be seen in more detail from the references, the three codes
allow for different applications of kinematic cuts due to their different
theoretical basis.

Since the other two programs (\ALIBABA\ and \ZSHAPE) perform only
Standard
Model calculations, we have restricted these comparisons to the
corresponding branch of \zf.
In this context, we have used the `recommended' \zf\ flags of
table~\ref{ta7}.
In addition, we perform comparisons using flag settings shown in
table~\ref{tbest}, which have been chosen such that the corrections
realized in \zf\ most closely resemble that of the other programs.

\begin{table}[htbp]\centering
\begin{tabular}{|c|c|c|c|c|c|c|c|c|}  \hline
{    CHFLAG}&\multicolumn{2}{c|}{    IVALUE}&
{    CHFLAG}&\multicolumn{2}{c|}{    IVALUE}&
{    CHFLAG}&\multicolumn{2}{c|}{    IVALUE}\\
\hline
{\tt AFBC} & \ 1 \ & 1 & {\tt ALPH} & \ 1 \ & 1 & {\tt ALST} & \ 1 \
& 1\\
{\tt AMT4} & 0 & 0 & {\tt BORN} & 0 & 0 & {\tt BOXD} & 0 & 1\\
{\tt CONV} & 1 & 1 & {\tt FINR} & 1 & 1 & {\tt FOT2} & 2 & 2\\
{\tt GAMS} & 1 & 1 & {\tt INCL} & 0 & 0 & {\tt INTF} & 0 & 1 \\
{\tt PART} & 0 & 0 & {\tt POWR} & 0 & 0 & {\tt PRNT} & - & - \\
{\tt QCDC} & 0 & 0 & {\tt QCD3} & 0 & 0 & {\tt VPOL} & 2 & 2 \\
{\tt WEAK} & 1 & 1 &            &   &   &            &   & \\
\hline
\end{tabular}
\caption
[
Flag settings in \zf\ for comparisons of cross sections and
asymmetries.]
{\it
Flag settings in \zf\ for comparisons of cross sections and
asymmetries; the values shown are:
first column - best agreement with \ZSHAPE,
second column -  best agreement with \ALIBABA.
}
\label{tbest}
\end{table}

We performed three series of comparisons:
\smallskip
\begin{itemize}
  \item[I.] A comparison of the total \zf\ and \ZSHAPE\ cross sections
        with an $s'$ cut and no angular acceptance cut.
  \item[II.] Comparisons of \zf\ and \ALIBABA\ muon
        pair production cross section
        and forward--backward asymmetry with $E_f^{\min}$, $\xi^{\max}$
        and
        angular acceptance cuts.
  \item[III.] As above, except that
  here the comparison is done for Bhabha scattering.
\end{itemize}

\begin{figure}[htbp]
%
%
\epsfysize=20cm
\epsffile{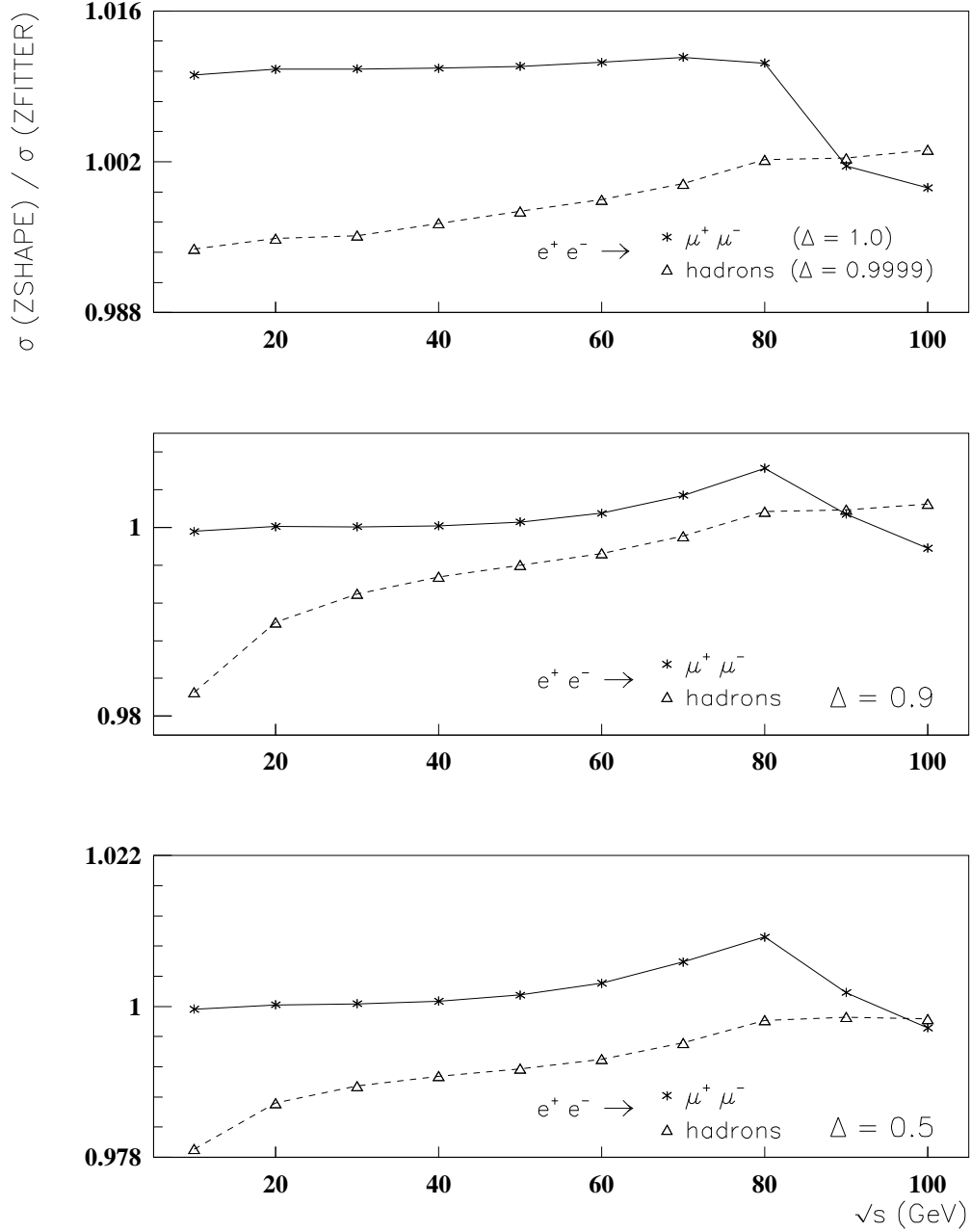}
\caption[
Cross section ratios from \zf\ and \ZSHAPE\
for muon and hadron production;
both programs with their `recommended' choice of flags.
]{\it
Ratio of cross section predictions from \zf\ and \ZSHAPE\
for muon and hadron production, as a function of the centre-of-mass
energy, for three different values of $\Delta = 1 - s'_{\min}/s$;
both programs with their `recommended' choice of flags.
}
\label{??1}
\end{figure}

For case I, we have compared quark and muon cross sections over a
large energy range, [10-100] GeV.
Due to \ZSHAPE\ limitations, only cross-section comparisons up to
\LEPI\ energies can be performed.
In fig.~\ref{??1}, we show the ratio of hadron and muon cross sections
as a function of the centre-of-mass energy for different values of $s'$.
As can be seen in the figure, the agreement at \LEPI\ energies is
excellent even though some higher-order weak corrections are not realized
in \ZSHAPE\footnote{
Very recently, a new version of \ZSHAPE\ was developed,
which now also contains higher-order electroweak corrections connected
with the t~quark.
The agreement of the program with \zf, with the corresponding
flag settings, is not worse than shown here~\cite{srwb}.}.
As the energy decreases from \LEPI, deviations
begin to appear, which reach
2\% in magnitude for the hadronic cross section.
For muons, the deviation goes to 1\% when $s'$ is at the kinematic
limit and 0.5\% otherwise, approaching in the latter case nearly exact
agreement at small energies where pure QED dominates.

\begin{figure}[htbp]
%
%
\epsfysize=20cm
\epsffile{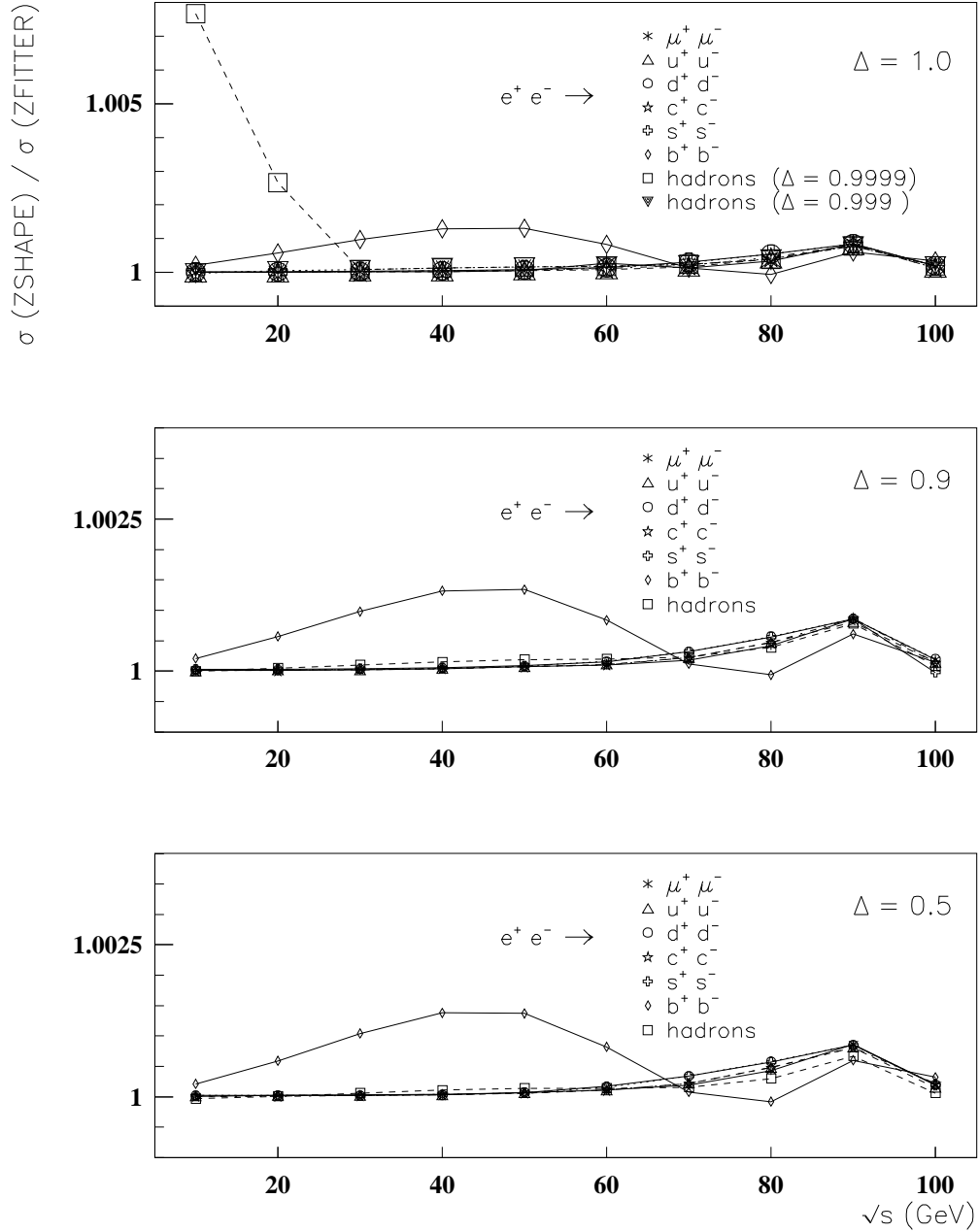}
\caption[
Cross section ratios from \zf\ and \ZSHAPE\
for muon and quark production;
flags are chosen such that
the theoretical assumptions of both programs are as similar as possible.
]{ \it
Ratio of cross-section predictions from \zf\ and \ZSHAPE\
for muon and quark production, as a function of the centre-of-mass
energy, as in fig.~\ref{??1}, but here flags are chosen such that
the theoretical assumptions of both programs are as similar as possible.
}
\label{2??}
\end{figure}

A considerable improvement in the agreement of these two programs can
be realized
through a judicious choice of \zf\ flags (see table~\ref{tbest}).
If various enhancements to \zf, which have been realized since the
1989 workshop, are inhibited, then a
dramatic improvement in the agreement
of the two programs is observed.
These enhancements are mainly concerned with QCD corrections, higher
order weak corrections and the handling of light-quark thresholds.
As can be seen from fig.~\ref{2??}, the disagreement shrinks to 0.1\%
for both muon and hadron production cross sections and for different
cuts.

\begin{figure}[htbp]
%
\epsfysize=9.cm
\epsffile{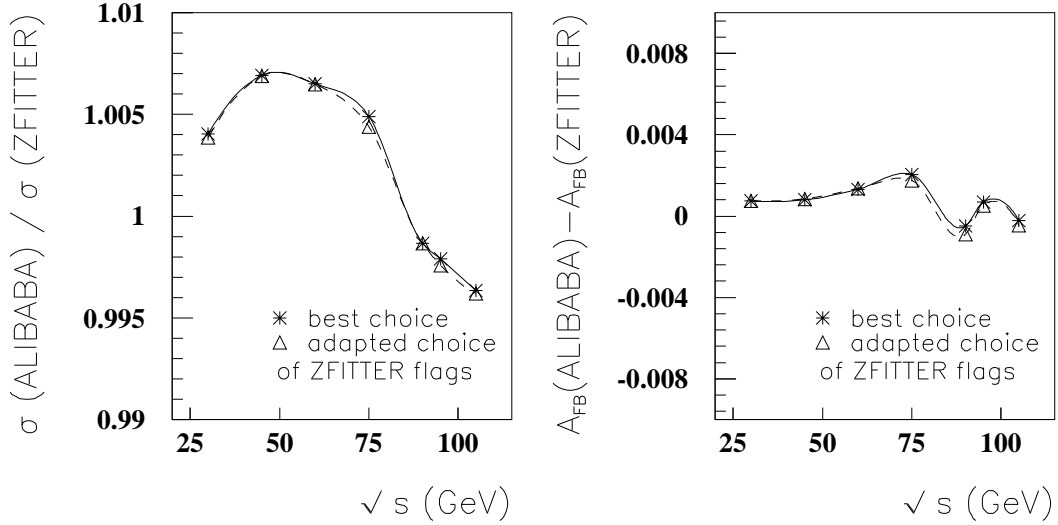}
\caption[
Comparisons of $\sigma^{\mu}$ and
$A_{FB}^{\mu}$
as predicted by \zf\ and \ALIBABA.
]{ \it
Ratio of cross sections, $\sigma^{\mu}$, and
difference of forward--backward asymmetries, $A_{FB}^{\mu}$,
as predicted by \zf\ and \ALIBABA,
as a function of the centre-of-mass energy.
An acceptance cut of
44$^{\circ} \leq$
$\vartheta$
$\leq$ 136$^{\circ}$,
an acollinearity cut of
$\xi \leq 25^{\circ}$
and a muon energy cut
of $E_f^{\min} = \linebreak[2] 5$ GeV have been employed.
The comparison is made for both the `recommended' \zf\ flag values
and those listed in table~\ref{tbest}.
}
\label{??3}
\end{figure}

\begin{figure}[htbp]
%
\epsfysize=9.cm
\epsffile{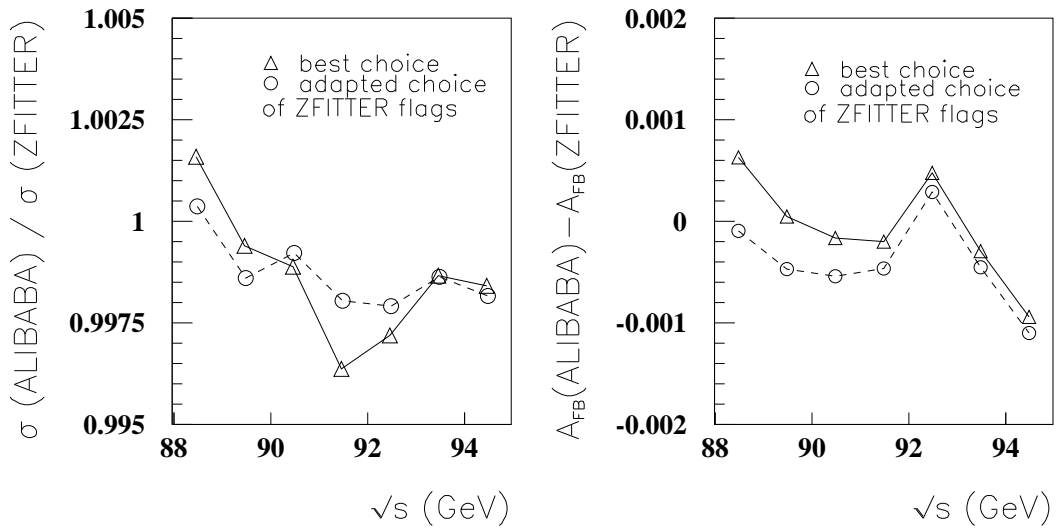}
\caption[
Same as in foregoing figure, at \LEPI\ energies.
]{ \it
Same as in fig.~\ref{??3}, for \LEPI\ energies explicitly.
}
\label{4??}
\end{figure}

In case II, we compare the predictions of \zf\ with \ALIBABA\ for the
muon production cross section and forward--backward asymmetry with cuts
on minimum fermion energy ($E_f^{\min}$) and acollinearity
($\xi^{\max}$).
We restrict this comparison to muons since \ALIBABA\ has no
hadron option.
In figs.~\ref{??3} and~\ref{4??}, we contrast the predictions for
these programs using the `recomended' \zf\ flags and another set of
flags (table~\ref{tbest}) chosen to minimize the differences in the
calculations performed by these programs.
The differences in the predictions, for the value of $m_t$ chosen, is
minor in both cases.
For the `recommended' flags, over the large energy interval covered
in fig.~\ref{??3}, the cross sections agree to within 0.7\%;
at \LEPI\ energies the agreement is good to within 0.2\%.
For the asymmetry, the difference of the two predictions is smaller than
0.2\% over the full energy range and well within 0.1\% at \LEPI\
energies.

\begin{figure}[thbp]
\epsfysize=98mm
\epsffile{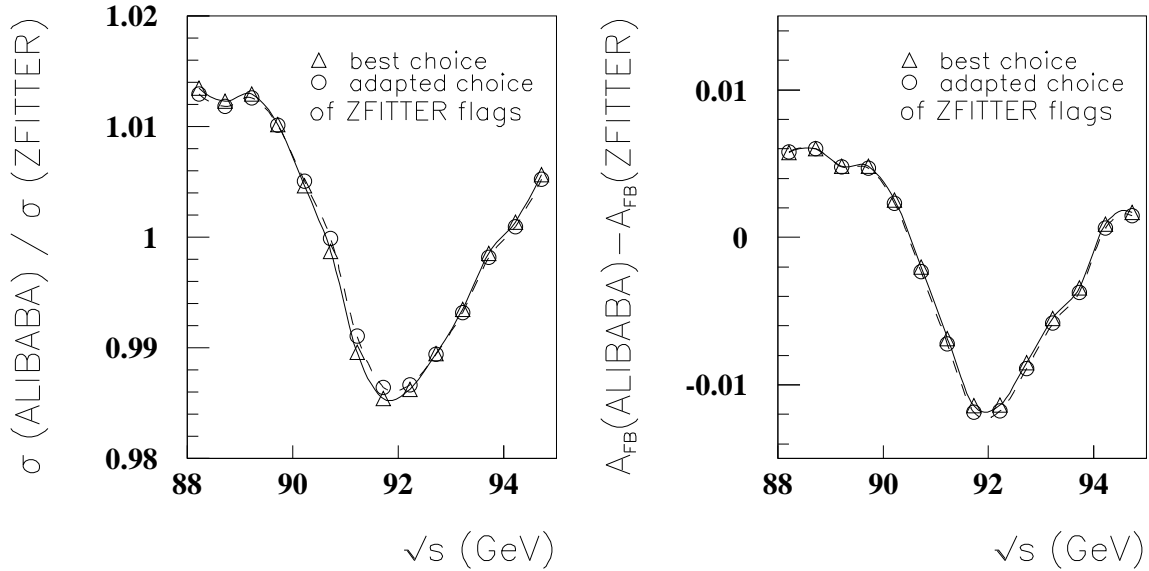}
\caption[
Same as in foregoing figure, for Bhabha scattering
at \LEPI\ energies.
]{\it
Same as in fig.~\ref{??3}, for Bhabha scattering
at \LEPI\ energies.
}
\label{5??}
\end{figure}

For case III, we compare in fig.~\ref{5??} the cross-section ratio and
forward--backward asymmetry for Bhabha scattering for \zf\ (via \BHANG)
and \ALIBABA\ with the same cuts as described above for case II.
Since \BHANG\ contains several approximations adapted to applications
at \LEPI, we restrict the energy range of the comparison
correspondingly.
As may be seen from the figure, the programs agree to within 1.5\%
for the cross-section ratio and within 1\% for the asymmetry difference.
\clearpage
\clearpage
\subsection
[Conclusions]
{Conclusions
\label{conclusions}}

There exists
a wealth of programs, of the semi-analytical and Monte Carlo variety,
 which can make predictions for the fermion pair production
process in $e^+e^-$ collisions.
The programs are varied in both their theoretical accuracy (\IE\ the
order to which the calculations are performed) and in the cuts that one
may apply.
\zf\ is a semi-analytic program with large inherent flexibility in
both these respects.
With a judicious selection of \zf\ flags, agreement with other programs
to the level of 0.5\% and better have been reached with the exception of
Bhabha scattering where the agreement is slightly worse.
\section*{Acknowledgements}
We thank S.~Kirsch for providing us with the \zf\ logo.
We are grateful to R.~Barbieri,
P.~Ch.~Christova, D.~Haidt, W.~Hollik, B.~Kniehl,
L.~Maiani, K.~Mazumdar for fruitful discussions and hints.
For allowing us to use unpublished programs,
thanks to W.~Hollik.

\vfill
\noindent
{\small
{\bf Note added in proof:}
\\
Recently, a new package has been published, which allows to
calculate the one-loop electroweak radiative corrections to two-fermion
production near the $Z$ resonance~\cite{stukn}. The test run output
for the unpolarized muon production cross section shown
in Sample~2 (p.~62 of~\cite{stukn}) reproduces the effective Born
approximation of \zf\ within 0.01\% at the resonance, and within
0.02\% in the full energy range covered by Sample~2.
The corresponding b-quark production cross section shown in Sample~3
does not agree with the \zf\ results.
} 

\clearpage


\newpage
\appendix
\section
[Common Blocks]
{Common Blocks
\label{app:cb}
}
%
%
\setcounter{equation}{0}
%

\subsection
[\zf\ Common Blocks]
{\zf\ Common Blocks
\label{zupars}}

\zf\ common blocks of potential interest to
the user are documented here.

\begin{verbatim}
      COMMON /ZUPARS/QDF,QCDCOR,QCDCOB,ALPHST,SIN2TW,S2TEFF(0:11),
     & WIDTHS(0:11)
\end{verbatim}

The common block /ZUPARS/ contains some \zf\ parameters:
\smallskip
\begin{description}
  \item[\tt QDF] is the final-state radiation factor $\frac{3}
  {4}\frac{\alpha}{\pi}$ introduced in (\ref{efast}) and (\ref{defrqed}).
  \item[\tt QCDCOR] is a QCD correction for all final quark states
  except \BB\ defined in (\ref{qcdcor}).
  \item[\tt QCDCOB] is a QCD correction for \BB\ final states
  defined in (\ref{qcdcob}).
  \item[\tt ALPHST] is $\alpha_s(M_Z^2)$ and is calculated
  as defined by flag {\tt ALST}.
  \item[\tt SIN2TW] is \Sw\ as in (\ref{defsw2}).
  \item[\tt S2TEFF] are the values of \SWE\ for each
  fermion channel (see (\ref{sw2f}) and table~\ref{indf}).
  Note that {\tt S2TEFF(10:11)} are not defined.
  \item[\tt WIDTHS] are the partial decay widths (\ref{defzwidth}) of the \Z\
  for fermion channels defined in table~\ref{indf}
  ({\tt WIDTHS(11)} is the total \Z\ width).
\end{description}

\begin{verbatim}
      COMMON/EWFORM/XALLCH(5,4),XFOTF
      COMPLEX*16 XALLCH,XFOTF
\end{verbatim}

Electroweak form factors are stored in the common block /EWFORM/.
\smallskip
\begin{description}
  \item[{\tt XALLCH(I,J)}] contains the form factors $\rho$,
  $\kappa_e$, $\kappa_f$\ and $\kappa_{ef}$ ({\tt J} = 1-4)
  for neutrinos, leptons, u~and c~quarks, d~and s~quarks, and b~quarks
  ({\tt I} = 1-5).
  These have been introduced in (\ref{mzff}).
  \item[{\tt XFOTF}] is $1+\Delta\alpha(s)$ as used in (\ref{dysfa}).
\end{description}

\begin{verbatim}
      COMMON/ZFCHMS/ALLCH(0:11),ALLMS(0:11)
\end{verbatim}

The common block /ZFCHMS/ contains the charges and masses of the fermions
(see table~\ref{indf}).
\smallskip
\begin{description}
  \item[\tt ALLCH] the fermion charges.
  \item[\tt ALLMS] the fermion masses.
\end{description}
Note that {\tt ALLCH(10)} and {\tt ALLMS(10)} are undefined.

We would also like to mention that the variables FAA, FZA, FZZ
which are introduced as DATA in subroutine {\tt EWCOUP}
allow us to switch on/off the $\gamma \gamma$, $\gamma Z$, $ZZ$ parts
of the cross sections, respectively.

\subsection
[DIZET Common Blocks]
{\DIZET\ Common Blocks
\label{dizetc}}

Two {\tt DIZET} internal common blocks of potential interest to the
user are documented here.

\begin{verbatim}
      COMMON/CDZRKZ/ARROFZ(0:10),ARKAFZ(0:10),ARVEFZ(0:10),ARSEFZ(0:10)
\end{verbatim}

Weak form factors, $\rho^Z_f$ and $\kappa^Z_f$, and vector couplings,
$\bar{v}^Z_f$
for partial \Z\ widths (\ref{defzwidth}) as calculated in subroutine
{\tt ZWRATE}.
The indices correspond to those of table~\ref{indf}.

\begin{description}
  \item[\tt ARROFZ] $\rho^Z_f$, introduced in~(\ref{defzwidth}).
  \item[\tt ARKAFZ] $\kappa^Z_f$, introduced in~(\ref{defzwidth}).
  \item[\tt ARVEFZ] $\bar{v}_f^Z$ as defined in~(\ref{vectcoup}).
  \item[\tt ARSEFZ] Effective weak mixing angles $s_W^{2,f}$
   as in~(\ref{sw2f}).
\end{description}
Note that the 10th element of these arrays is undefined.

\begin{verbatim}
      COMMON/CDZXKF/XROKF
\end{verbatim}

This variable is the ratio of two different definitions of the
weak mixing angle as defined in~(\ref{sw2ms}):
\ba
\Re e  \, \mbox{\tt XROKF}  =  \frac {
\sin^2 \theta_W^{\overline{\mathrm{MS}}}
} {\sin^2 \theta_W}.  \nonumber
\label{xrokf}
\ea
%
%
%
%


\section
[Subroutine ZFTEST]
{Subroutine {\tt ZFTEST}
\label{zftest} }
%
%
\setcounter{equation}{0}
%

The \zf\ distribution package includes subroutine {\tt ZFTEST} which
serves essentially three purposes:
\begin{itemize}
  \item[1.] It is an example of how to use \zf.
  \item[2.] It is an internal consistency check of the different \zf\
   branches.
  \item[3.] It allows one to check that \zf\ has been properly installed on
   the machine.
\end{itemize}
The routine creates a table of cross sections and asymmetries as
a function of \RS\ near the \Z\ peak.

To run {\tt ZFTEST} the user needs to create the following main
program:

\begin{verbatim}
      PROGRAM ZFMAIN
      CALL ZFTEST
      END
\end{verbatim}

After compiling and linking it to
{\tt ZFITR4\_5}, \DIZET\ and \BHANG\ the results
presented in appendix~\ref{results} should be obtained.
The corresponding Fortran files may be found at ZFITTER@CERNVM.
\newpage
\subsection
[Subroutine ZFTEST]
{Subroutine {\tt ZFTEST}
\label{szftest}}
\small
\begin{verbatim}
      SUBROUTINE ZFTEST
*     ========== ======
************************************************************************
*
*     SUBR. ZFTEST
*
*     Example program to demonstrate the use of the ZFITTER package.
*
************************************************************************
*
      IMPLICIT REAL*8(A-H,O-Z)
      COMPLEX*16 XVPOL
      DIMENSION XS(0:11,5),AFB(0:11,4),TAUPOL(2),TAUAFB(2)
*
* constants
*
      PARAMETER(GMU=1.166388D-5,ALFAI=137.0359895D0,ALFA=1.D0/ALFAI,
     +          CONS=1.D0)
      PARAMETER(ZMASS=91.175D0,TMASS=140.D0,HMASS=300.D0)
      PARAMETER(AME=0.511D-3,AMU=0.106D0,AMT=1.784D0,ALFAS=.120D0)
      PARAMETER(RSMN=87.D0,DRS=1.D0,NRS=9)
      PARAMETER(ANG0=35D0,ANG1=145D0)
      PARAMETER(QE=-1.D0,AE=-.5D0,QU= 2.D0/3.D0,AU= .5D0,
     +                            QD=-1.D0/3.D0,AD=-.5D0)
*
* ZFITTER common blocks
*
      COMMON /ZUPARS/QDF,QCDCOR,QCDCOB,ALPHST,SIN2TW,S2TEFF(0:11),
     & WIDTHS(0:11)
      COMMON /CDZRKZ/ARROFZ(0:10),ARKAFZ(0:10),ARVEFZ(0:10),ARSEFZ(0:10)
      COMMON /EWFORM/XALLCH(5,4),XFOTF
      COMPLEX*16 XALLCH,XFOTF
*
*-----------------------------------------------------------------------
*
* initialize
*
      CALL ZUINIT
*
* set ZFITTER flags and print flag values
*
      CALL ZUFLAG('PRNT',1)
      CALL ZUINFO(0)
*
* do weak sector calculations
*
      CALL ZUWEAK(ZMASS,TMASS,HMASS,ALFAS)
*
* define cuts for fermion channels and print cut values
*
      CALL ZUCUTS( 1,0,15.D0,20.D0,0.D0,ANG0,ANG1)
      CALL ZUCUTS( 2,0,15.D0,20.D0,0.D0,ANG0,ANG1)
      CALL ZUCUTS( 3,0,15.D0,20.D0,0.D0,ANG0,ANG1)
      CALL ZUCUTS(11,0,15.D0,20.D0,0.D0,ANG0,ANG1)
      CALL ZUINFO(1)
*
* make table of cross sections and asymmetries
*
      PI   = DACOS(-1.D0)
      GAMZ = WIDTHS(11)/1000.
      GAME = WIDTHS( 1)/1000.
      GAE  = SQRT(ARROFZ(1))/2.
      GVE  = ARVEFZ(1)*GAE
      DO I = 1,NRS
        RS = RSMN+REAL(I-1)*DRS
* table header
        PRINT *,' SQRT(S) = ',REAL(RS)
        PRINT *,'      <----------- Cross Section ---------->',
     +   '  <----- Asymmetry ---->','  <---TauPol--->'
        PRINT *,'INDF  ZUTHSM  ZUXSEC   ZUXSA  ZUXSA2  ZUSMAT',
     +   '  ZUTHSM   ZUXSA  ZUXSA2  ZUTPSM   ZUTAU'
* loop over fermion indices
        DO INDF = 0,11
         S=RS**2
* standard model interf. (INTRF=1)
         CALL ZUTHSM(INDF,RS,ZMASS,TMASS,HMASS,ALFAS,
     +     XS(INDF,1),AFB(INDF,1))
         IF(INDF.EQ.3) CALL ZUTPSM(RS,ZMASS,TMASS,HMASS,ALFAS,
     +     TAUPOL(1),TAUAFB(1))
* cross section interf. (INTRF=2)
           GAMF = WIDTHS(INDF)/1000.
         IF(INDF.EQ.11) GAMF = WIDTHS( 1)/1000.
           CALL ZUXSEC(INDF,RS,ZMASS,GAMZ,GAME,GAMF,XS(INDF,2))
* cross section & forward--backward asymmetry interf. (INTRF=3)
         IF(INDF.NE.0 .AND. INDF.NE.10) THEN
           GAF = SQRT(ARROFZ(INDF))/2.
           GVF = ARVEFZ(INDF)*GAF
           IF(INDF.EQ.11) THEN
             GAF = SQRT(ARROFZ(1))/2.
             GVF = ARVEFZ(1)*GAF
           ENDIF
           CALL ZUXSA(INDF,RS,ZMASS,GAMZ,0,GVE,GAE,GVF,GAF,
     +     XS(INDF,3),AFB(INDF,3))
         ENDIF
* tau polarization interf. (INTRF=3)
         IF(INDF.EQ.3) CALL ZUTAU(RS,ZMASS,GAMZ,0,GVE,GAE,GVF,GAF,
     +     TAUPOL(2),TAUAFB(2))
* cross section & forward--backward asymmetry interf. for gv**2 and
* ga**2 (IBRA=4)
         IF((INDF.GE.1 .AND. INDF.LE.3) .OR. INDF.EQ.11) THEN
           GVF2 = GVF**2
           GAF2 = GAF**2
           CALL ZUXSA2(INDF,RS,ZMASS,GAMZ,0,GVF2,GAF2,
     +     XS(INDF,4),AFB(INDF,4))
         ENDIF
* S-matrix interf. (INTRF=5)
* Parameters are fitted with code FITSMA FORTRAN.
         IF(INDF.EQ.2 .OR. INDF.EQ.10) THEN
           IF(INDF.EQ.2) THEN
             AMZS =91.14132D0
             GAMZS= 2.48354D0
             RR=0.14159D0
             RI=0.15092D0
             RG=1.13310D0
               ELSE
             AMZS =91.14126D0
             GAMZS= 2.48406D0
             RR=2.94028D0
             RI=3.15125D0
             RG=2.81462D0
           ENDIF
C The parameters AMZS, GAMZS, RR, RI, RG correspond to:
C           AMZS   = ZMASS-GAMZ**2/2.D0/ZMASS
C           GAMZS  = GAMZ -GAMZ**3/2.D0/ZMASS**2
C           VE     =  -.5D0+2.D0*SIN2TW
C           VU     =  0.5D0-4.D0/3.D0*SIN2TW
C           VD     = -0.5D0+2.D0/3.D0*SIN2TW
C           AKAPPA = GMU*AMZS1*AMZS1/(SQRT(2.D0)*2.D0*PI*ALFA)
C           XVPOL  = 1.D0/(2.D0-XFOTF)
C           IF(INDF.LE.3) THEN
C             RZ = CONS*AKAPPA**2*(AE**2+VE**2)**2*(1.D0+.75D0*ALFA/PI)
C             SZ = CONS*AKAPPA*VE**2*(1.D0+.75D0*ALFA/PI)
C             RG = CONS*CDABS(XVPOL)**2*(1.D0+.75D0*ALFA/PI)
C           ELSE
C             RZ = CONS*AKAPPA**2*(AE**2+VE**2)*3.D0*
C    +         (2.D0*(AU**2+VU**2)+3.D0*(AD**2+VD**2))
C             SZ = CONS*AKAPPA*3.D0*QE*VE*(2.D0*VU*QU+3.D0*VD*QD)
C             RG = CONS*CDABS(XVPOL)**2*3.D0*11.D0/9.D0
C           ENDIF
C           RR = RZ
C           RI = RZ+2.D0*SZ*DREAL(XVPOL)
c
           R1 = 0d0
           R2 = 0d0
           R3 = 0d0
         CALL ZUSMAT(INDF,RS,AMZS,GAMZS,RR,RI,R0,R1,R2,RG,XS(INDF,5))
         ENDIF
* results
         IF(INDF.EQ.0) THEN
           PRINT 9000,INDF,(XS(INDF,J),J=1,2)
         ELSEIF(INDF.EQ.1 .OR. INDF.EQ.11) THEN
           PRINT 9010,INDF,(XS(INDF,J),J=1,4),AFB(INDF,1),
     +      (AFB(INDF,J),J=3,4)
         ELSEIF(INDF.EQ.2) THEN
           PRINT 9005,INDF,(XS(INDF,J),J=1,5),AFB(INDF,1),
     +      (AFB(INDF,J),J=3,4)
         ELSEIF(INDF.EQ.3) THEN
           PRINT 9015,INDF,(XS(INDF,J),J=1,4),AFB(INDF,1),
     +      (AFB(INDF,J),J=3,4),(TAUPOL(J),J=1,2)
         ELSEIF(INDF.EQ.10) THEN
           PRINT 9025,INDF,(XS(INDF,J),J=1,2),XS(INDF,5)
         ELSE
           PRINT 9020,INDF,(XS(INDF,J),J=1,3),AFB(INDF,1),
     +      AFB(INDF,3)
          ENDIF
        ENDDO
        PRINT *
      ENDDO
      RETURN
 9000 FORMAT(1X,I4,2F8.4)
 9005 FORMAT(1X,I4,9F8.4)
 9010 FORMAT(1X,I4,4F8.4,8X,3F8.4)
 9015 FORMAT(1X,I4,4F8.4,8X,5F8.4)
 9020 FORMAT(1X,I4,3F8.4,16X,2F8.4)
 9025 FORMAT(1X,I4,2F8.4,16X,F8.4)
*                                                             END ZFTEST
      END
\end{verbatim}
\normalsize

\newpage
\subsection
[ZFTEST Results]
{{\tt ZFTEST} Results
\label{results}}

\footnotesize
\begin{verbatim}
 ******************************************************
 ******************************************************
 **           This is ZFITTER version 4.5            **
 **                   92/04/19                       **
 ******************************************************
 ** The authors of the ZFITTER package are:          **
 **                                                  **
 **   D.Bardin      (Dubna)                          **
 **   M.Bilenky     (Dubna)                          **
 **   A.Chizhov     (Dubna)                          **
 **   A.Olshevsky   (Dubna)                          **
 **   S.Riemann     (Zeuthen)                        **
 **   T.Riemann     (Zeuthen)                        **
 **   M.Sachwitz    (Zeuthen)                        **
 **   A.Sazonov     (Dubna)                          **
 **   Yu.Sedykh     (Dubna)                          **
 **   I.Sheer       (UC San Diego)                   **
 **                                                  **
 ******************************************************
 ** Questions and comments to ZFITTER@CERNVM.CERN.CH **
 ******************************************************

 ZUINIT> ZFITTER defaults:

 ZFITTER flag values:
 AFBC: 1 ALPH: 0 ALST: 1 AMT4: 3 BORN: 0
 BOXD: 0 CONV: 0 FINR: 1 FOT2: 1 GAMS: 1
 INCL: 1 INTF: 1 DUMY: 0 PART: 0 POWR: 1
 PRNT: 0 QCD3: 1 QCDC: 1 VPOL: 3 WEAK: 1


 ZFITTER cut values:
   INDF  ICUT    ACOL      EMIN      S_PR    ANG0    ANG1
      0    -1    0.00    0.0000    0.0000    0.00  180.00
      1    -1    0.00    0.0000    0.0000    0.00  180.00
      2    -1    0.00    0.0000    0.0000    0.00  180.00
      3    -1    0.00    0.0000    0.0000    0.00  180.00
      4    -1    0.00    0.0000    0.0000    0.00  180.00
      5    -1    0.00    0.0000    0.0000    0.00  180.00
      6    -1    0.00    0.0000    0.0000    0.00  180.00
      7    -1    0.00    0.0000    0.0000    0.00  180.00
      8    -1    0.00    0.0000    0.0000    0.00  180.00
      9    -1    0.00    0.0000    0.0000    0.00  180.00
     10    -1    0.00    0.0000    0.0000    0.00  180.00
     11    -1    0.00    0.0000    0.0000    0.00  180.00


 ZFITTER flag values:
 AFBC: 1 ALPH: 0 ALST: 1 AMT4: 3 BORN: 0
 BOXD: 0 CONV: 0 FINR: 1 FOT2: 1 GAMS: 1
 INCL: 1 INTF: 1 DUMY: 0 PART: 0 POWR: 1
 PRNT: 1 QCD3: 1 QCDC: 1 VPOL: 3 WEAK: 1
\end{verbatim}
\newpage
\begin{verbatim}
  ZMASS =   91.17500;  TMASS =  140.00000
  HMASS =  300.00000;  ALFAS =    0.12000
 ALPHST =    0.12000; SIN2TW =    0.22817
 QCDCOR =    1.03954; QCDCOB =    1.04020

 CHANNEL        WIDTH
 -------        -----
 nu,nubar       166.6
 e+,e-           83.6
 mu+,mu-         83.6
 tau+,tau-       83.4
 u,ubar         296.6
 d,dbar         382.9
 c,cbar         296.2
 s,sbar         382.9
 t,tbar           0.0
 b,bbar         375.7
 hadron        1734.2
 total         2484.7

 ZFITTER cut values:
   INDF  ICUT    ACOL      EMIN      S_PR    ANG0    ANG1
      0    -1    0.00    0.0000    0.0000    0.00  180.00
      1     0   15.00   20.0000    0.0000   35.00  145.00
      2     0   15.00   20.0000    0.0000   35.00  145.00
      3     0   15.00   20.0000    0.0000   35.00  145.00
      4    -1    0.00    0.0000    0.0000    0.00  180.00
      5    -1    0.00    0.0000    0.0000    0.00  180.00
      6    -1    0.00    0.0000    0.0000    0.00  180.00
      7    -1    0.00    0.0000    0.0000    0.00  180.00
      8    -1    0.00    0.0000    0.0000    0.00  180.00
      9    -1    0.00    0.0000    0.0000    0.00  180.00
     10    -1    0.00    0.0000    0.0000    0.00  180.00
     11     0   15.00   20.0000    0.0000   35.00  145.00

\end{verbatim}

\newpage

\begin{verbatim}
   SQRT(S) =   87.0000000
       <----------- Cross Section ---------->  <----- Asymmetry ---->
 INDF  ZUTHSM  ZUXSEC   ZUXSA  ZUXSA2  ZUSMAT  ZUTHSM   ZUXSA  ZUXSA2
    0  0.2363  0.2364
    1  0.0935  0.0935  0.0935  0.0935         -0.3521 -0.3521 -0.3521
    2  0.0954  0.0954  0.0954  0.0954  0.0954 -0.3522 -0.3521 -0.3521
    3  0.0961  0.0961  0.0961  0.0961         -0.3525 -0.3524 -0.3524
    4  0.4504  0.4504  0.4505                 -0.1802 -0.1802
    5  0.5480  0.5480  0.5481                 -0.0097 -0.0097
    6  0.4414  0.4414  0.4415                 -0.1837 -0.1837
    7  0.5472  0.5472  0.5473                 -0.0097 -0.0097
    8  0.0000  0.0000  0.0000                  0.0000  0.0000
    9  0.5346  0.5346  0.5347                 -0.0103 -0.0103
   10  2.5217  2.5217                  2.5217
   11  0.4405  0.4404  0.4404  0.4406          0.6259  0.6258  0.6258
       <---TauPol--->
 INDF  ZUTPSM   ZUTAU
    3  -0.0825  -0.0826

\end{verbatim}
\begin{verbatim}
 SQRT(S) =   88.0000000
       <----------- Cross Section ---------->  <----- Asymmetry ---->
 INDF  ZUTHSM  ZUXSEC   ZUXSA  ZUXSA2  ZUSMAT  ZUTHSM   ZUXSA  ZUXSA2
    0  0.3826  0.3827
    1  0.1461  0.1461  0.1461  0.1461         -0.2723 -0.2722 -0.2722
    2  0.1489  0.1489  0.1490  0.1490  0.1489 -0.2724 -0.2723 -0.2723
    3  0.1500  0.1500  0.1501  0.1501         -0.2727 -0.2726 -0.2726
    4  0.7092  0.7091  0.7093                 -0.1241 -0.1240
    5  0.8831  0.8831  0.8832                  0.0149  0.0149
    6  0.7000  0.6999  0.7001                 -0.1255 -0.1255
    7  0.8824  0.8823  0.8825                  0.0149  0.0149
    8  0.0000  0.0000  0.0000                  0.0000  0.0000
    9  0.8635  0.8633  0.8635                  0.0145  0.0145
   10  4.0381  4.0378                  4.0381
   11  0.5124  0.5122  0.5122  0.5124          0.5569  0.5568  0.5569
       <---TauPol--->
 INDF  ZUTPSM   ZUTAU
    3  -0.0951  -0.0952

\end{verbatim}
\newpage
\begin{verbatim}

  SQRT(S) =   89.0000000
       <----------- Cross Section ---------->  <----- Asymmetry ---->
 INDF  ZUTHSM  ZUXSEC   ZUXSA  ZUXSA2  ZUSMAT  ZUTHSM   ZUXSA  ZUXSA2
    0  0.7007  0.7007
    1  0.2609  0.2608  0.2609  0.2609         -0.1870 -0.1868 -0.1868
    2  0.2658  0.2658  0.2659  0.2659  0.2658 -0.1870 -0.1869 -0.1869
    3  0.2677  0.2677  0.2678  0.2678         -0.1873 -0.1871 -0.1871
    4  1.2733  1.2731  1.2734                 -0.0662 -0.0661
    5  1.6126  1.6124  1.6127                  0.0388  0.0388
    6  1.2634  1.2633  1.2635                 -0.0666 -0.0666
    7  1.6118  1.6117  1.6120                  0.0388  0.0388
    8  0.0000  0.0000  0.0000                  0.0000  0.0000
    9  1.5792  1.5789  1.5792                  0.0385  0.0385
   10  7.3403  7.3394                  7.3402
   11  0.6562  0.6558  0.6561  0.6562          0.4592  0.4591  0.4593
       <---TauPol--->
 INDF  ZUTPSM   ZUTAU
    3  -0.1071  -0.1072


  SQRT(S) =   90.0000000
       <----------- Cross Section ---------->  <----- Asymmetry ---->
 INDF  ZUTHSM  ZUXSEC   ZUXSA  ZUXSA2  ZUSMAT  ZUTHSM   ZUXSA  ZUXSA2
    0  1.4868  1.4867
    1  0.5452  0.5451  0.5453  0.5453         -0.0993 -0.0992 -0.0992
    2  0.5555  0.5553  0.5555  0.5555  0.5554 -0.0994 -0.0992 -0.0992
    3  0.5593  0.5592  0.5593  0.5593         -0.0996 -0.0994 -0.0994
    4  2.6703  2.6698  2.6704                 -0.0086 -0.0085
    5  3.4173  3.4169  3.4175                  0.0617  0.0617
    6  2.6586  2.6581  2.6587                 -0.0086 -0.0085
    7  3.4166  3.4161  3.4167                  0.0617  0.0617
    8  0.0000  0.0000  0.0000                  0.0000  0.0000
    9  3.3503  3.3495  3.3501                  0.0614  0.0615
   10 15.5131 15.5104                 15.5119
   11  0.9678  0.9669  0.9677  0.9679          0.3302  0.3301  0.3303
       <---TauPol--->
 INDF  ZUTPSM   ZUTAU
    3  -0.1181  -0.1182

\end{verbatim}
\newpage
\begin{verbatim}
 SQRT(S) =   91.0000000
       <----------- Cross Section ---------->  <----- Asymmetry ---->
 INDF  ZUTHSM  ZUXSEC   ZUXSA  ZUXSA2  ZUSMAT  ZUTHSM   ZUXSA  ZUXSA2
    0  2.8218  2.8214
    1  1.0295  1.0292  1.0295  1.0295         -0.0155 -0.0153 -0.0153
    2  1.0484  1.0481  1.0485  1.0485  1.0484 -0.0155 -0.0153 -0.0153
    3  1.0556  1.0552  1.0556  1.0556         -0.0156 -0.0154 -0.0154
    4  5.0483  5.0472  5.0483                  0.0453  0.0454
    5  6.4869  6.4858  6.4870                  0.0826  0.0827
    6  5.0335  5.0323  5.0335                  0.0454  0.0455
    7  6.4862  6.4850  6.4862                  0.0826  0.0827
    8  0.0000  0.0000  0.0000                  0.0000  0.0000
    9  6.3635  6.3617  6.3629                  0.0825  0.0825
   10 29.4183 29.4120                 29.4150
   11  1.3601  1.3581  1.3600  1.3601          0.2053  0.2052  0.2054
       <---TauPol--->
 INDF  ZUTPSM   ZUTAU
    3  -0.1274  -0.1275


 SQRT(S) =   92.0000000
       <----------- Cross Section ---------->  <----- Asymmetry ---->
 INDF  ZUTHSM  ZUXSEC   ZUXSA  ZUXSA2  ZUSMAT  ZUTHSM   ZUXSA  ZUXSA2
    0  2.3301  2.3295
    1  0.8526  0.8523  0.8525  0.8525          0.0530  0.0532  0.0532
    2  0.8681  0.8677  0.8680  0.8680  0.8681  0.0529  0.0531  0.0531
    3  0.8739  0.8735  0.8738  0.8738          0.0529  0.0531  0.0531
    4  4.1794  4.1783  4.1793                  0.0884  0.0885
    5  5.3632  5.3620  5.3630                  0.0993  0.0994
    6  4.1661  4.1650  4.1660                  0.0886  0.0887
    7  5.3624  5.3612  5.3622                  0.0994  0.0994
    8  0.0000  0.0000  0.0000                  0.0000  0.0000
    9  5.2618  5.2601  5.2611                  0.0993  0.0994
   10 24.3329 24.3266                 24.3329
   11  0.9720  0.9701  0.9719  0.9719          0.1624  0.1624  0.1626
       <---TauPol--->
 INDF  ZUTPSM   ZUTAU
    3  -0.1341  -0.1341

\end{verbatim}
\newpage
\begin{verbatim}

 SQRT(S) =   93.0000000
       <----------- Cross Section ---------->  <----- Asymmetry ---->
 INDF  ZUTHSM  ZUXSEC   ZUXSA  ZUXSA2  ZUSMAT  ZUTHSM   ZUXSA  ZUXSA2
    0  1.3333  1.3329
    1  0.4914  0.4912  0.4913  0.4913          0.1013  0.1015  0.1015
    2  0.5002  0.5000  0.5001  0.5001  0.5002  0.1013  0.1015  0.1015
    3  0.5035  0.5033  0.5035  0.5035          0.1013  0.1015  0.1015
    4  2.4063  2.4056  2.4062                  0.1182  0.1183
    5  3.0742  3.0734  3.0740                  0.1110  0.1111
    6  2.3957  2.3950  2.3956                  0.1186  0.1187
    7  3.0735  3.0727  3.0733                  0.1111  0.1111
    8  0.0000  0.0000  0.0000                  0.0000  0.0000
    9  3.0156  3.0145  3.0151                  0.1111  0.1112
   10 13.9653 13.9612                 13.9658
   11  0.5607  0.5595  0.5606  0.5606          0.2113  0.2113  0.2115
       <---TauPol--->
 INDF  ZUTPSM   ZUTAU
    3  -0.1381  -0.1380


  SQRT(S) =   94.0000000
       <----------- Cross Section ---------->  <----- Asymmetry ---->
 INDF  ZUTHSM  ZUXSEC   ZUXSA  ZUXSA2  ZUSMAT  ZUTHSM   ZUXSA  ZUXSA2
    0  0.8343  0.8340
    1  0.3102  0.3100  0.3101  0.3101          0.1358  0.1360  0.1360
    2  0.3157  0.3156  0.3157  0.3157  0.3157  0.1357  0.1359  0.1359
    3  0.3178  0.3177  0.3178  0.3178          0.1357  0.1359  0.1359
    4  1.5173  1.5168  1.5172                  0.1390  0.1391
    5  1.9272  1.9267  1.9270                  0.1193  0.1194
    6  1.5081  1.5076  1.5080                  0.1397  0.1398
    7  1.9266  1.9260  1.9264                  0.1194  0.1194
    8  0.0000  0.0000  0.0000                  0.0000  0.0000
    9  1.8899  1.8892  1.8895                  0.1195  0.1196
   10  8.7692  8.7663                  8.7692
   11  0.3862  0.3854  0.3861  0.3862          0.3006  0.3006  0.3008
       <---TauPol--->
 INDF  ZUTPSM   ZUTAU
    3  -0.1404  -0.1403
\end{verbatim}
\newpage
\begin{verbatim}


  SQRT(S) =   95.0000000
       <----------- Cross Section ---------->  <----- Asymmetry ---->
 INDF  ZUTHSM  ZUXSEC   ZUXSA  ZUXSA2  ZUSMAT  ZUTHSM   ZUXSA  ZUXSA2
    0  0.5808  0.5805
    1  0.2180  0.2178  0.2179  0.2179          0.1617  0.1619  0.1619
    2  0.2218  0.2217  0.2218  0.2218  0.2219  0.1615  0.1617  0.1617
    3  0.2233  0.2232  0.2233  0.2233          0.1615  0.1617  0.1617
    4  1.0652  1.0648  1.0650                  0.1542  0.1542
    5  1.3442  1.3438  1.3441                  0.1255  0.1256
    6  1.0567  1.0563  1.0566                  0.1552  0.1553
    7  1.3436  1.3432  1.3434                  0.1256  0.1257
    8  0.0000  0.0000  0.0000                  0.0000  0.0000
    9  1.3177  1.3171  1.3174                  0.1258  0.1259
   10  6.1274  6.1252                  6.1271
   11  0.3061  0.3056  0.3060  0.3061          0.3902  0.3902  0.3904
       <---TauPol--->
 INDF  ZUTPSM   ZUTAU
    3  -0.1417  -0.1416
\end{verbatim}
\normalsize


\begin{thebibliography}{99}
\bibitem{alpha4}
   Subroutines {\tt ALPHA4, ALPHA5} are based on~\cite{mar:pr,mrs:pr};
a coding by B. Kniehl is used.
\end{thebibliography}
\end{document}